\documentclass[12pt]{article}
\usepackage{jheppub}

\pdfoutput=1
\usepackage{slashed}
\usepackage{color, verbatim}
\usepackage{latexsym}
\usepackage{epsfig}
\usepackage{amsmath,accents}

\usepackage{amssymb}
\usepackage{graphicx}
\usepackage{bm}

\usepackage[font={small}]{caption}

\usepackage{hyperref}
\usepackage{epstopdf}
\definecolor{myred}{rgb}{0.7, 0, 0}
\definecolor{myblue}{rgb}{0, 0, 0.7}
\definecolor{mygreen}{rgb}{0.04, 0.7, 0.5}
\hypersetup{colorlinks,citecolor=myred,linkcolor=myblue,urlcolor=myblue,linktocpage=true}

\makeatletter

\@addtoreset{equation}{section}
\makeatother

\newcommand{\be}{\begin{equation}}
\newcommand{\ee}{\end{equation}}
\newcommand{\bea}{\begin{eqnarray}}
\newcommand{\eea}{\end{eqnarray}}

\newcommand{\diag}{\operatorname{diag}}

\begin{document}

\thispagestyle{empty}

\begin{center}


\begin{center}

\vspace{.5cm}

{\Large\bf Lepton-flavor universality violation in\\ \vspace{0.3cm}
$R_K$ and $R_{D^{(\ast)}}$ from warped space}

\end{center}

\vspace{1.cm}

\textbf{
Eugenio Meg\'ias$^{\,a}$,  Mariano Quir\'os$^{\,b}$, Lindber Salas$^{\,b}$
}\\

\vspace{.1cm}
${}^a\!\!$ {\em {Departamento de F\'{\i}sica Te\'orica, Universidad del Pa\'{\i}s Vasco UPV/EHU, \\ Apartado 644,  48080 Bilbao, Spain}}

\vspace{.1cm}
${}^b\!\!$ {\em {Institut de F\'{\i}sica d'Altes Energies (IFAE),\\ The Barcelona Institute of  Science and Technology (BIST),\\ Campus UAB, 08193 Bellaterra (Barcelona) Spain}}


\end{center}

\vspace{0.8cm}

\centerline{\bf Abstract}
\vspace{2 mm}
\begin{quote}\small
Some anomalies in the processes $b\to s\ell\ell$ ($\ell=\mu,e$) and $b\to c \ell\bar\nu_\ell$ ($\ell=\tau,\mu,e$), in particular in the observables $R_K$ and $R_{D^{(\ast)}}$, have been found by the BaBar, LHCb and Belle Collaborations, leading to a possible lepton flavor universality violation. If these anomalies were confirmed they would inevitably lead to physics beyond the Standard Model. In this paper we try to accommodate the present anomalies in an extra dimensional theory, solving the naturalness problem of the Standard Model by means of a warped metric with a strong conformality violation near the infra-red brane. The $R_K$ anomaly can be accommodated provided that the left-handed bottom quark and muon lepton have some degree of compositeness in the dual theory. The theory is consistent with all electroweak and flavor observables, and with all direct searches of Kaluza-Klein electroweak gauge bosons and gluons. The fermion spectrum, and fermion mixing angles, can be reproduced by mostly elementary right-handed bottom quarks, and tau and muon leptons. Moreover the $R_{D^{(\ast)}}$ anomaly requires a strong degree of 
compositeness for the left-handed tau leptons, which turns out to be in tension with experimental data on the $g_{\tau_L}^Z$ coupling, possibly unless some degree of fine-tuning is introduced in the fixing of the CKM matrix.

\end{quote}

\vfill

\newpage

\tableofcontents

\newpage
\section{Introduction}
\label{introduction}
While direct signals of new physics seem to be elusive up to now at the Large Hadron Collider (LHC), there exist anomalies showing up at the LHC, mainly by the LHCb Collaboration, as well as at electron collider $B$-factories, in particular by the BaBar and Belle Collaborations at SLAC and KEK, respectively. In the absence of direct experimental signatures of theories restoring the Standard Model naturalness, a legitimate attitude is to figure out which are the natural theories whose \textit{direct detection should be hidden} from the actual experimental conditions, but that can accommodate possible explanations of (part of) the existing anomalies. This one is the point of view we will adopt in this paper.

There are two main ultra-violet (UV) completions of the Standard Model which can restore its naturalness and solve the Higgs hierarchy problem: \textbf{i)} Supersymmetry, where the Higgs mass is protected by a (super)symmetry; and, \textbf{ii)} Extra dimensional theories with a warped extra dimension, by which the Planck scale is warped down to the TeV scale along the extra dimension~\cite{Randall:1999ee}, or its dual, where the Higgs is composite and melts beyond the condensation scale at the TeV. 

In this paper we will use the latter set of theories. In particular we will consider a set of warped theories with a strong deformation of conformality towards the infra-red (IR) brane~\cite{Cabrer:2009we,Cabrer:2010si,Cabrer:2011fb,Cabrer:2011vu,Cabrer:2011mw,Carmona:2011ib,Cabrer:2011qb,Quiros:2013yaa,Megias:2015ory,Megias:2015qqh,Megias:2016jcw}, such that the Standard Model can propagate in the bulk of the fifth dimension, consistently with all measured electroweak observables. The theory is characterized by the superpotential
\be
W(\phi)=6 k(1+e^{a_0\phi})^{b_0}
\label{superp}
\ee
where $a_0$ and $b_0$ are real (dimensionless) parameters which govern the back reaction on the gravitational metric $A(y)$, $\phi$ is the (dimensionless) scalar field stabilizing the fifth dimension and $k$ is a parameter with mass dimension providing the curvature along the fifth dimension. We will not  specify here the details of the five-dimensional (5D) model, as they were widely covered in the literature, Refs.~\cite{Cabrer:2009we,Cabrer:2010si,Cabrer:2011fb,Cabrer:2011vu,Cabrer:2011mw,Carmona:2011ib,Cabrer:2011qb,Quiros:2013yaa,Megias:2015ory,Megias:2015qqh,Megias:2016jcw}, which we refer the reader to~\footnote{For reviews see e.g.~Refs.~\cite{Davoudiasl:2009cd,Quiros:2013yaa}.}. In this paper we will consider the superpotential of Eq.~(\ref{superp}), with the particular values of the parameters
\be
b_0=2\,, \quad a_0=0.15 \,,
\label{metrica}
\ee
although somewhat similar results could equally well be obtained with different values. As we will see, these particular values minimize the impact of Kaluza-Klein (KK) modes in the electroweak observables and thus leave more room to accommodate possible anomalies.

In our model the Standard Model fermions $f_{L,R}$ propagate in the bulk of the extra dimension and their zero mode wave function, as determined by appropriate boundary conditions and the 5D Dirac mass $M_{f_{L,R}}(y)=\mp c_{f_{L,R}} W(\phi)$, depend on the real parameters $c_{L,R}$ which, in turn, determine the degree of compositeness of the corresponding field in the dual theory: composite (elementary) fermions are localized towards the IR (UV) brane and their corresponding parameter satisfies the relation $c_{f_{L,R}}<0.5$ ($c_{f_{L,R}}>0.5$). In particular their wave function is given by
\be
f_{L,R}(y,x)=\frac{e^{(2-c_{L,R})A(y)}}{\displaystyle\left(\int dy\, e^{A(1-2 c_{L,R})} \right)^{1/2}} f_{L,R}(x)\,,
\label{fermion}
\ee
where $f_{L,R}(x)$ is the four-dimensional (4D) spinor.

Our choice of the 5D gravitational metric guarantees that the correction to
the universal (oblique) observables, encoded in the Peskin-Takeuchi variables $S,T,U$~\cite{Peskin:1991sw}, and the non-universal ones, in particular the shifts in the couplings $Z\bar ff$ where $f=b,\tau,\mu,e$, stay below their experimental values as we now show.
 
\subsubsection*{Oblique observables}

\begin{figure}[!htb]
\centering
\includegraphics[width=10cm]{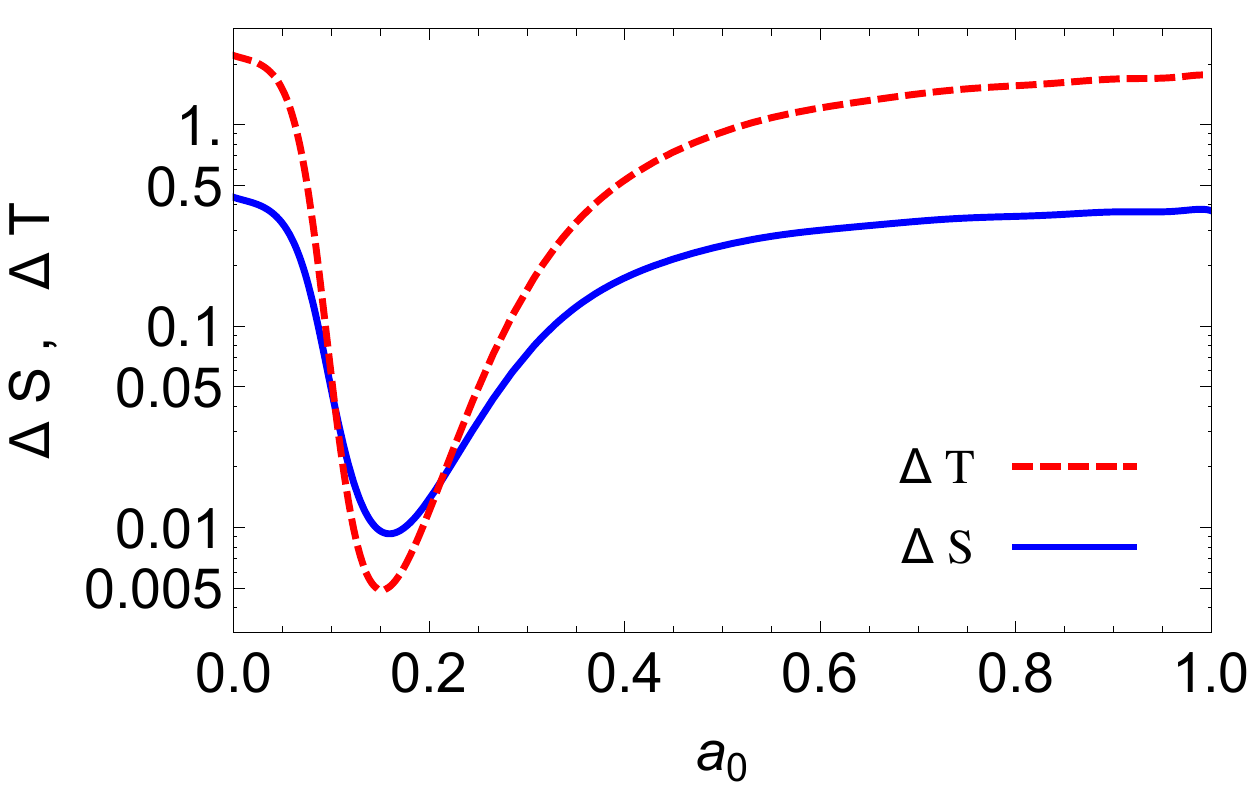}
\caption{\it Contribution to the $S$ and $T$ parameters from the gauge KK modes as a function of $a_0$. We have considered $b_0=2$ and $m_{KK}= 2~\textrm{TeV}$.} 
\label{fig:ST}
\end{figure}

In our model they are given by the following expressions~\cite{Cabrer:2011fb}
\begin{align}
\alpha_{EM} \Delta T & =s^2_W \frac{m_Z^2}{\rho^2}k^2 y_1\int_{0}^{y_1}
\left[1-\Omega_h(y)\right]^2e^{2A(y)-2A(y_1)} dy\,,
\nonumber\\
\alpha_{EM} \Delta S & =8c^2_Ws^2_W \frac{m_Z^2}{\rho^2}k^2 y_1\int_{0}^{y_1}
\left(1-\frac{y}{y_1}\right)\left[1-\Omega_h(y)\right]e^{2A(y)-2A(y_1)} dy\,,
\end{align}
and $\alpha_{EM} \Delta U \simeq 0$, where $\rho=ke^{-A(y_1)}$,
$
\Omega_h(y)=\frac{\omega(y)}{\omega(y_1)}$, and $\omega(y)=\int_{0}^{y}
h^2(\bar y)e^{-2A(\bar y)} d\bar y 
$, where the Higgs profile is taken to be $h(y)=h(0) \exp(\alpha y)$ with $\alpha=2 A(y_1)/ky_1$.
The present experimental bounds on the $S$ and $T$ parameters are given by~\cite{Olive:2016xmw}
\be
\Delta S=0.07\pm 0.08,\quad \Delta T=0.10\pm 0.07\qquad (\rho\simeq 0.90)\,.
\ee
We show in Fig.~\ref{fig:ST} plots of $\Delta S$ and $\Delta T$, as functions of $a_0$, for fixed value of $m_{KK}=2$ TeV, and $b_0=2$. 
We can see that the contribution to the $S$ and $T$ parameters implies the $2\sigma$ interval
\be
0.1\lesssim a_0\lesssim  0.3 \,.
\ee
In order to minimize the contribution to oblique parameters we will choose, from here on, the value $a_0=0.15$.

\subsubsection*{The $Z\overline{f} f$ coupling}
\begin{figure}[htb]
\centering
\includegraphics[width=12.cm]{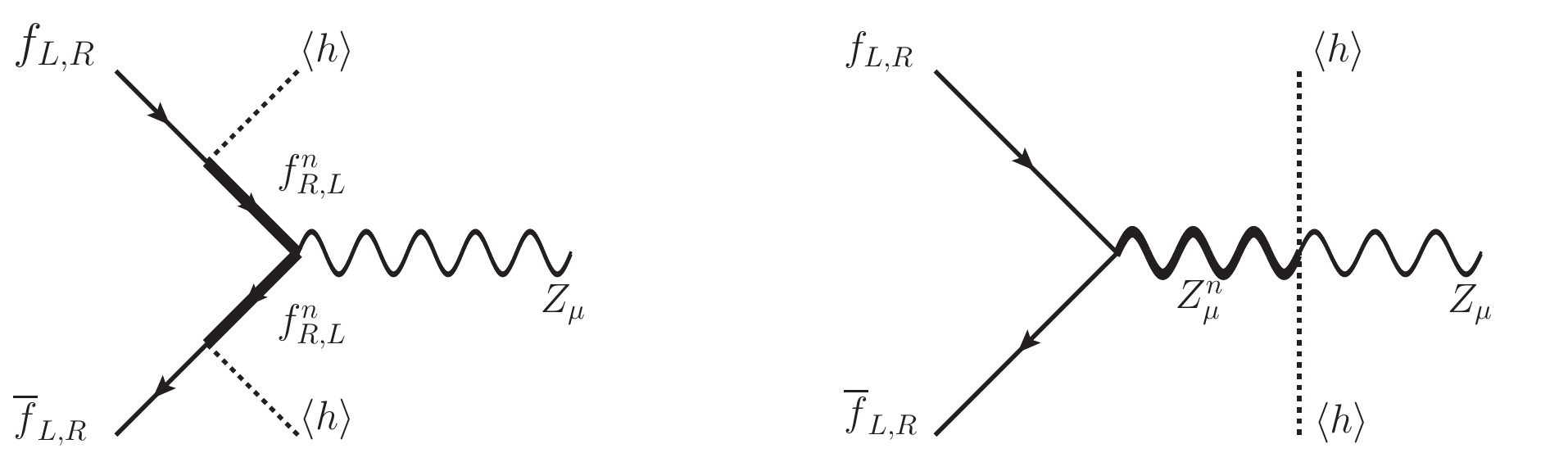} 
\caption{\it 
Diagrams contributing to $\delta g_{f_{L,R}}/g_{f_{L,R}}$.}
\label{fig:Zff}
\end{figure}

 The $Z$ boson coupling to SM fermions $f_{L,R}$ with a
sizeable degree of compositeness can be modified by two independent effects: one coming from the vector KK modes and the other from the fermion KK excitations.
The distortion in the couplings can be straightforwardly written as a sum over the contributions of the various KK modes, as shown in Fig.~\ref{fig:Zff},
thus obtaining the full result~\cite{Cabrer:2011qb,Megias:2016bde}
\begin{equation}\label{eq:delta_g}
\delta g_{f_{L,R}}= - g_{f_{L,R}}^{SM}m_Z^2\widehat \alpha_{f_{L,R}}\pm g\frac{v^2}{2}\widehat\beta_{f_{L,R}}\,,
\end{equation}
where $g_{f_{L,R}}^{SM}$ denotes the (tree-level) $Z$ coupling to the $f_{L,R}$ fields in the SM, while
\begin{align}
\widehat\alpha_{f_{L,R}}= & y_1\int_0^{y_1} e^{2A}\left(\Omega_h-\frac{y}{y_1}\right)\left(\Omega_{f_{L,R}}-1\right)\,,\nonumber\\
\widehat\beta_{f_{L,R}}= & Y_f^2\int_0^{y_1} e^{2A}\left( \frac{d \Omega_{f_{R,L}}}{dy}\right)^{-1}\left(\Gamma_f-\Omega_{f_{R,L}}\right)^2\,,\label{eq:defs1}
\end{align}
with $Y_f$ the 4D Yukawa coupling and
\be\label{eq:defs2}
\Omega_{f_{L,R}}(y)=\frac{\displaystyle \int_0^y e^{(1-2c_{f_{L,R}})A}}{\displaystyle \int_0^{y_1 }e^{(1-2c_{f_{L,R}})A}}\,,\qquad
\Gamma_f(y)=\frac{\displaystyle \int_0^y he^{-(c_{f_L}+c_{f_R})A}}{\displaystyle \int_0^{y_1} he^{-(c_{f_L}+c_{f_R})A}}\,.
\ee
It is easy to recognize that the two terms in Eq.~(\ref{eq:delta_g}) correspond, respectively, to the effects of the massive vector and fermion KK modes.
\begin{figure}[htb]
\centering
\includegraphics[width=7.1cm]{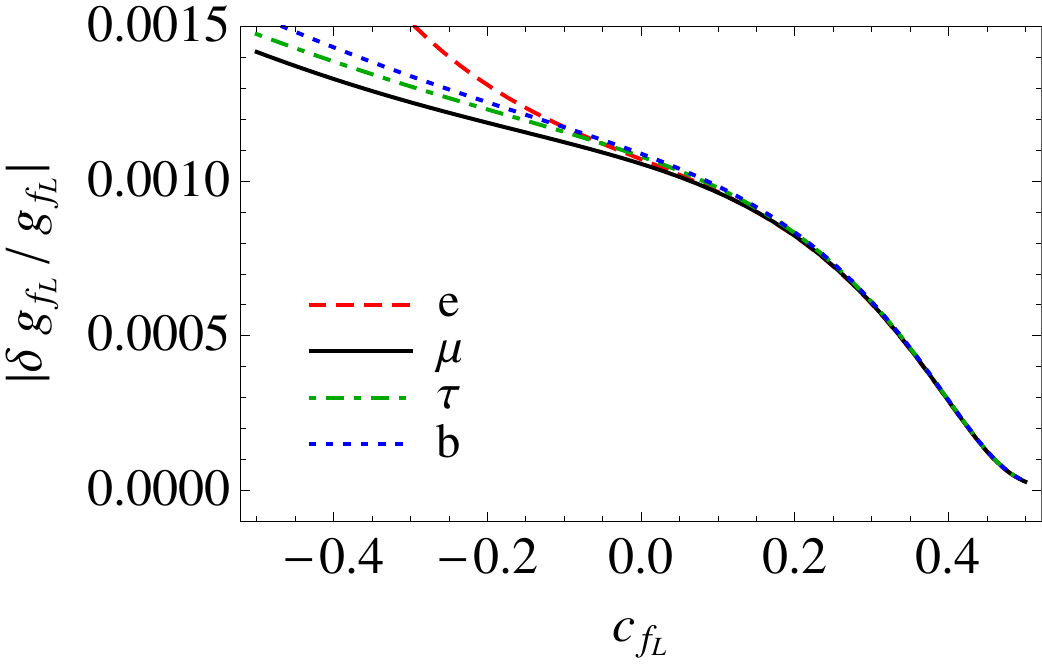} \hfill
\includegraphics[width=7.1cm]{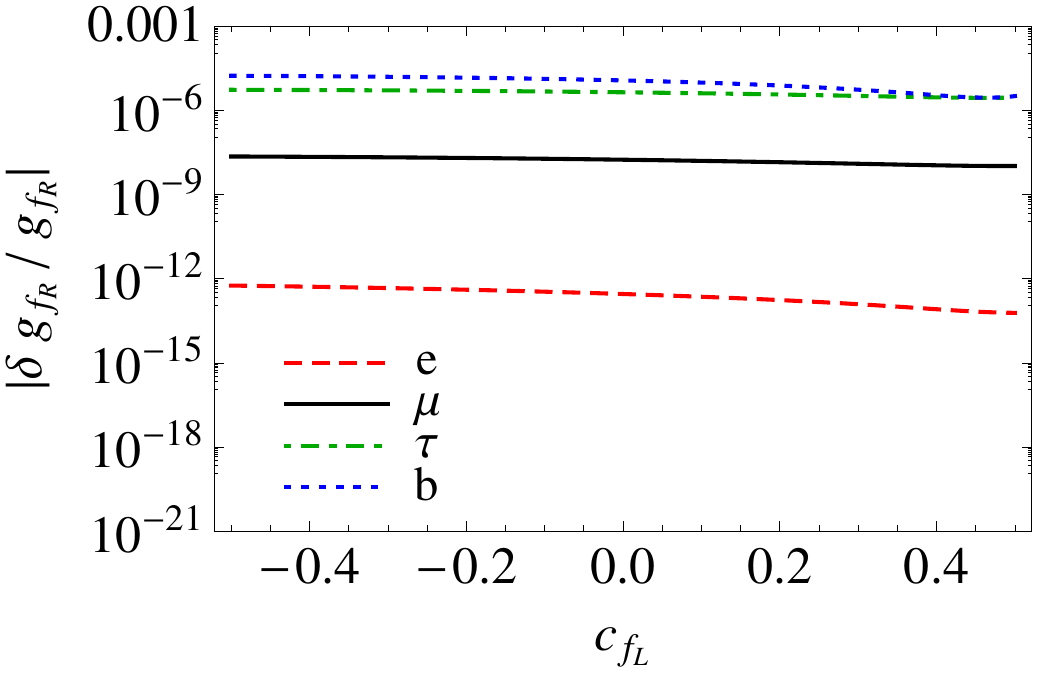} \hfill
\caption{\it 
Contribution to $|\delta g_{f_L}/g_{f_L}|$ (left panel) and $|\delta g_{f_R}/g_{f_R}|$ (right panel) from KK modes for the electron (dashed red line), muon (solid black line), tau lepton (dot-dashed green line) and bottom quark (dotted blue line). The allowed region corresponds to the regime $|\delta g_{f_{L,R}}/g_{f_{L,R}}| \lesssim 10^{-3}$. We have considered $(c_{e_R},c_{\mu_R},c_{\tau_R},c_{b_R}) = (0.85, 0.65, 0.55,0.55)$.}
\label{fig:deltagL}
\end{figure} 

We plot in Fig.~\ref{fig:deltagL} the value of $|\delta g_{f_L}/g_{f_L}|$ (left panel) and $|\delta g_{f_R}/g_{f_R}|$ (right panel) as a function of $c_{f_L}$ for $f= e, \mu, \tau, b$ and $(c_{e_R},c_{\mu_R},c_{\tau_R},c_{b_R}) = (0.85, 0.65, 0.55,0.55)$. We can see that in all cases the constraint $|\delta g_{f_L}/g_{f_L}|\lesssim 10^{-3}$~\cite{Olive:2016xmw} implies the mild constraint $c_{f_L}\gtrsim -0.5$. In particular from the values of $|\delta g_{\ell_{L,R}}/g_{\ell_{L,R}}|$ for $\ell=e,\mu,\tau$ we see that for $c_{\ell_L}\gtrsim -0.5$ no lepton flavor universality breaking appears at the $Z$-pole in agreement with the very strong LEP bounds on lepton non-universal couplings~\cite{Olive:2016xmw}.

From Eq.~(\ref{fermion}) it is easily seen that the coupling of electroweak and strong KK gauge bosons to a fermion $f$ with $c_f=0.5$ vanishes due to the orthonormality of KK modes. Therefore if we assume that first generation quarks $(f=u,d)$ are such that $c_f\simeq 0.5$, it follows that Drell-Yan production of electroweak and strong KK gauge bosons from light quarks vanishes, or at least is greatly suppressed. Likewise the production of KK gluons by gluon fusion, or electroweak KK gauge bosons by vector-boson fusion, vanishes by orthonormality of KK modes, which can therefore only be produced by pairs, an energetically disfavored process. Therefore our theory satisfies our original strategy that~\textit{direct detection can be hidden}, depending on the degree of compositeness (or elementariness) of the Standard Model fermions.

On the experimental side, lepton flavor universality violation (LFUV) has been recently observed by the BaBar, Belle and LHCb Collaborations in the observables $R_{D^{(\ast)}}$~\cite{Lees:2012xj,Lees:2013uzd,Huschle:2015rga,Sato:2016svk,Abdesselam:2016xqt,Hirose:2016wfn,Aaij:2015yra} and $R_K$~\cite{Aaij:2014ora}. In the present paper we will attempt to accommodate in our theory the actual experimental data exhibiting LFUV. The relevant involved fermions are $b_L$, $\tau_L$ and $\mu_L$, characterized by the constants $c_{b_L}$, $c_{\tau_L}$ and $c_{\mu_L}$. We will see that explaining all anomalies would require some degree of compositeness for the above fermions, a feature which is not motivated (as usually assumed for the Standard Model fermions) by the value of their masses, as it is e.g.~the case of the top quark $t_R$. The required degree of compositeness of these not-so-heavy fermions has phenomenological consequences which, on the one hand, must be in agreement with all present and past experimental data, and on the other hand could trigger new phenomena to be searched for at present and future colliders. 

Previous analyses in the literature have considered various \textit{ad hoc} extensions of the Standard Model suitable to accommodate the anomalies, in particular including new gauge bosons~\cite{Altmannshofer:2013foa,Gauld:2013qba,Sierra:2015fma,Crivellin:2015era,Celis:2015ara,Falkowski:2015zwa,Descotes-Genon:2015uva,Allanach:2015gkd,Buttazzo:2016kid,Biancofiore:2013ki,
Descotes-Genon:2016hem}, leptoquarks~\cite{Kosnik:2012dj,Sakaki:2013bfa,Hiller:2014yaa,Gripaios:2014tna,Sahoo:2015wya,Becirevic:2015asa,Alonso:2015sja,Dumont:2016xpj,Das:2016vkr,Sahoo:2016pet,Bhattacharya:2016mcc,Alonso:2016oyd,Celis:2016azn,Altmannshofer:2017wqy,Chen:2017hir} and general effective field theory frameworks~\cite{Greljo:2015mma,Alok:2016qyh,Ivanov:2016qtw,Faroughy:2016osc,Feruglio:2016gvd}~\footnote{In particular in Ref.~\cite{Feruglio:2016gvd} the effects of lepton flavor violation, as well as the renormalization group running, on lepton flavor universality violation have been explored.}. Anomalies in $R_K$, and $b\to s\ell\ell$ processes, have also been addressed in Randall-Sundrum~\cite{Biancofiore:2014wpa,Biancofiore:2014uba,Carmona:2015ena} and flat space~\cite{GarciaGarcia:2016nvr} extra dimensional scenarios.  On the other hand, our approach is based on a \textit{lepton flavor conserving} minimal model solving the naturalness problem of the Standard Model \textit{without invoking any extra physics}.

The contents of this paper are as follows. The analysis of the $R_K$ anomaly, as well as some comments about $R_{K^\ast}$, is performed in Sec.~\ref{RK}. As the result depends on the unitary transformations diagonalizing the quark mass matrices, and in the absence of a particular UV theory predicting the 5D Yukawa matrices, we will consider for the diagonalizing matrices $V_{u_{L,R}}$ and $V_{d_{L,R}}$ generic Wolfenstein-like parametrizations satisfying the relation $V_{u_L}^\dagger V_{d_L}=V_{\rm CKM}$. Without making a statistical analysis of the parameter space we will assign generic values to the parameters which optimize the results. 
In Sec.~\ref{Bs}
 we impose constraints on the (almost) elementary electrons from the branching fraction of $\bar B\to \bar K ee$, as compared to its Standard Model value, and 
we adjust other observables, as e.g.~$B_s\to \mu^+\mu^-$, which appear in the $b\to s\mu^+\mu^-$ decay process. We present in Sec.~\ref{constraints} the result of imposing the different constraints, including electroweak observables, direct searches and flavor constraints. All together they restrict the available region of parameters where the anomalies can be accommodated. An overproduction, with respect to the Standard Model prediction, in the branching ratios $\mathcal B(\bar B\to \bar K\tau\tau)$ and $\mathcal B(\bar B\to\bar K\nu\bar\nu)$ can generically appear. This issue, as well as the region allowed by present data, is analyzed in Sec.~\ref{processes}.
In Sec.~\ref{RD} we consider the $R_{D^{(\ast)}}$ anomaly and we contrast it with the non-observation of flavor universality violation effects in the $\mu/e$ sector and with lepton flavor universality tests in tau decays. We will prove that the $R_{D^{(\ast)}}$ anomaly, along with a strict Wolfenstein-like parametrization of diagonalizing unitary matrices, is in tension with electroweak observables, in particular with experimental data on the coupling $g_{\tau_L}^Z$. As we will point out this problem can be resolved by somehow slightly giving up on the Wolfenstein-like structure of diagonalizing matrices and thus allowing a small amount of fine-tuning when fixing the CKM matrix. Finally our conclusions and outlook are presented in Sec.~\ref{conclusions}.

\section{Lepton-flavor universality violation in $R_K$}
\label{RK}
The LHCb Collaboration has determined the ratio of branching ratios $\mathcal B(\bar B\to \bar K \ell\ell)$ for muons over electrons yielding the result~\cite{Aaij:2014ora}
\be
R_K\equiv R_K^{\mu/e}=\frac{\mathcal B(\bar B\to\bar K \mu\mu)}{\mathcal B(\bar B\to \bar K ee)}=0.745^{+0.090}_{-0.074}\pm 0.036
\ee
which, by combining systematic and statistical uncertainties in quadrature, implies a deviation $\sim 2.6\sigma$ with respect to the Standard Model prediction $R_K^{\rm SM}=1.0003\pm 0.0001$~\cite{Bobeth:2007dw,Bordone:2016gaq}.

One can interpret this result by using an effective description given by the $\Delta F=1$ Lagrangian
\be
\mathcal L_{eff}=\frac{4G_F}{\sqrt{2}}\,\frac{\alpha}{4\pi} V^*_{ts}V_{tb}\sum_i C_i\,\mathcal O_i\,,
\label{efflagrangian}
\ee  
where the Wilson coefficients $C_i=C_i^{\rm SM}+\Delta C_i$, are the sum of a SM contribution $C_i^{\rm SM}$ and a new-physics one $\Delta C_i$. The sum in Eq.~(\ref{efflagrangian}) includes the operators
\begin{equation}
\begin{array}{l@{\qquad}l}
\mathcal O_9^\ell =(\bar s_L\gamma_{\mu}  b_L)(\bar\ell \gamma^\mu\ell)\,, & \mathcal O_{10}^\ell=(\bar s_L\gamma_{\mu}  b_L)(\bar\ell \gamma^\mu\gamma_5\ell)\,,\\
\rule{0pt}{1.25em}\mathcal O_9^{\prime\ell} =(\bar s_R\gamma_{\mu}  b_R)(\bar\ell \gamma^\mu\ell)\,, & \mathcal O_{10}^{\prime\ell}=(\bar s_R\gamma_{\mu}  b_R)(\bar\ell \gamma^\mu\gamma_5\ell)\,.
\end{array}
\label{operators}
\end{equation}
for $\ell=\mu,e$.

After electroweak symmetry breaking the mass matrices for $u$ and $d$-type quarks are diagonalized by the unitary matrices $V_{u_{L,R}}$ and $V_{d_{L,R}}$, and so their matrix elements, unlike those of the CKM matrix, are not measured experimentally and moreover are model dependent. In the absence of a general (UV) theory, providing the 5D Yukawa couplings $\widehat Y_{u,d}$, we will just consider the general form for these matrices by assuming they reproduce the physical CKM matrix $V$, i.e.~they satisfy the condition $V\equiv V_{u_L}^\dagger V_{d_L}$.

Given the hierarchical structure of the quark mass spectrum and mixing angles, we will then assume for the matrices $V_{d_L}$ and $V_{u_L}$ Wolfenstein-like parametrizations as 
\be
V_{d_L}=\begin{pmatrix} 1-\frac{1}{2}\lambda_0^2 & \lambda_0 & A \lambda^2\lambda_0(1-r)(\rho_0-i \eta_0)\\
-\lambda_0 & 1-\frac{1}{2}\lambda_0^2 & A \lambda^2(1-r)\\
A\lambda^2\lambda_0(1-r)(1-\rho_0-i\eta_0) &\quad -A \lambda^2(1-r) &1
\end{pmatrix}\ ,
\label{Vd}
\ee
with values of the parameters $(r,\lambda_0,\rho_0,\eta_0)$ consistent with the hierarchical structure of the matrix, and
\begin{equation}
V_{u_L}=
\begin{pmatrix} 1-\frac{1}{2}(\lambda-\lambda_0)^2 & (\lambda_0-\lambda)\left(1+\frac{1}{2}\lambda_0\lambda \right) & (V_{u_L})_{13}\\
-(\lambda_0-\lambda)\left(1+\frac{1}{2}\lambda_0\lambda \right) & 1-\frac{1}{2}(\lambda-\lambda_0)^2 & -A\lambda^2\,r\\
(V_{u_L})_{31} & A \lambda^2\, r &1
\end{pmatrix}\ ,
\label{Vu}
\end{equation}
where 
\begin{align}
(V_{u_L})_{31}&=A\lambda^3(\rho+i\eta)+A\lambda^2(1-r)\left[\lambda_0(1-\rho_0-i\eta_0)-\lambda\right]\nonumber\\
(V_{u_L})_{13}&=A \lambda^3(1-\rho+i \eta)+A\lambda^2 \lambda_0[(1-r)(\rho_0-i\eta_0)-1]\ , \\
\end{align}
and where~\cite{Olive:2016xmw}
\be
\lambda=0.225,\quad A=0.811,\quad \rho=0.124,\quad \eta=0.356
\ee
are the parameters of the CKM matrix $V$ in the Wolfenstein parametrization
\be
V=\begin{pmatrix} 1-\frac{1}{2}\lambda^2 & \lambda & A \lambda^3(\rho-i\eta)\\
-\lambda & 1-\frac{1}{2}\lambda^2 & A \lambda^2\\
A\lambda^3(1-\rho-i\eta) & -A \lambda^2 &1
\end{pmatrix} \,.
\label{CKM}
\ee
The matrix forms of (\ref{Vd}) and (\ref{Vu}) guarantee the precise determination of the CKM matrix elements in (\ref{CKM}).
In particular, in numerical calculations, we will make the particular choice

\be
\lambda_0\simeq\mathcal O( \lambda),\quad \eta_0\simeq\mathcal O( \eta),\quad 0\lesssim r\lesssim 1,\quad 0\lesssim\rho_0\lesssim 1
\ee
which guarantees the Wolfenstein-like structure of the matrices $V_{d_L}$ and $V_{u_L}$.

\begin{figure}[htb]
\centering
\vspace{0.1cm}
\includegraphics[width=8.cm]{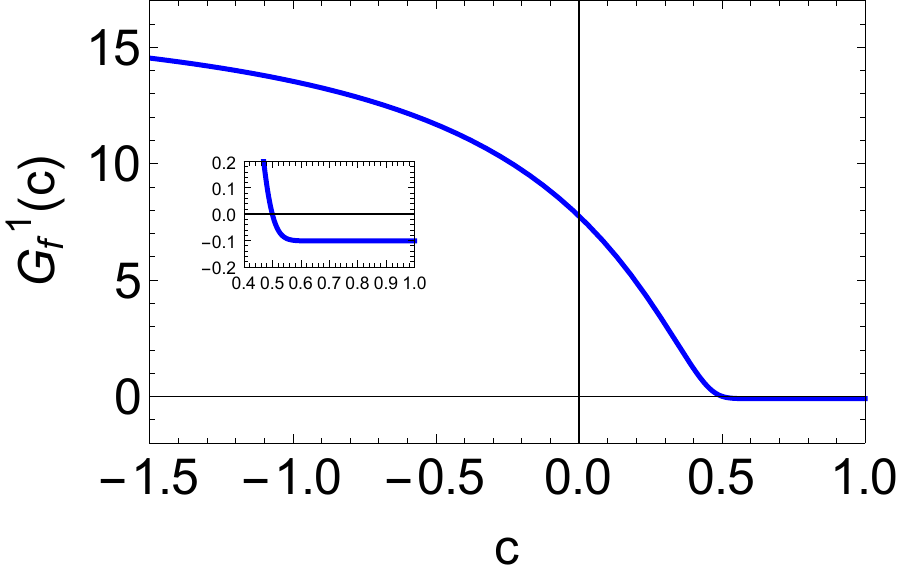}
\caption{\it Coupling (normalized with respect to the 4D coupling) of a fermion zero-mode with the $n=1$ KK gauge field, $W^{(n)}_\mu$, as a function of the fermion localization parameter $c$ [cf. Eq.~(\ref{integral})]. }
\label{fig:Gf}
\end{figure}

Our theory contains the neutral current interaction Lagrangian
\be
\mathcal L=\frac{g}{c_W}\sum_{X=Z,\gamma}\sum_n X_n^\mu\left(g_{f_L}^{X_n}\bar f_L \gamma_\mu f_L+g_{f_R}^{X_n}\bar f_R \gamma_\mu f_R \right) \,,
\label{intlagrangian}
\ee
where $g^{X_n}_{f_{L,R}}=g^{X}_{f_{L,R}} G_{f_{L,R}}^{\,n}$ with
\begin{align}
g_{f_L}^Z&=(T_{3f}-Q_f s^2_W),\ g_{f_R}^Z=-Q_f s^2_W \,, \nonumber\\
g_{f_L}^\gamma&=Q_f s_W c_W,\quad g_{f_R}^\gamma=Q_f s_W c_W 
\end{align}
and the couplings $G^n_{f}$ defined as
\be
G^n_{f}=\frac{{\displaystyle \sqrt{y_1} \int e^{-3A} f_A^n(y) f^2(y)}}{\sqrt{\int [f_A^n(y)]^2 } \int e^{-3A}  f^2(y)  }  \,,
\label{integral}
\ee
where $f_A^n(y)$ is the profile of the gauge boson $n$-KK mode and $f(y)$ the profile of the corresponding fermion zero-mode, as given by Eq.~(\ref{fermion}). The plot of $G_f^n(c)$ (for $n=1$) as a function of the parameter $c$, which determines the localization of the fermion zero mode, is shown in Fig.~\ref{fig:Gf}. Notice that it vanishes for $c=0.5$, as anticipated in Sec.~\ref{introduction}, while it grows in the IR, and stabilizes itself around $ -0.1$ in the UV.

In the following we will assume that the first and second generation quarks respect the universality condition. This implies an approximate accidental
$U(2)_{q_L} \otimes U(2)_{u_R} \otimes U(2)_{d_R}$ global flavor symmetry, which is only broken by the Yukawa couplings~\cite{Megias:2016bde}. 
For simplicity in our numerical analysis we will moreover choose $c_{q^{1}_L}=c_{q^{2}_L}\equiv c_{q_L}$, as well as
$c_{u_R}=c_{c_R}=c_{d_R}=c_{s_R}\equiv c_{q_R}$. The values $r=0.75$ and $m_{KK}=2$ TeV have been chosen, and will be adopted, without explicit mention, in the rest of the paper.

In our model, contact interactions can be obtained by the exchange of KK modes of the $Z$ $(Z_n)$ and the photon $(\gamma_n)$. They give rise to the Wilson coefficients~\footnote{Notice that the Wilson coefficients in Eq.~(\ref{C9}) differ by a factor $(1-r)$ with respect to those in Ref.~\cite{Megias:2016bde}, where moreover we were assuming $V_{u_L}\simeq 1_3$ and $V_{d_L}=V$.}~\cite{Megias:2016bde}
\begin{equation}
\begin{array}{l@{\hspace{.9em}}l}
\displaystyle \Delta C_9^{(\prime)\ell}=-(1-r)\sum_{X=Z,\gamma}\,\sum_n\frac{2\pi g^2 g_{\ell_V}^{X_n}\left(g^{X_n}_{b_{L(R)}}-g^{X_n}_{q_{L(R)}}\right)}{\sqrt{2} G_F\alpha c_W^2 M^2_n}\,,
\\
\rule{0pt}{2.em}\displaystyle \Delta C_{10}^{(\prime)\ell}=(1-r) \sum_{X=Z,\gamma}\,\sum_n\frac{2\pi g^2 g_{\ell_A}^{X_n}\left(g^{X_n}_{b_{L(R)}}-g^{X_n}_{q_{L(R)}}\right)}{\sqrt{2} G_F\alpha c_W^2 M^2_n}\,.
\end{array}
\label{C9}
\end{equation}
\begin{figure}[htb]
\centering
\includegraphics[width=7.5cm]{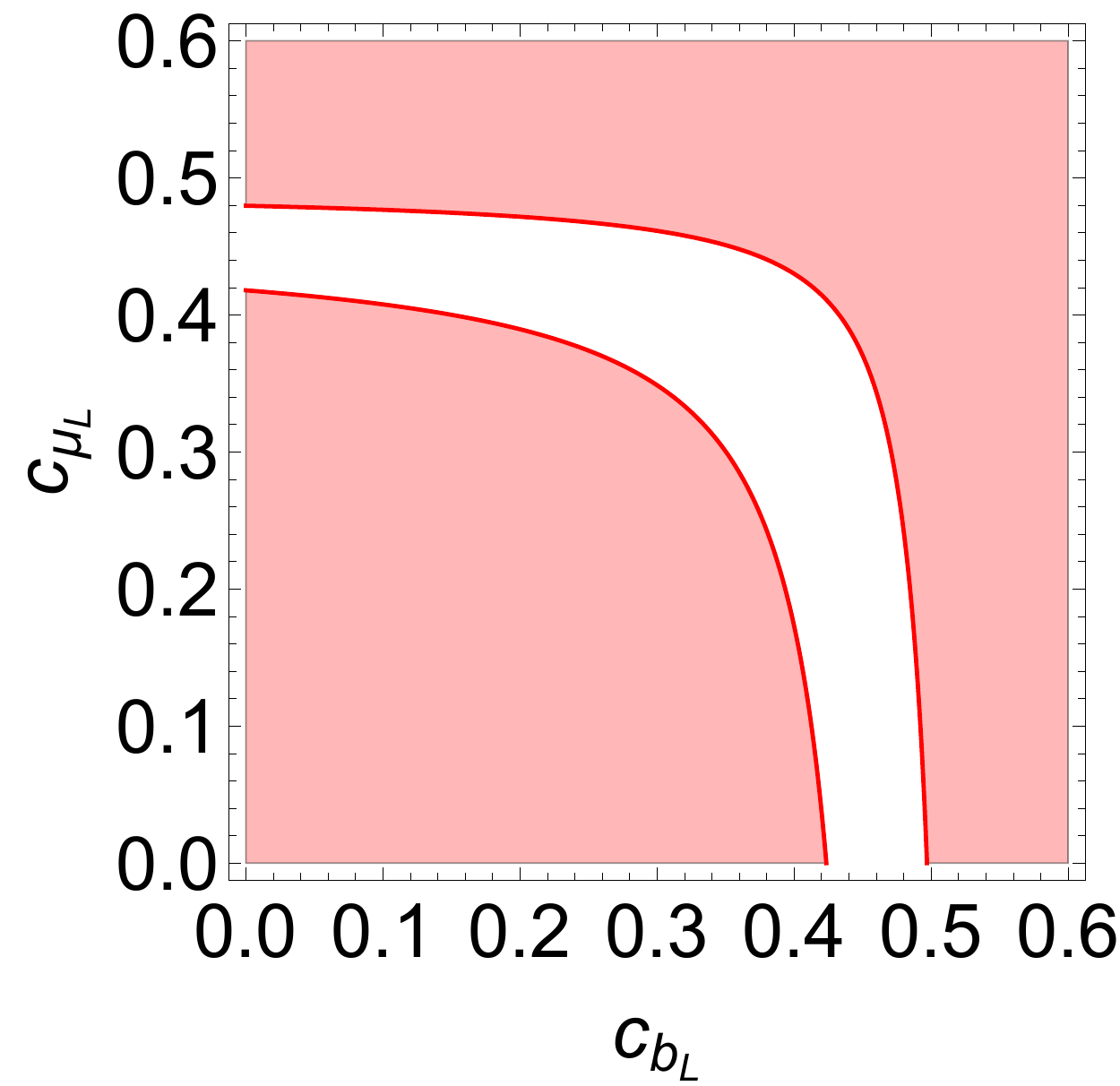}
\caption{\it Region in the $(c_{b_L},c_{\mu_L})$ plane that accommodates the $2\sigma$ region $R_K \in [0.580,0.939]$. We display the result for $c_{e_L}=0.5$. 
}
\label{fig:RK2}
\end{figure} 
where $g^{X_n}_{f_{V,A}}=g^{X_n}_{f_{L}}\pm g^{X_n}_{f_{R}}$.
Using now the Standard Model prediction $C_{\rm 9} ^{\rm SM}=-C_{\rm 10}^{\rm SM}\simeq 4.2$ at the $m_b$ scale, and following Ref.~\cite{Das:2014sra},  we find the $2\sigma$ interval
$0.580<R_K<0.939$, 
where we have combined statistical and systematic uncertainties in quadrature, while the
observable $R_K$, in terms of the Wilson coefficients, is given by~\cite{Das:2014sra,Hiller:2014yaa}
\be
R_K=\frac{\left|C_9^{\rm SM}+\Delta C_9^\mu+\Delta C_9^{\prime\,\mu} \right|^2+\left|C_{10}^{\rm SM}+\Delta C_{10}^\mu+\Delta C_{10}^{\prime\,\mu} \right|^2}{\left|C_9^{\rm SM}+\Delta C_9^e+\Delta C_9^{\prime\, e} \right|^2+\left|C_{10}^{\rm SM}+\Delta C_{10}^e+\Delta C_{10}^{\prime\, e} \right|^2} \,.
\ee
In Fig.~\ref{fig:RK2} we show in the $(c_{b_L},c_{\mu_L})$ plane the $2\sigma$ region allowed by the experimental data on $R_K$, the region between the solid red lines, where we use the values $c_{e_L}=0.5$, $c_{q_L}=c_{q_R}=0.8$, $c_{b_R}=0.55$ and $c_{\mu_R}=0.65$.
 As we can see from this plot, both fermions $b_L$ and $\mu_L$ must be localized towards the IR, and thus have to exhibit some degree of compositeness in the dual theory. Here we obtain the mild constraints 
$$c_{b_L}\lesssim 0.50 \qquad \textrm{ and } \qquad c_{\mu_L}\lesssim 0.49\,.$$ 
In fact as we can see from the plot of Fig.~\ref{fig:RK2} the degrees of compositeness of $b_L$ and $\mu_L$ are inversely proportional to each other. 
 
 To conclude this section, we would like to mention that the model prediction for the related observable $R_{K^\ast}$ recently measured by the LHCb Collaboration~\cite{Aaij:2015oid} is $R_{K^\ast}\simeq R_K$. In fact a measurement of  $R_{K^\ast}$ in agreement with the Standard Model prediction $R_{K^{\ast}}^{\rm SM}\simeq 1$~\cite{Bordone:2016gaq} would be in tension with our explanation of the $R_K$ anomaly.

\section{Other $b\to s\ell^+\ell^-$ processes}
\label{Bs}

\begin{figure}[htb]
\centering
\includegraphics[width=7.5cm]{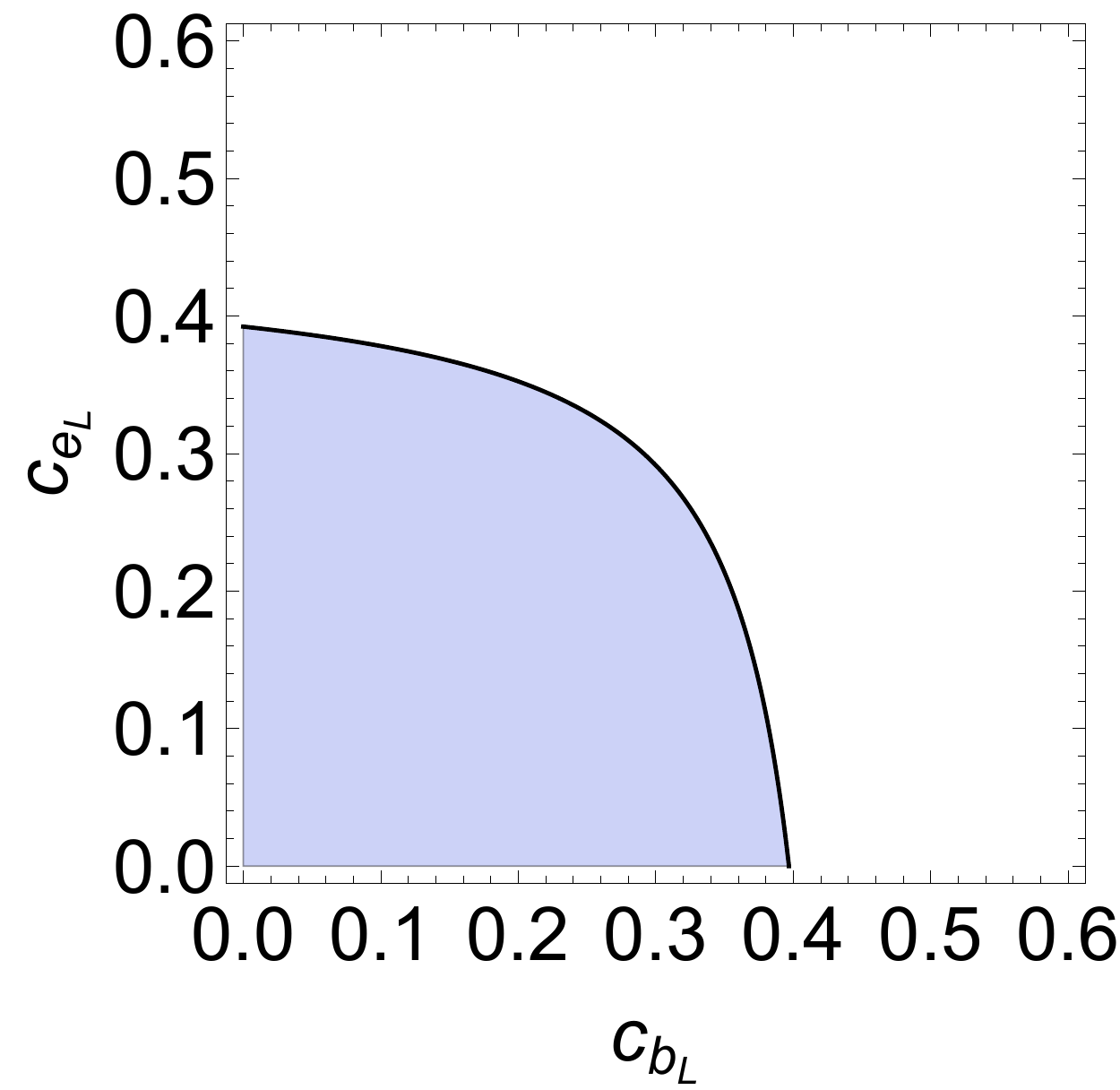} \hfill
\includegraphics[width=7.5cm]{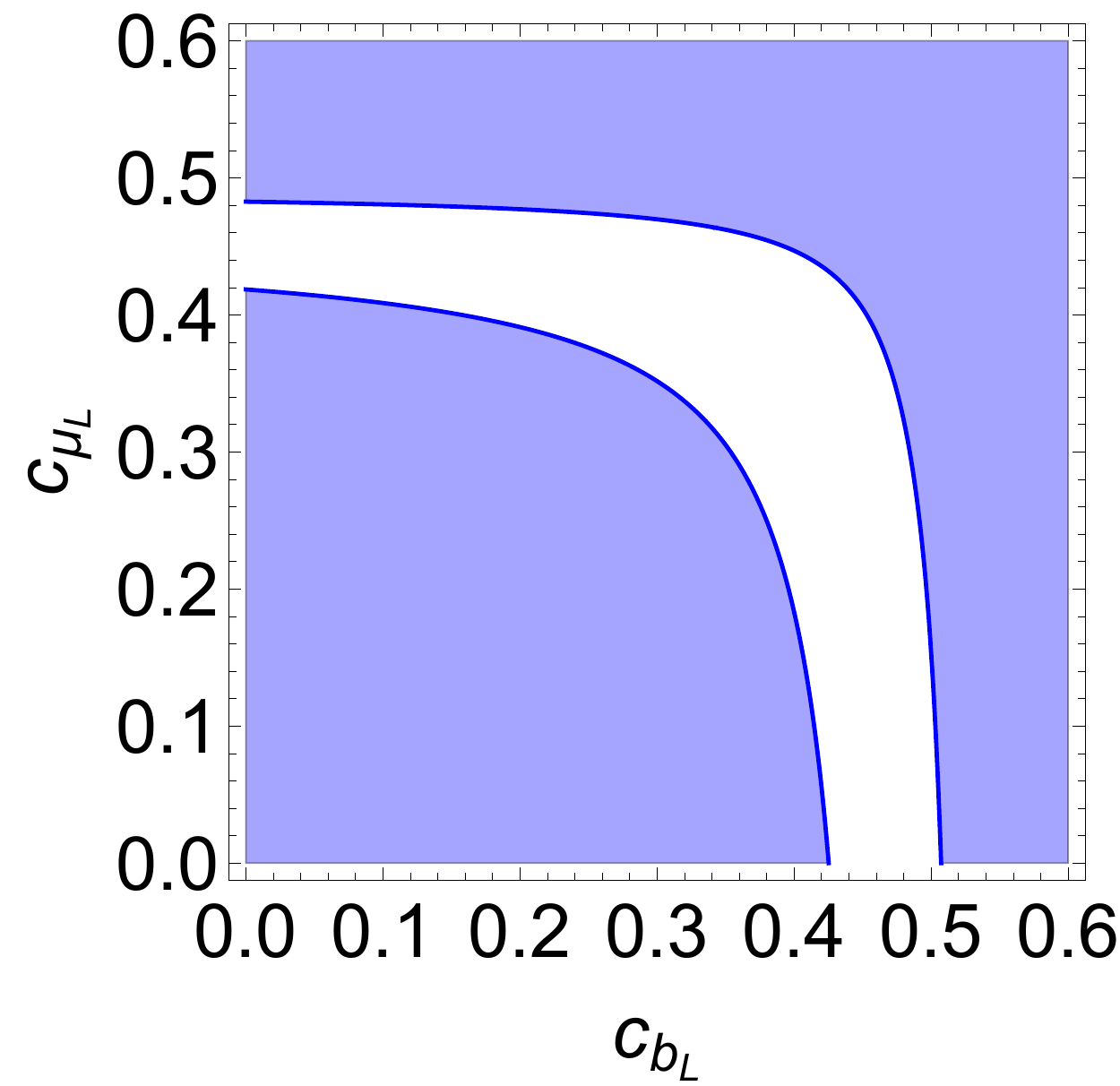}
\caption{\it Left panel: Region in the plane $(c_{b_L},c_{e_L})$ that accommodates the constraint on ${\mathcal B(\bar B\to \bar K ee)}$ given by Eq.~(\ref{eq:BKee}). Right panel: Region in the plane $(c_{b_L},c_{\mu_L})$ that accommodates the $1\sigma$ region $R_0 \in [0.594,0.962]$. 
}
\label{fig:R0}
\end{figure}

The values of $c_{e_{L,R}}$ are constrained by the LHCb measurement~\cite{Aaij:2014ora} of the branching ratio $\mathcal B(\bar B\to \bar K ee)$ and the $2\sigma$ result~\cite{Hiller:2014yaa}
\be
0.41\lesssim R_K^{\,e}\equiv\frac{\mathcal B(\bar B\to \bar K ee)}{\mathcal B(\bar B\to \bar K ee)_{\rm SM}}\lesssim 1.25  \label{eq:BKee}
\ee 
where
\be
R_K^{\,e}=
\frac{\left|C_9^{\rm SM}+\Delta C_9^e+\Delta C_9^{\prime\, e} \right|^2+\left|C_{10}^{\rm SM}+\Delta C_{10}^e+\Delta C_{10}^{\prime\,e} \right|^2}{2\left|C_9^{\rm SM} \right|^2}\ .
\label{cociente}
\ee
The corresponding allowed region in the plane $(c_{b_L},c_{e_L})$ is shown in the left panel of Fig.~\ref{fig:R0}, where we use the values $c_{q_L}=c_{q_R}=0.8$ and $c_{b_R}=0.55$. We can see from the plot that values of $c_{e_L}$ around $c_{e_L}=0.5$ allow any value of $c_{b_L}$, as in particular for such value of $c_{e_L}$ we have that $\Delta C_9^e=\Delta C_{10}^e$ and $\Delta C_9^{\prime e}=\Delta C_{10}^{\prime e}$, and the ratio (\ref{cociente}) is one to linear order in the Wilson coefficients. Moreover for $c_{e_{L,R}}=0.5$ the coupling of electrons to the KK modes of gauge bosons vanishes and there is no contribution to observables involving the electron. On the other hand this is not the case for muons, which accomplishes in our model the mechanism of lepton flavor universality violation from the new physics mediated by the KK modes of electroweak gauge bosons.

The rare flavor-changing neutral current decay $B_s\to\mu^+\mu^-$ has been recently observed by the LHCb Collaboration with a branching fraction~\cite{CMS:2014xfa}
\be
\mathcal B(B_s\to\mu^+\mu^-)=\left( 2.8^{+0.7}_{-0.6} \right)\times 10^{-9} \,,
\ee
pretty consistent with the Standard Model prediction~\cite{Bobeth:2013uxa}
\be
\mathcal B(B_s\to\mu^+\mu^-)_{\rm SM}=\left( 3.66\pm 0.23 \right)\times 10^{-9} \,.
\ee
By combining the experimental and theoretical uncertainties in quadrature we can write the ratio
\be
R_0=\frac{\mathcal B(B_s\to\mu^+\mu^-)}{\mathcal B(B_s\to\mu^+\mu^-)_{\rm SM}}=0.765^{+0.197}_{-0.171}
\ee
while, in terms of the effective operator Wilson coefficients in Eq.~(\ref{C9}), we have~\cite{Altmannshofer:2017wqy}
\be
R_0=\left|\frac{C_{10}^{\rm SM}+\Delta C_{10}^\mu-\Delta C_{10}^{\prime\mu}}{C_{10}^{\rm SM}}   \right|^2  \,.
\ee
The $1\sigma$ region allowed by $R_0$ is shown in the right panel plot of Fig.~\ref{fig:R0}.

\begin{figure}[htb]
\centering
\includegraphics[width=7.5cm]{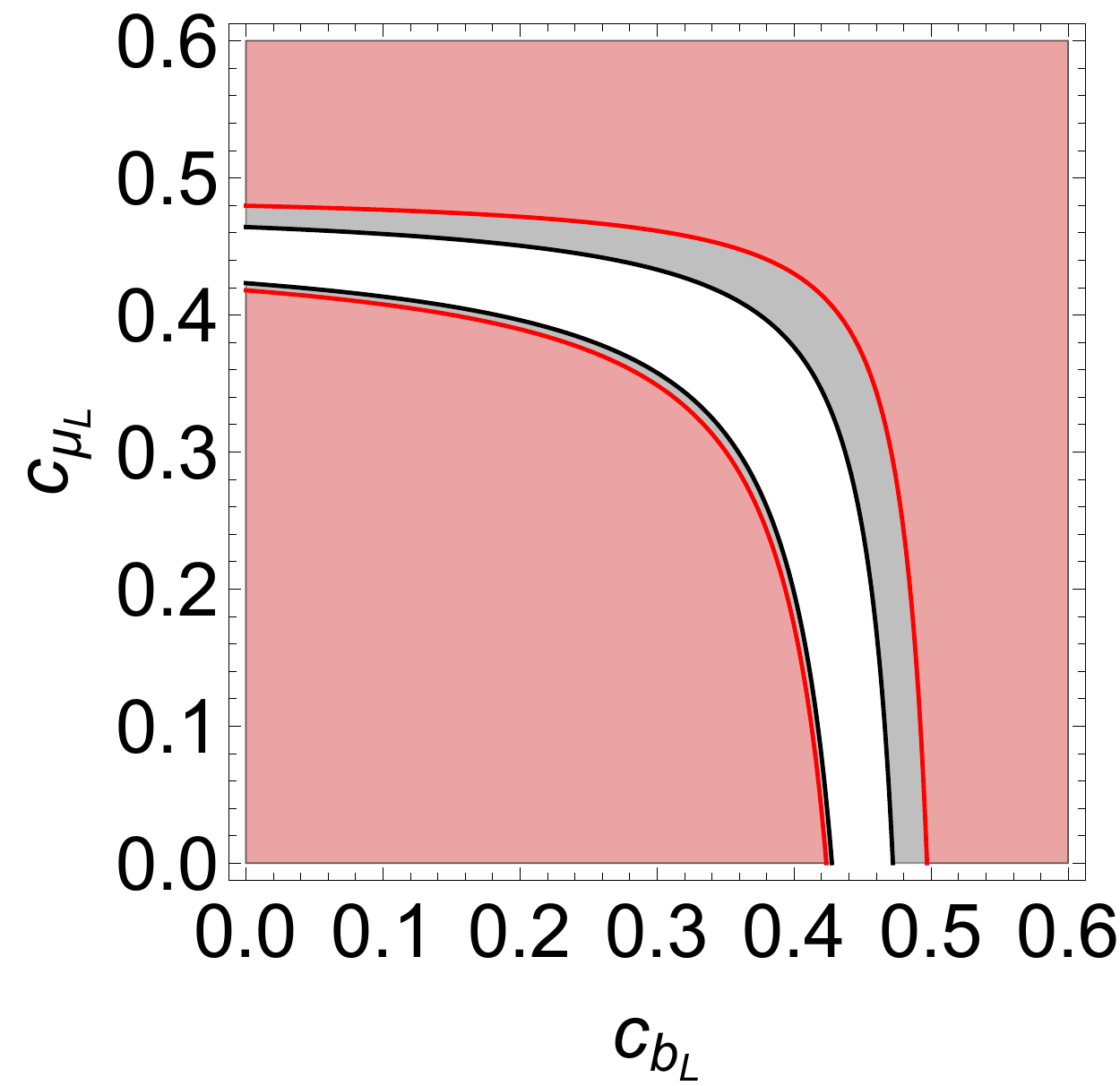}
\caption{\it Region in the $(c_{b_L},c_{\mu_L})$ plane that accommodates the $2\sigma$ region $\Delta C_9 \in [-0.93,-0.31]$ from the fit of Ref.~\cite{Altmannshofer:2017fio} (blank region inside gray bands). We overlap as well the allowed region coming from the $R_K$ anomaly (blank region inside red bands).
}
\label{fig:C9}
\end{figure}

Global fits to the Wilson coefficients $\Delta C_{9,10}^{(\prime)\,\mu}$ have also been performed
in the literature using a set of observables, including the branching ratios for $B\to K^* \ell\ell$, $B_s\to\phi\mu\mu$ and $B_s\to\mu\mu$, in Refs.~\cite{Hurth:2016fbr,Mahmoudi:2016mgr,Capdevila:2016ivx,Capdevila:2016fhq}. However, as observed in Refs.~\cite{Hurth:2016fbr,Mahmoudi:2016mgr}, removing the data on $R_K$ from the fits, lepton universality can be restored at a slightly larger deviation than $1\sigma$. As in our model we have the approximate relation $\Delta C_9\simeq -\Delta C_{10}$, using the recent multi-observable fit (which includes $R_{K^{(\ast)}}$) from Ref.~\cite{Altmannshofer:2017fio} we get the $2\sigma$ interval $\Delta C_9\in [-0.93,-0.31]$. We shown in the plot of Fig.~\ref{fig:C9} the region in the $(c_{b_L},c_{\mu_L})$ plane that accommodates the previous constraint on $C_9$, where we also superimpose the plot from $R_K$ in Fig.~\ref{fig:RK2}. As we can see the plot on the fitted value of $C_9$ slightly deviates from the plot in Fig.~\ref{fig:RK2} on the experimental value of $R_K$. We can conclude that at present $R_K$ is the main driving force for lepton flavor non-universality in the $\mu/e$ sector. For that reason, as our paper deals mainly with NP effects on lepton flavor non-universality, we will just consider in our analysis $R_K$ data.  

\section{Constraints}
\label{constraints}

As we have seen in the previous sections lepton flavor non-universality, mainly in the observable $R_K$, imply different degree of compositeness mainly for the fermions $b_L$ and $\mu_L$,  all of them localized towards the IR brane. This fact triggers modifications in the couplings of fermions with the $Z$ gauge boson, which are very constrained by experimental data. In particular, the KK modes of electroweak gauge bosons can trigger, through the mixing with electroweak gauge bosons after electroweak symmetry breaking, a modification of the universal (oblique) observables which were already considered in Sec.~\ref{introduction}. Moreover KK modes of the gluon can trigger $\Delta F=2$ flavor violating effective operators, which are also very constrained by the experimental data. Finally, direct searches of electroweak gauge boson KK modes decaying into muons and taus, and direct searches of gluon KK modes decaying into top quarks, by Drell-Yan processes, do depend on the couplings of fermions to KK modes, which in turn depend on the constants $c_{b_L}$, $c_{\tau_L}$ and $c_{\mu_L}$, as we have seen in Fig.~\ref{fig:Gf}. All these constraints will be considered in this section.

\subsection{Radiative corrections to the $Z$-couplings}
\label{radiative}
Our fundamental theory contains the interaction Lagrangian of Eq.~(\ref{intlagrangian}).
%
%
%
Upon integration of the KK modes $Z_n$ and $\gamma_n$ we obtain the effective Lagrangian
\be
\mathcal L_{eff}=\frac{C^{t\ell}_n}{M_n^2}  (\bar t_L\gamma_\mu t)(\ell_L \gamma^\mu \ell_L)
\ee
where~\footnote{In the language of Ref.~\cite{Feruglio:2016gvd} we have
$$C_{3}^{t_L\ell_L}=-\frac{g^2}{4} G_{b_L}^nG_{\ell_L}^n,\quad C_{1}^{t_L\ell_L}=
\frac{g^2 s_W^2}{12 c_W^2} G_{b_L}^nG_{\ell_L}^n \,.$$}
\be
C^{t\ell}_n = -\frac{g^2}{c_W^2}\left( g_{u_L}^Z g_{\ell_L}^Z+g_{u_L}^\gamma g_{\ell_L}^\gamma \right)G_{b_L}^{\,n}G_{\ell_L}^{\,n} \,.
\ee

\begin{figure}[htb]
\centering
\includegraphics[width=7.5cm]{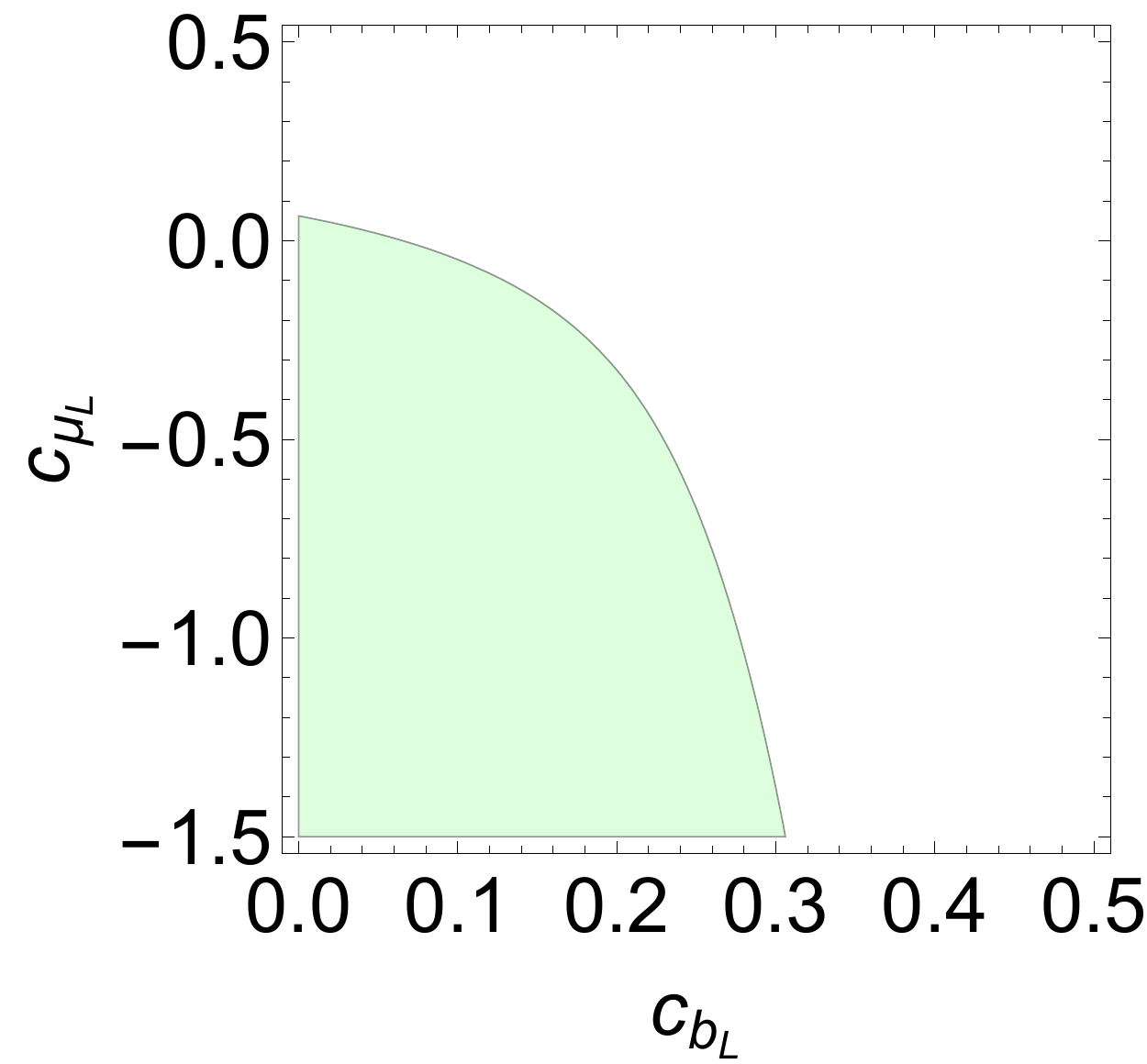}
\caption{\it Region in the plane $(c_{b_L},c_{\mu_L})$ that accommodates $\Delta g_{\mu_L}^Z+\delta g_{\mu_L}^Z$ as shown in Eq.~(\ref{eq:amu}).}
\label{fig:Deltagmu}
\end{figure}

Using the formalism of Ref.~\cite{Feruglio:2016gvd} the RG evolution of the operator $(\bar t_L\gamma_\mu t_L)(\bar\ell_L \gamma_\mu \ell_L)$ gives rise to the operator $(H^\dagger D_\mu H)(\bar\ell_L \gamma^\mu \ell_L)$, which in turn generates the shift $g_{\ell_L}^Z\to g_{\ell_L}^Z+\Delta g_{\ell_L}^Z$ with~\footnote{We are neglecting here the contribution from Yukawa couplings other than the top quark.}
\be
\Delta g_{\ell_L}^Z=\frac{v^2}{M_n^2}\frac{1}{16\pi^2}\left[3y_t^2C^{t\ell}_n \log\frac{M_n}{m_t}
-\frac{g^4}{4 }\left(1-\frac{s_W^4}{9 c_W^4}\right)G_{b_L}^n G_{\ell_L}^n \log\frac{M_n}{m_Z}\right] \,.
\label{eq:Delta_g}
\ee
We can now use the fit from experimental data in Ref.~\cite{ALEPH:2005ab} 
\begin{equation}
g_{\mu_L}^Z=-0.2689\pm 0.0011 
\end{equation}
leading to the result~\footnote{
The recent fit from Ref.~\cite{Falkowski:2017pss} yields $\Delta g_{\mu_L}^Z+\delta g_{\mu_L}^Z=(0.1\pm 1.2)\times 10^{-3}$ fully consistent with Eq.~(\ref{eq:amu}).
}
\be
\Delta g_{\mu_L}^Z+\delta g_{\mu_L}^Z=(0.49\pm 1.1)\times 10^{-3}
\label{eq:amu}
\ee
where $\delta g_{\mu_L}^Z$ stands for the tree-level contribution from the $Z$ and fermion KK-modes in Eq.~(\ref{eq:delta_g}). The resulting $2\sigma$ allowed (white) region is shown in the plot of Fig.~\ref{fig:Deltagmu} in the $(c_{b_L},c_{\mu_L})$ plane. We can see that the permitted region is not in conflict with the plot of Fig.~\ref{fig:RK2}, where the allowed region consistent with the data on $R_K$ was exhibited.

\subsection{LHC Drell-Yan dilepton resonance searches}
\label{dimuons}
\begin{figure}[htb]
\centering
\includegraphics[width=8.cm]{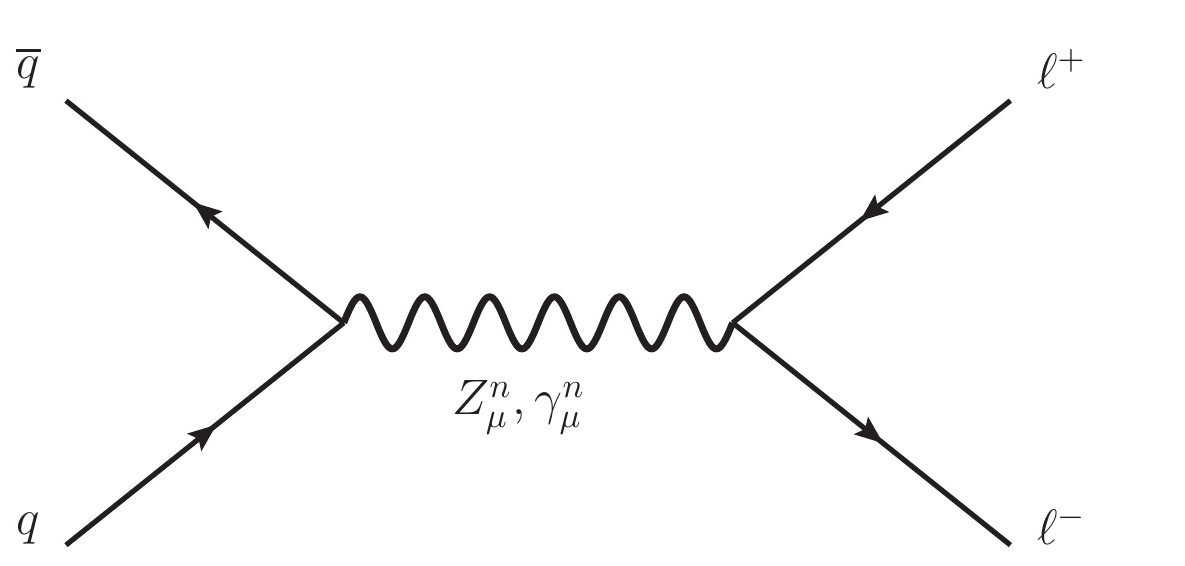} 
\caption{\it 
Diagrams contributing to the process $\sigma(pp\to Z_n/\gamma_n\to \ell^+\ell^-)$.}
\label{fig:dileptons}
\end{figure}

An additional experimental constraint comes from direct searches for high-mass resonances decaying
into dilepton final states. The resonances $Z^n_\mu$ and $\gamma^n_\mu$ can be produced by Drell-Yan processes and decay into a pair of leptons as in Fig~\ref{fig:dileptons}. 

In the narrow width approximation the cross-section for the process $pp\to Z^n/\gamma^n \to \ell^+\ell^-$ approximately scales as
\begin{align}
&\sigma(pp\to Z_n/\gamma_n\to \ell^+\ell^-)\propto A_\ell=\sum_{X=Z,\gamma}A^{X^n}_\ell\,,\nonumber\\
A^{X^n}_\ell&=\frac{(g_{\ell_L}^2+g_{\ell_R}^2)\left(2g_{u_L}^2+2g_{u_R}^2+g_{d_L}^2+g_{d_R}^2\right)}
{\sum_f(g^2_{f_L} + g^2_{f_R})} \,,
\label{brappmu}
\end{align}
where all couplings refer to the $g_{f_{L,R}}^{X^n}$ couplings, and for simplicity we have omitted the superscript $X^n$. In the denominator the sum over $f$ covers the three generation of quarks and leptons in the Standard Model. As this process is flavor conserving we are neglecting here the small correction from mixing angles.

\begin{figure}[!htb]
\centering
\includegraphics[width=7.5cm]{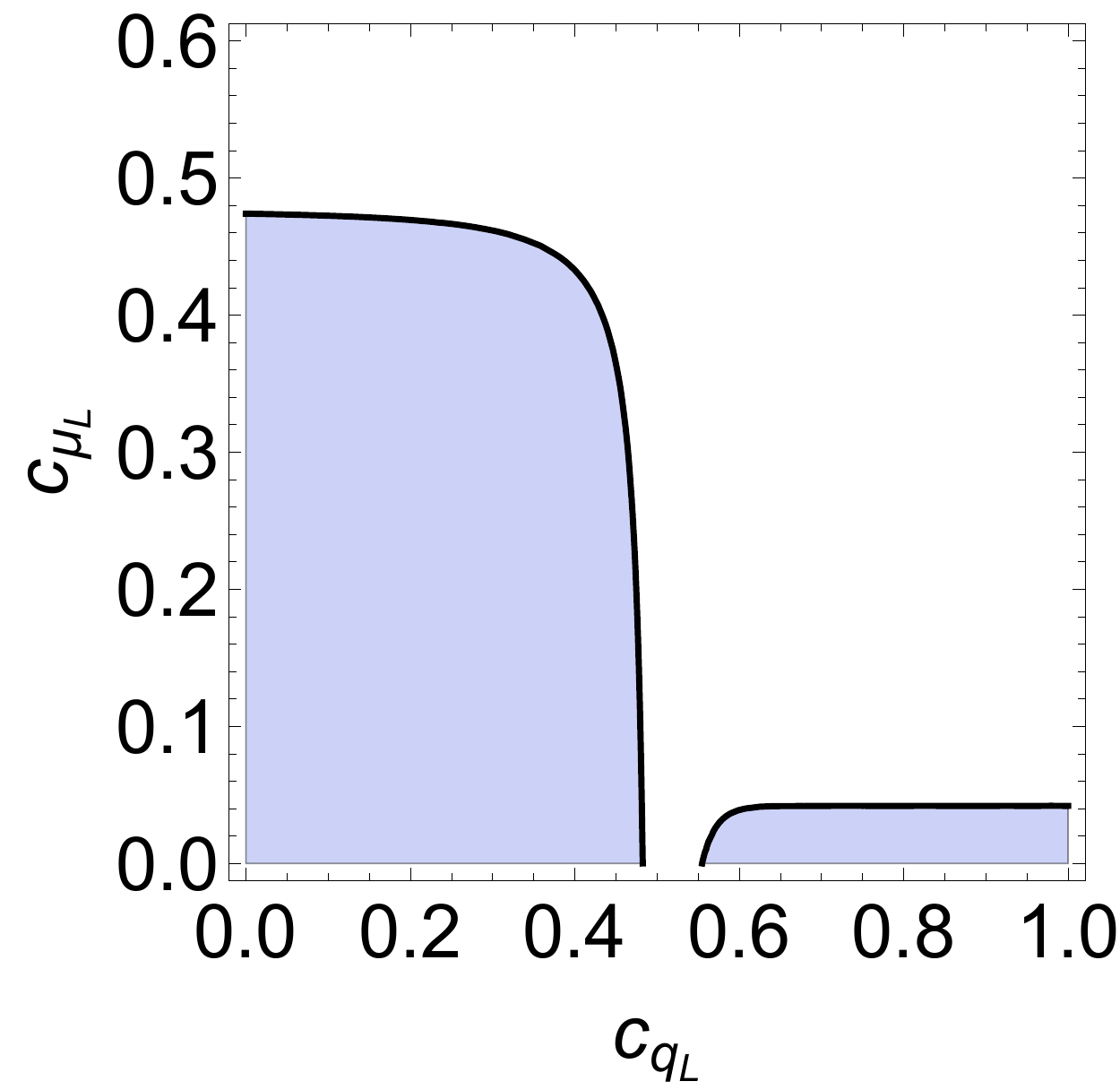}  \hfill
\includegraphics[width=7.5cm]{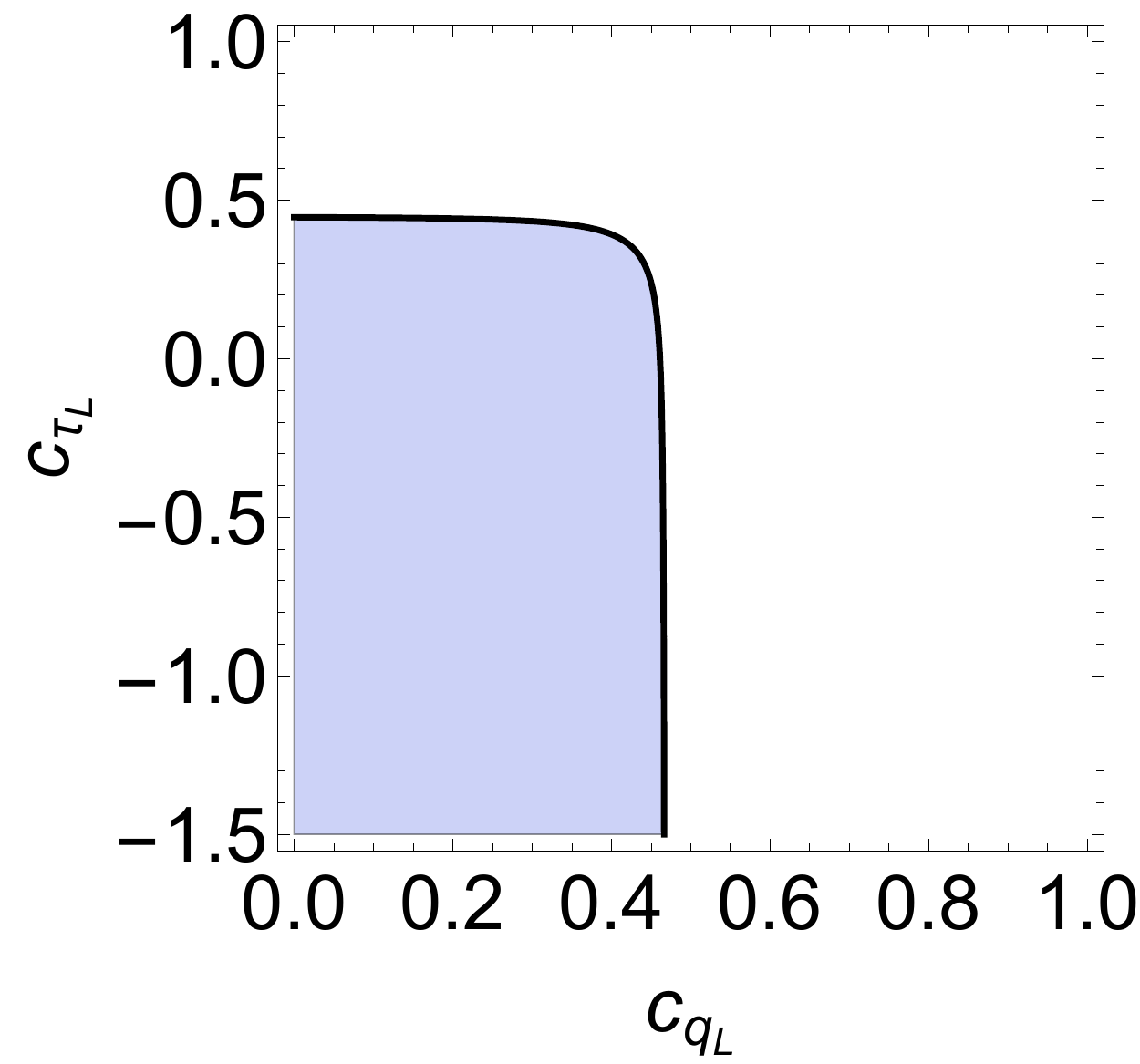} 
\caption{\it
Exclusion region in the plane $(c_{q_L},c_{\mu_L,\tau_L})$ coming from the searches of massive resonances decaying into di-muons $A_\mu < 0.003$ (left panel), and decaying into di-taus $A _\tau< 0.022$ (right panel),  for $M_n = 2$ TeV. We have used $c_{q_R}=0.8$, $c_{\mu_R}=0.65$, $c_{\tau_R}=0.55$, $c_{b_L}=0.2$, $c_{b_R}=0.55$ and $c_{t_R}=0.45$. In the left panel we have considered $c_{\tau_L}=0.1$, and in the right panel~$c_{\mu_L}=0.44$.}
\label{fig:dimuon}
\end{figure}

The best bounds on dimuon resonances have been given by the ATLAS Collaboration~\cite{Aaboud:2016cth} based on 3.2 fb$^{-1}$ data at $\sqrt{s}=13$ TeV. ATLAS obtained a 95\% CL bound on the sequential Standard Model (SSM) $Z^\prime_{SSM}$ gauge boson mass as 
$M_{Z^\prime_{SSM}}\gtrsim 3.36$ TeV. After rescaling the bound we get for our 2 TeV KK-mode the bound $A_\mu\lesssim 0.003$. The allowed region in the plane $(c_{q_L},c_{\mu_L})$ is shown in the left panel of Fig.~\ref{fig:dimuon}. Similarly the strongest bounds on ditau resonances have been obtained by the CMS Collaboration~\cite{Khachatryan:2016qkc} based on 2.2 fb$^{-1}$ data at $\sqrt{s}=13$ TeV. CMS got the 95\% CL bound $M_{Z^\prime_{SSM}}\gtrsim 2.1$ TeV. After rescaling this result it translates into $A_\tau\lesssim 0.022$. The allowed region in the plane $(c_{q_L},c_{\tau_L})$ is shown in the right panel of Fig.~\ref{fig:dimuon}.

Note that for values $c_{\mu_L} > 0.474$ and $c_{\tau_L} > 0.446$ there is no bound on $c_{q_L}$. Alternatively, as we can see from both panels of Fig.~\ref{fig:dimuon}, for $c_{\mu_L}\gtrsim 0.04$ and any value of $c_{\tau_L}$ we obtain the mild bound  $c_{q_L}\gtrsim 0.48$. In summary,
 the constraints on the production of dilepton resonances imply that the first generation of quarks is mostly UV localized (elementary) as expected from their mass spectrum. 

\subsection{Direct Drell-Yan KK gluon searches}
\label{direct}

\begin{figure}[!htb]
\centering
\includegraphics[width=7.cm]{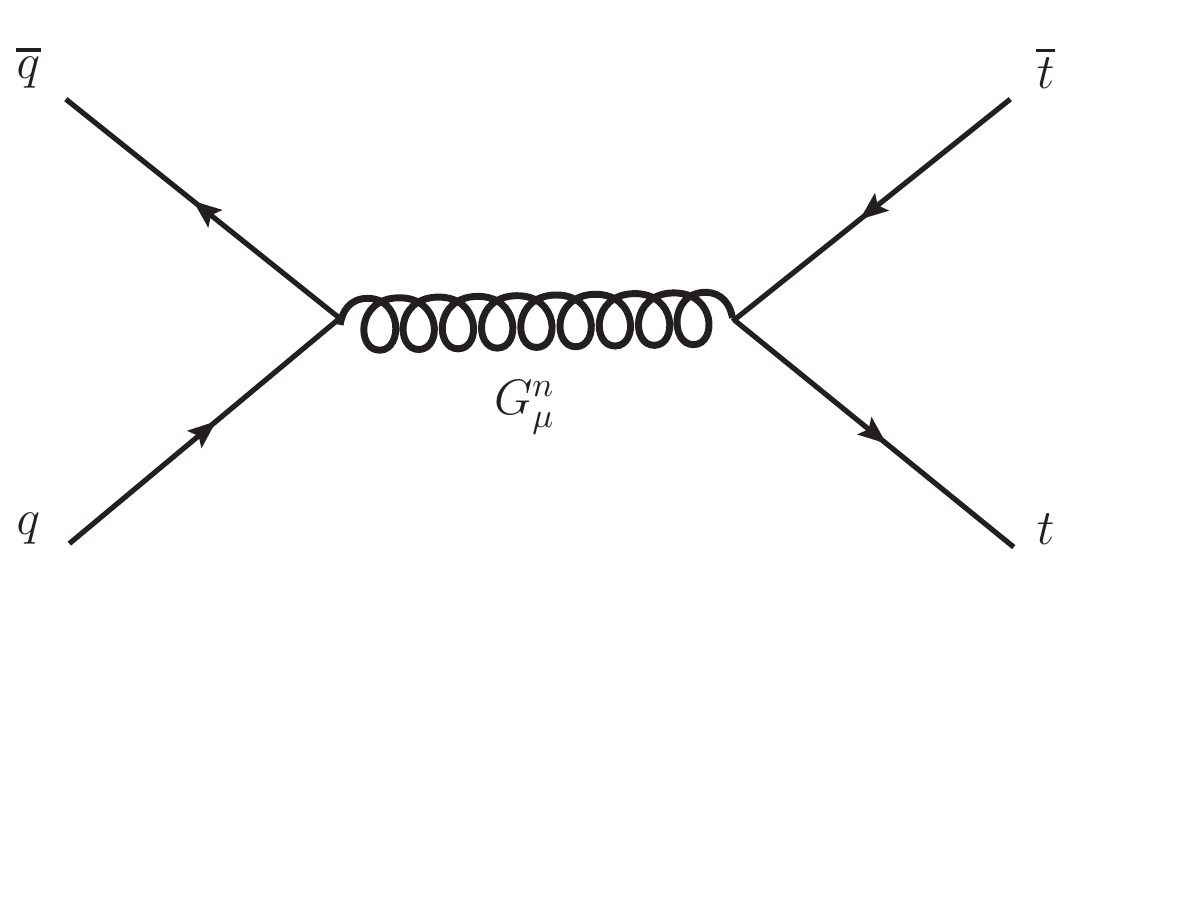} 
\includegraphics[width=7.cm]{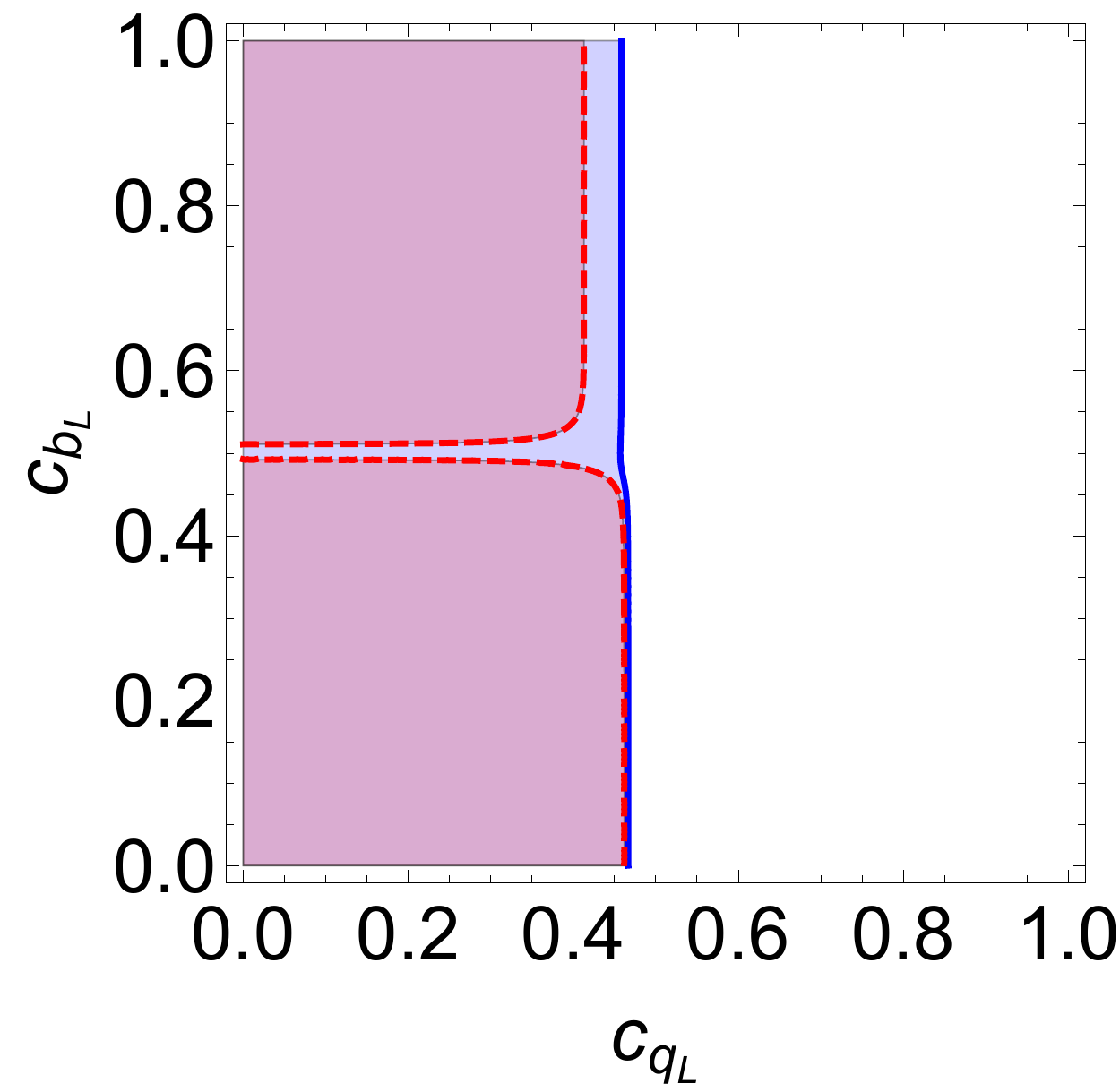} 
\caption{\it Left panel: Diagrams contributing to the process $\sigma(pp\to G^n\to t\bar t)$.
Right panel: Exclusion in the plane $(c_{q_L},c_{b_L})$ coming from ATLAS (red) and CMS (blue) searches of KK gluons decaying into $t\bar t$. We have used $c_{q_R}=0.8$, $c_{b_R}=0.55$ and $c_{t_R}=0.45$.
}
\label{fig:KKgluon}
\end{figure}

Single KK gluons $G^n_\mu$ can be produced at LHC by Drell-Yan processes~\footnote{The vertex $GGG^n$ vanishes by orthonormality of wave functions so that $G^n$ cannot be produced by gluon fusion, unless $G^n$ is emitted by a top-quark loop in which case the production is loop suppressed.}, and decay into top quarks as in the  left panel diagram of Fig.~\ref{fig:KKgluon}.
ATLAS and CMS have considered KK-gluon production in Randall-Sundrum theories~\cite{Randall:1999ee} by the Drell-Yan mechanism. ATLAS~\cite{Aad:2015fna} uses the formalism in Ref.~\cite{Lillie:2007yh}, where they consider $G_{q_{L,R}}^1\simeq -0.2$, for $(q=u,d,c,s)$, $G_{b_R}^1\simeq -0.2$, $G_{t_L}^1\simeq 0.95$ and $G_{t_R}^1\simeq 1.98$. From data at $\sqrt{s}=8$ TeV corresponding to an integrated luminosity of 20.3 fb$^{-1}$ they obtain the 95\% CL lower bound $M^{ATLAS}_1\gtrsim 2.2$ TeV. CMS~\cite{Chatrchyan:2013lca} uses data at $\sqrt{s}=8$ TeV corresponding to an integrated luminosity of 19.7 fb$^{-1}$. Using the formalism of Ref.~\cite{Agashe:2006hk}, where they consider $G_{q_{L,R}}^1\simeq -0.2$, for $(q=u,d,c,s)$, $G_{b_R}^1\simeq -0.2$, $G_{t_L}^1\simeq 1$ and $G_{t_R}^1\simeq 5$, they obtain the 95\% CL lower bound $M^{CMS}_1\gtrsim 2.5$ TeV.

The coupling of the KK-gluon with the fermion $f$ has vector and axial components (unlike the coupling of the gluon zero mode to fermions) and is given by
\be
g_{f\bar f G_n}=g_s \left(G_{f_L}^nP_L+G_{f_R}^nP_R\right)\gamma^\mu t^A
\ee
where $g_s$ is the 4D strong coupling, $t^A$ the $SU(3)$ generators in the triplet representation, $P_{L,R}$ the chirality projectors, and the functions $G_{f_{L,R}}^n$ are defined in Eq.~(\ref{integral}). Therefore the production cross-section, assuming $c_{u_{L,R}}=c_{d_{L,R}}=c_{c_{L,R}}=c_{s_{L,R}}\equiv c_{q_{L,R}}$, scales as
\be
\sigma(pp\to G_n\to t \bar t)\propto\sum_n \frac{1}{M_n^4}\frac{[(G_{q_L}^n)^2+(G_{q_R}^n)^2][(G_{t_L}^n)^2+G_{t_R}^n)^2]}{\sum_{f}[(G_{f_L}^n)^2+(G_{f_R}^n)^2]}
\ee
where the sum over $f$ goes over the three generations of Standard Model quarks, and we are again neglecting the small correction from mixing angles.
By comparison with our model parameters and couplings we can translate the ATLAS and CMS bounds into the exclusion plot in the plane $(c_{q_L},c_{b_L})$, as shown in the right panel of Fig.~\ref{fig:KKgluon}. Notice that given the values of the considered couplings in the ATLAS and CMS models, the CMS bound provides the strongest limit.
In particular, and independently of the value of $c_{b_L}$, searches for KK-gluons lead to the bound $c_{q_L}\gtrsim 0.47$.

\subsection{Dimuon resonance from bottom-bottom fusion}
\begin{figure}[!htb]
\centering
\includegraphics[width=7.cm]{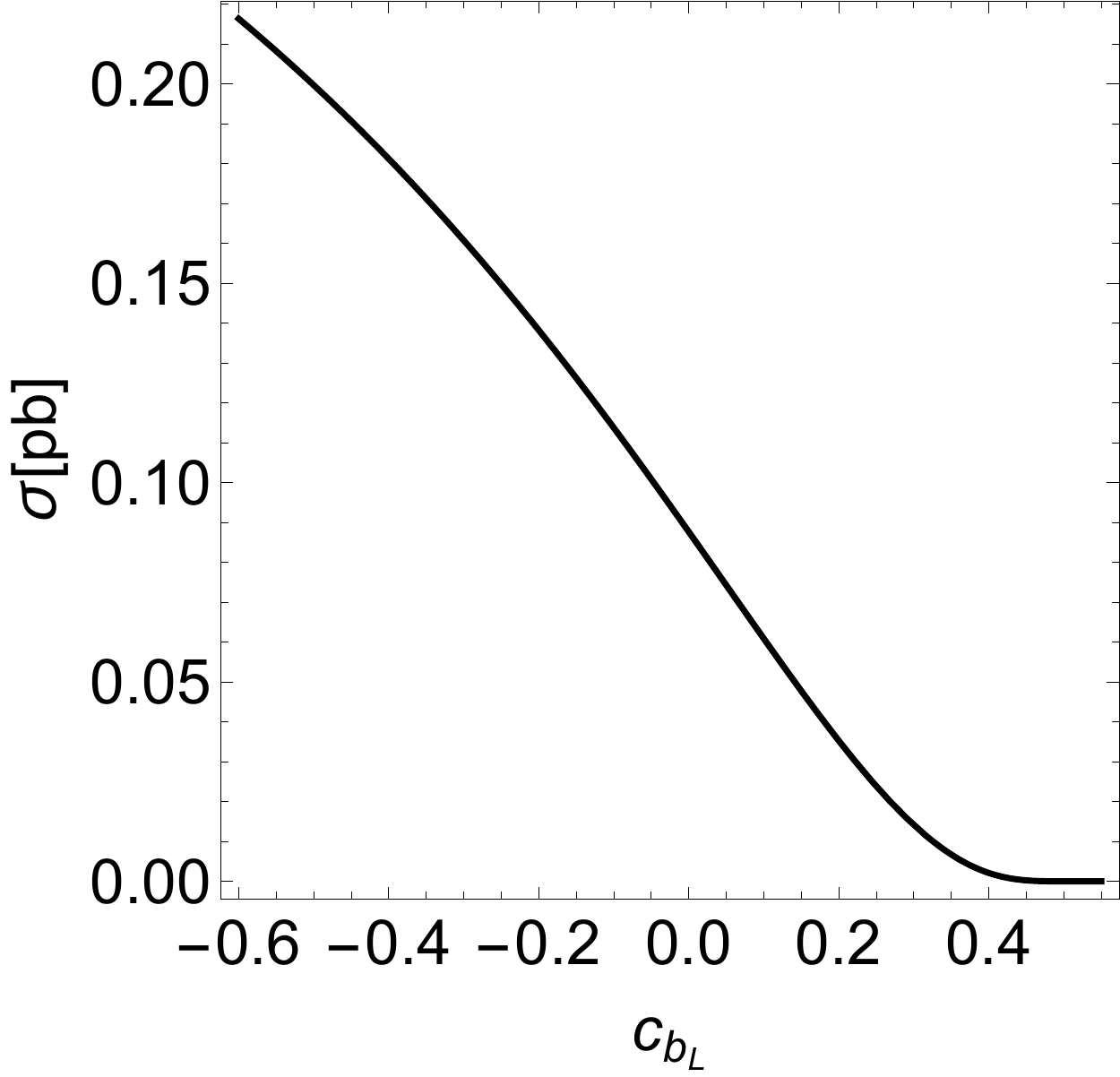} \hspace{0.5cm}\hspace{0.3cm}
\includegraphics[width=7.cm]{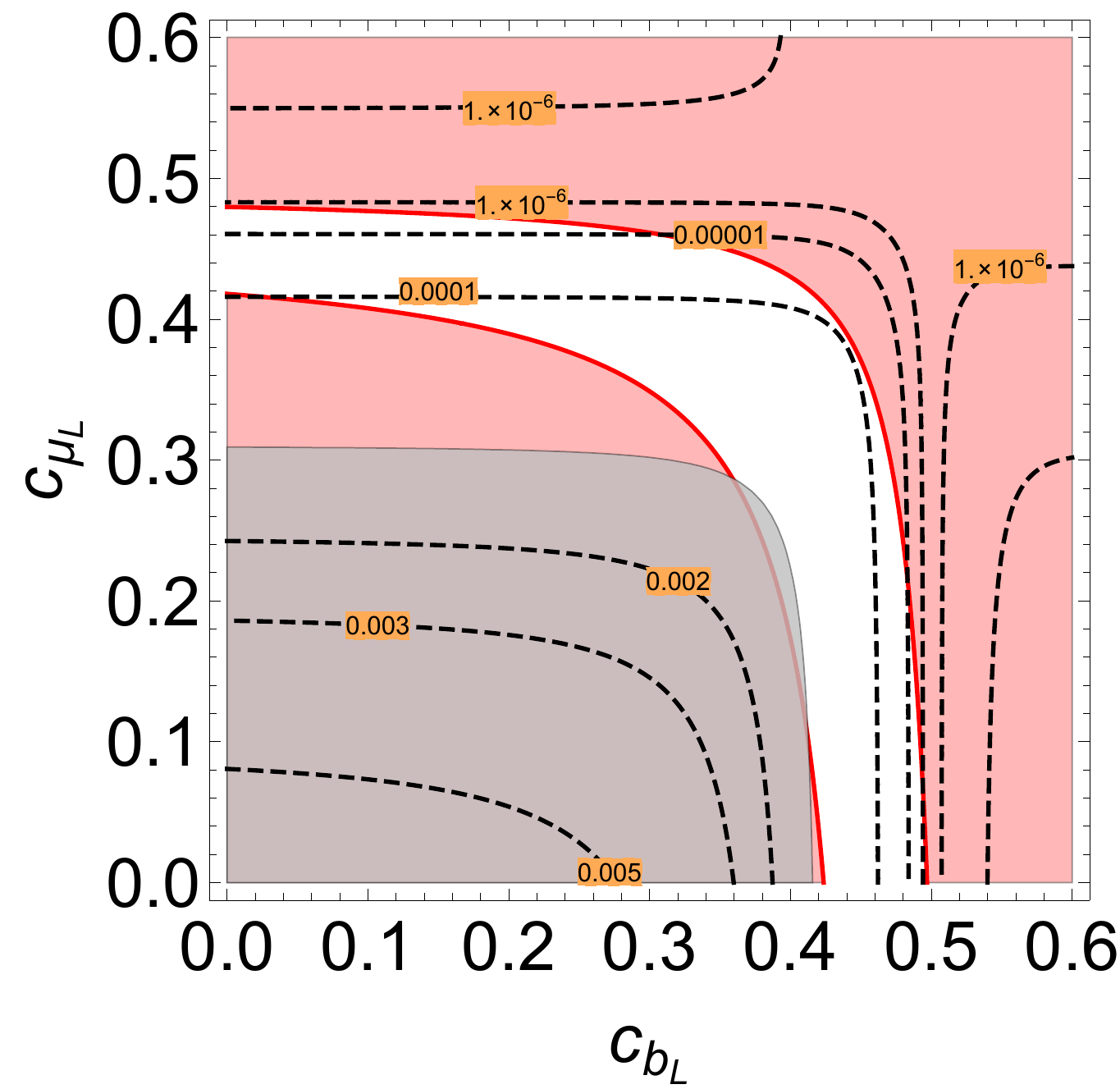} 
\caption{\it Left panel: Cross-section production (in pb) in our model for $Z^n$ ($n=1$) from bottom-bottom fusion in as a function of $c_{b_L}$.
Right panel:  Contour plot of $\sigma \cdot \mathcal B[Z^n \to \mu^+ \mu^-]$ for $n=1$ (in pb) in the plane $(c_{b_L},c_{\mu_L})$. We show as a gray band in the bottom part of the figure the experimentally excluded region $\sigma \cdot \mathcal B > 10^{-3}$.  We overlap as well the allowed region coming from the $R_K$ anomaly. We have considered $c_{\tau_L} = 0.35$.
%
}
\label{fig:dimuons}
\end{figure}
In view of the strong constraints imposed by the $R_{K^{(\ast)}}$ observables on the parameters $c_{b_L}$ and $c_{\mu_L}$, $\mu^+\mu^-$ production from heavy flavor (bottom) annihilation in the colliding protons ($b\bar b\to\mu^+\mu^-$) can be sizeable in spite of its suppression by the small PDFs~\footnote{In fact we can assume here that, for composite enough $b_L$ quarks, and elementary first and second generation quarks, with $c_{q_{L,R}}\gtrsim 0.5$, bottom-bottom fusion can be dominant over the Drell-Yan mechanism for $Z^n$ production at the LHC.}. This issue has been thoroughly analyzed in Refs.~\cite{Faroughy:2016osc,Greljo:2017vvb}. In particular, using the results from Ref.~\cite{Faroughy:2016osc} the cross section for $Z^n$ (with $M_n=2$ TeV, $n=1$) production from bottom-bottom fusion $\sigma(b\bar b\to Z^1$) is shown in the left panel of Fig.~\ref{fig:dimuons} as a function of $c_{b_L}$.
Contour plots of $\sigma\cdot\mathcal B(Z^1\to\mu^+\mu^-)$ are shown in the right panel of Fig.~\ref{fig:dimuons}. The experimental bounds from the ATLAS dilepton search at $\sqrt{s}=13$ TeV and 3.2 fb$^{-1}$~\cite{Aaboud:2016cth} for a vector resonance with 2 TeV mass correspond to $\sigma\mathcal B\lesssim 10^{-3}$ pb at 95\% CL. We can see the corresponding exclusion region in the right panel plot of Fig.~\ref{fig:dimuons}, which we overlap with the region allowed by the $R_K$ anomaly. As we can see from Fig.~\ref{fig:dimuons} most of the space allowed by the $R_K$ anomaly is also allowed by the present LHC bounds on the production of KK $Z$ resonances decaying into dimuons.

\subsection{Flavor observables}
\label{flavor}
 
 New physics contributions to $\Delta F=2$  processes  come  from  the  exchange  of  
 gluon  KK  modes. The leading flavor-violating couplings of the KK gluons $G_{n\mu}^A$ involving the down quarks are given by
\begin{align}
\mathcal L_s 
=& g_sG_{n\mu}^A\Big[
\,\bar d_i\gamma^\mu t_A\left\{ (V_{d_L}^*)_{3i} (V_{d_L})_{3j}\left(G_{b_{L}}^n-G_{q_{L}}^n  \right)P_L\right.\nonumber\\
+&\left.(V_{d_R}^*)_{3i} (V_{d_R})_{3j}\left(G_{b_{R}}^n-G_{q_{R}}^n  \right)P_R\right\} d_j + {\rm h.c.} \Big]\,,\label{eq:lagrangian_QCD}
\end{align}
where $t_A$ are the $SU(3)$ generators in the triplet representation.
After integrating out the massive KK gluons, the couplings in Eq.~(\ref{eq:lagrangian_QCD}) give rise to the following set of $\Delta F=2$ dimension-six operators~\cite{Megias:2016bde}
\begin{align}
{\cal L}_{\Delta F = 2} = \sum_n \Bigg\{&\frac{c_{dij}^{LL(n)}}{M_n^2} (\overline d_{iL} \gamma^\mu d_{jL}) (\overline d_{iL} \gamma_\mu d_{jL})
+ \frac{c_{dij}^{RR(n)}}{M_n^2} (\overline d_{iR} \gamma^\mu d_{jR}) (\overline d_{iR} \gamma_\mu d_{jR})\nonumber\\
&+ \frac{c_{dij}^{LR(n)}}{M_n^2} (\overline d_{iR} d_{jL}) (\overline d_{iL} d_{jR})\Bigg\}\,,
\end{align}
where 
\begin{align}
c^{LL,RR(n)}_{dij} & = \frac{g_s^2}{6}\left[(V_{d_L}^*)_{3i} (V_{d_L})_{3j}\right]^2\left(G_{b_{L,R}}^n-G_{q_{L,R}}^n  \right)^2\,,\nonumber\\
\rule{0pt}{1.5em} c^{LR(n)}_{dij} & = g_s^2\left[ (V_{d_L}^*)_{3i} (V_{d_L})_{3j}\right]\left[ (V_{d_R}^*)_{3i} (V_{d_R})_{3j}\right]\left(G_{b_{L}}^n-G_{q_{L}}^n  \right)
\left(G_{b_{R}}^n-G_{q_{R}}^n  \right)\,.
\end{align}

\begin{figure}[!htb]
\centering
\includegraphics[width=7.cm]{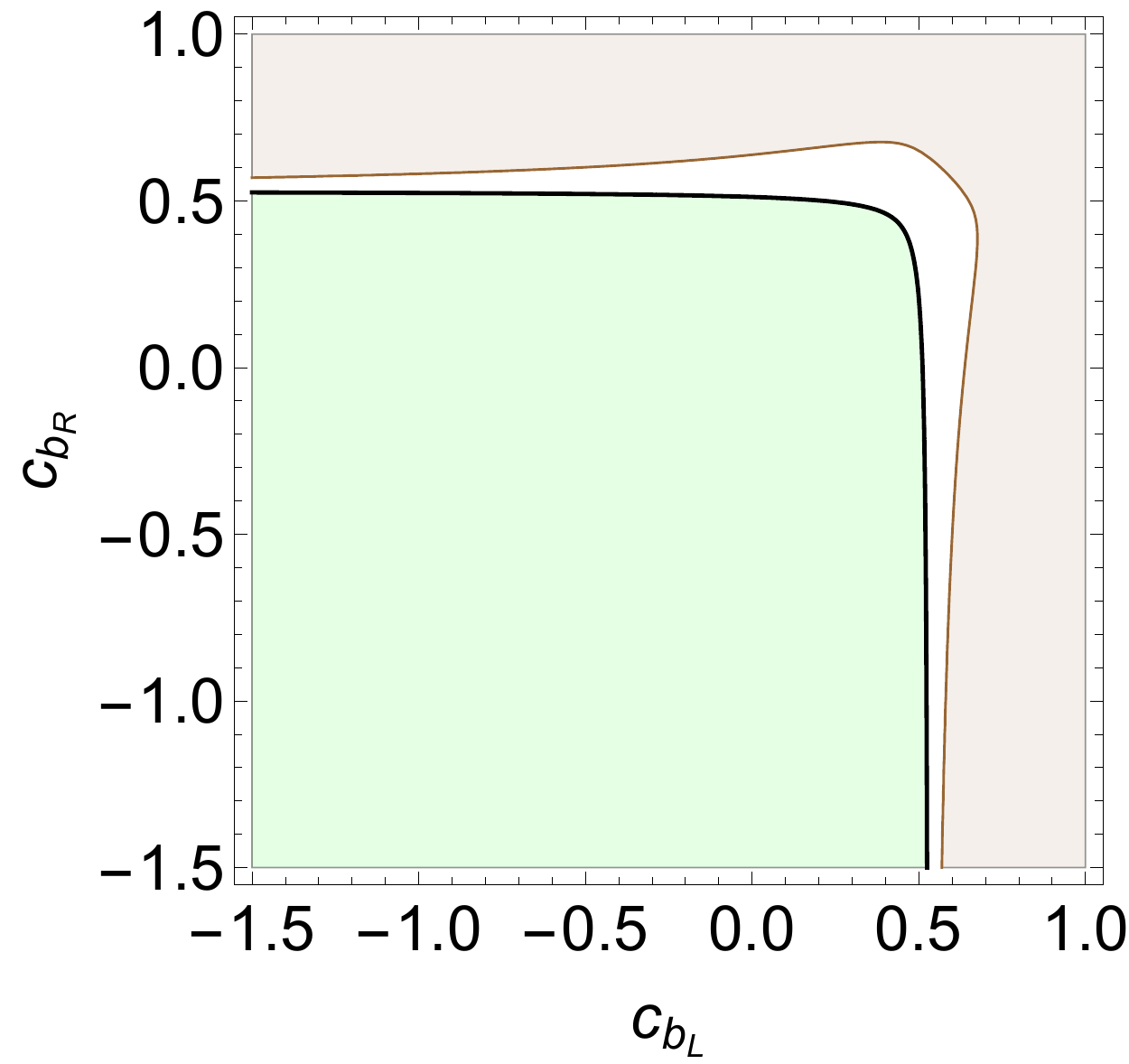} \hspace{0.5cm}
\includegraphics[width=7.cm]{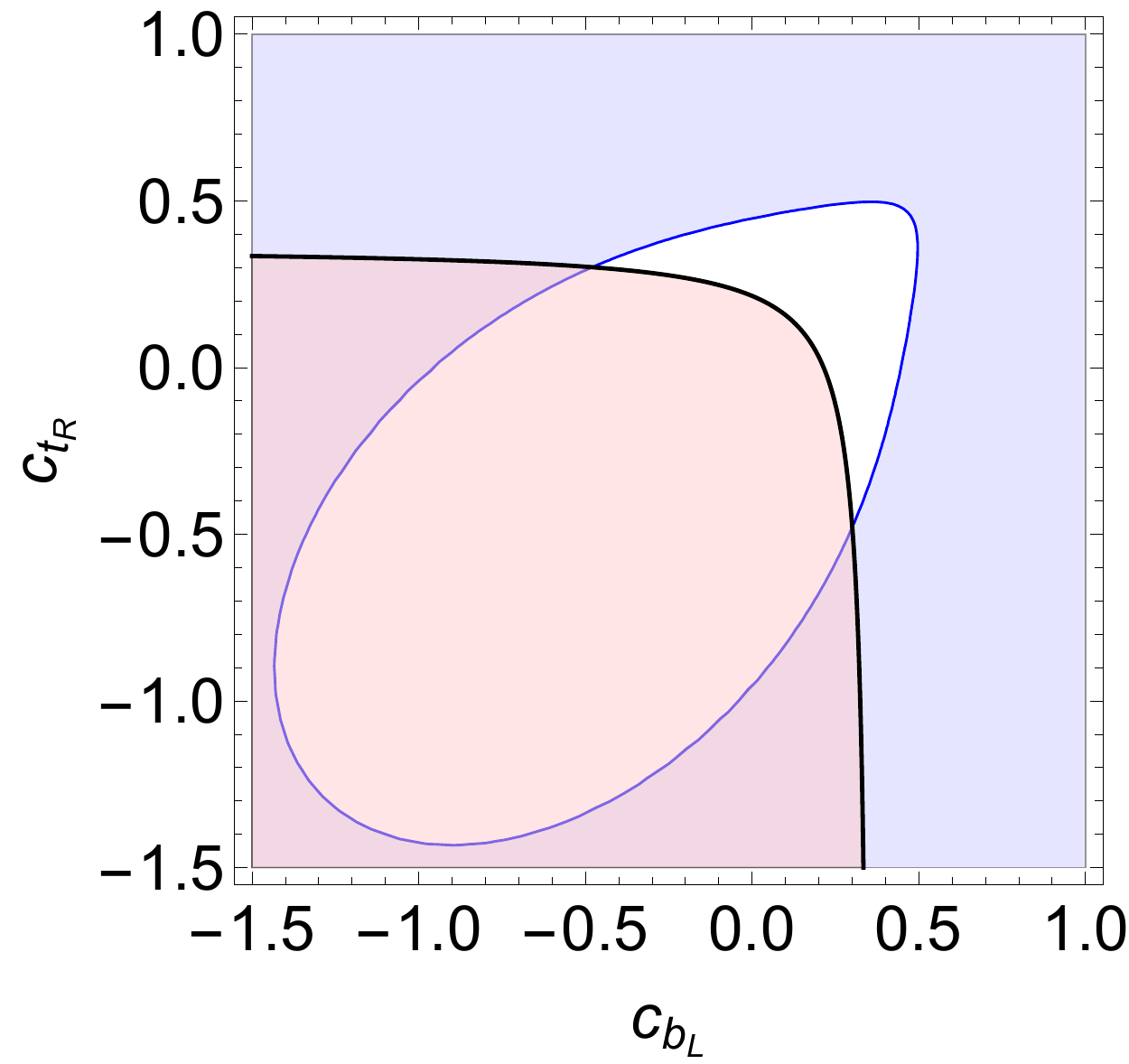} 
\caption{\it Left panel:  Region in the plane $(c_{b_L},c_{b_R})$ compatible with the flavor constraints for down-type quarks, Eqs.~(\ref{eq:bound1Re})-(\ref{eq:boundBsIm}), and the value of the bottom quark mass for  $\sqrt{k}\widehat Y_b<4$. Green region represents the excluded regime from the constraint~(\ref{eq:boundBsIm}).
Right panel: Region in the plane $(c_{b_L},c_{t_R})$ compatible with the flavor constraints for up-type quarks, Eqs.~(\ref{eq:bounduRe})-(\ref{eq:bounduIm}), and top quark mass for $\sqrt{k}\widehat Y_t<4$. Red region represents the excluded regime from the constraint of the second equation in~(\ref{eq:bounduRe}). 
}
\label{fig:flavor2}
\end{figure}

We will assume for the matrices $V_{d_R}$ and $V_{u_R}$ the same structure as for the matrices $V_{d_L}$ and $V_{u_L}$, respectively. 
The strongest current bounds on the $\Delta F=2$ operators come from the operators $(\bar s_{L,R}\gamma^\mu d_{L,R})^2$ and $(\bar s_R d_L)(\bar s_L d_R)$ which contribute to the observables $\Delta m_K$ and $\epsilon_K$ respectively~\cite{Isidori:2015oea}. For the matrix configuration of Eqs.~(\ref{Vd}) and (\ref{Vu}) the experimental bounds on $\Delta m_K$ and $\epsilon_K$ can be translated into the constraints
\begin{align}
\sum_n \frac{(G_{b_{L,R}}^n-G_{q_{L,R}}^n)^2}{M_n^2[{\rm TeV}]} & \leq \frac{1.8}{\lambda_0^2(1-r)^4 \left| (1-\rho_0)^2 - \eta_0^2 \right|} \,, \label{eq:bound1Re} \\
\sum_n \frac{\left(G_{b_{L}}^n-G_{q_{L}}^n  \right)\left(G_{b_{R}}^n-G_{q_{R}}^n  \right)}{M_n^2[{\rm TeV}]} &\leq \frac{0.0023}{\lambda_0^2(1-r)^4 \left| (1-\rho_0)^2 - \eta_0^2 \right|} \,, \label{eq:boundBsRe} \\
\sum_n \frac{(G_{b_{L,R}}^n-G_{q_{L,R}}^n)^2}{M_n^2[{\rm TeV}]} & \leq \frac{0.0034}{\eta_0 \lambda_0^2 (1-r)^4 |1-\rho_0|} \,, \label{eq:bound1Im} \\
\sum_n \frac{\left(G_{b_{L}}^n-G_{q_{L}}^n  \right)\left(G_{b_{R}}^n-G_{q_{R}}^n  \right)}{M_n^2[{\rm TeV}]} &\leq \frac{4.3 \times10^{-6}}{\eta_0 \lambda_0^2 (1-r)^4 |1-\rho_0|} \,, \label{eq:boundBsIm} 
\end{align}
corresponding to the constraints on $\textrm{Re} \, c^{LL,RR(n)}_{d21}$, $\textrm{Re} \, c^{LR(n)}_{d21}$,  $\textrm{Im} \, c^{LL,RR(n)}_{d21}$ and $\textrm{Im} \, c^{LR(n)}_{d21}$ respectively. We display in the left panel of Fig.~\ref{fig:flavor2} these constraints in the plane $(c_{b_L},c_{b_R})$. We have considered for the parameters $\lambda_0$, $\eta_0$ and $\rho_0$ the values
\be
\lambda_0=\lambda \,, \quad \rho_0=0.5 \,, \quad \eta_0 = \eta \,, 
\label{lambda0rho0}
\ee
although other choices would lead to similar constraints.
We display as the green shaded region the constraint from Eq.~(\ref{eq:boundBsIm}). The constraints from Eqs.~(\ref{eq:bound1Re})-(\ref{eq:bound1Im}) are outside the plot range and thus do not interfere with the available region. In this analysis we are taking $c_{q_{L,R}}=0.8$. The white region is where the flavor bounds from Eqs.~(\ref{eq:bound1Re})-(\ref{eq:boundBsIm}) are satisfied, and the bottom mass can be fixed with a 5D Yukawa coupling $\sqrt{k}\widehat Y_b\lesssim 4$.

Moreover, flavor-violating couplings of the KK gluons $G_{n\mu}^A$ involving the up-type quarks are similarly given by
\begin{align}
\mathcal L_s 
=& g_sG_{n\mu}^A\Big[
\,\bar u_i\gamma^\mu t_A\left\{ (V_{u_L}^*)_{3i} (V_{u_L})_{3j}\left(G_{t_{L}}^n-G_{q_{L}}^n  \right)P_L\right.\nonumber\\
+&\left.(V_{u_R}^*)_{3i} (V_{u_R})_{3j}\left(G_{t_{R}}^n-G_{q_{R}}^n  \right)P_R\right\} u_j + {\rm h.c.} \Big]\,.\label{eq:lagrangian_QCD2}
\end{align}
After integrating out the KK gluons, operators as $(\bar c_{L,R}\gamma^\mu u_{L,R})^2$ and $(\bar c_R u_L)(\bar c_L u_R)$, which contribute to the observables $\Delta m_D$ and $\phi_D$~\cite{Isidori:2015oea}, are generated with Wilson coefficients 
\begin{align}
c^{LL,RR(n)}_{uij} & = \frac{g_s^2}{6}\left[(V_{u_L}^*)_{3i} (V_{u_L})_{3j}\right]^2\left(G_{t_{L,R}}^n-G_{q_{L,R}}^n  \right)^2\,,\nonumber\\
\rule{0pt}{1.5em} c^{LR(n)}_{uij} & = g_s^2\left[ (V_{u_L}^*)_{3i} (V_{u_L})_{3j}\right]\left[ (V_{u_R}^*)_{3i} (V_{u_R})_{3j}\right]\left(G_{t_{L}}^n-G_{q_{L}}^n  \right)
\left(G_{t_{R}}^n-G_{q_{R}}^n  \right)\,.
\end{align}
By again assuming that $V_{u_R}$ has the same structure as $V_{u_L}$, the experimental data translate into the bounds
\begin{align}
\sum_n \frac{(G_{t_{L,R}}^n-G_{q_{L,R}}^n)^2}{M_n^2[{\rm TeV}]} & \leq \frac{22.1}{r^2 F},\  
&\sum_n \frac{\left(G_{t_{L}}^n-G_{q_{L}}^n  \right)\left(G_{t_{R}}^n-G_{q_{R}}^n  \right)}{M_n^2[{\rm TeV}]} &  \leq \frac{0.375}{r^2 F} \,, \label{eq:bounduRe}  \\
\sum_n \frac{(G_{t_{L,R}}^n-G_{q_{L,R}}^n)^2}{M_n^2[{\rm TeV}]} & \leq \frac{1.97}{r^2 G},  &
\sum_n \frac{\left(G_{t_{L}}^n-G_{q_{L}}^n  \right)\left(G_{t_{R}}^n-G_{q_{R}}^n  \right)}{M_n^2[{\rm TeV}]} &\leq \frac{0.036}{r^2 G} \,,  \label{eq:bounduIm}
\end{align}
corresponding to the constraints on $\textrm{Re} \, c^{LL,RR(n)}_{u21}$ and $\textrm{Re} \, c^{LR(n)}_{u21}$ in Eq.~(\ref{eq:bounduRe}), and  $\textrm{Im} \, c^{LL,RR(n)}_{u21}$ and $\textrm{Im} \, c^{LR(n)}_{u21}$ in Eq.~(\ref{eq:bounduIm}), respectively. In these equations the functions $F$ and $G$ are defined by
\begin{align}
F&=\left| (1-r)^2 \left[ \left( 1 - \frac{\lambda_0}{\lambda}(1-\rho_0)\right)^2 - \left( \frac{\lambda_0}{\lambda} \eta_0 \right)^2 \right] \right. \nonumber\\   
  &
  +\left. 2(1-r) \left[ \eta_0 \eta \frac{\lambda_0}{\lambda} - \rho \left( 1 - \frac{\lambda_0}{\lambda}(1-\rho_0) \right)  \right] +  \rho^2 - \eta^2  \right| \label{F}\\
  G&= \left| (1-r)^2 \eta_0 \frac{\lambda_0}{\lambda} \left[1 - \frac{\lambda_0}{\lambda} (1-\rho_0) \right] \right. \nonumber\\
  & - \left. (1-r) \left[\eta \left(1 + \frac{\lambda_0}{\lambda}(-1+\rho_0)\right) + \eta_0 \frac{\lambda_0}{\lambda} \rho \right]   + \eta \rho \right| \label{G} \,.
\end{align}
We show in the right panel of Fig.~\ref{fig:flavor2} these constraints in the plane $(c_{b_L},c_{t_R})$ for $r=0.75$ and the values of $\lambda_0$, $\rho_0$ and $\eta_0$ from Eq.~(\ref{lambda0rho0}). 
The constraints from the first Eq.~(\ref{eq:bounduRe}) and Eqs.~(\ref{eq:bounduIm}) are out of the plot range in this case. The white area is the region that can accommodate the top quark mass for 5D Yukawa couplings $\sqrt{k}\widehat Y_t\lesssim 4$.

\section{The $b\to s \nu\overline\nu$ and $b\to s \tau\tau$ modes}
\label{processes}
If there is a contribution to the process $\bar B\to \bar K \mu\mu$, contributions to the processes $\bar B\to \bar K \tau\tau$ and $\bar B\to \bar K \nu\bar\nu$ will also be generated. 
We will start by considering the process $\bar B\to \bar K \nu\bar\nu$ and define
\be
R_K^{\,\nu} =\frac{\mathcal B(\bar B\to \bar K \nu\bar\nu)}{\mathcal B(\bar B\to \bar K \nu\bar\nu)_{\rm SM}} \,.
\ee
This process is encoded by the effective operators
\begin{align}
\mathcal O^{ij}_ \nu&=(\bar s_L\gamma^\mu b_L)(\bar \nu^i  \gamma_\mu(1-\gamma_5)\nu^j) \,, \nonumber\\
\mathcal O^{\prime ij}_ \nu&=(\bar s_R\gamma^\mu b_R)(\bar \nu^i  \gamma_\mu(1-\gamma_5)\nu^j) \,, 
 \end{align}
 generated by the Lagrangian
 \be
 \mathcal L=\frac{g}{4 c_W}Z^n_\mu \bar\nu U^\dagger\gamma_\mu (1-\gamma_5) G^n U\nu \,,
 \ee
where $U$ is the Pontecorvo-Maki-Nakagawa-Sakata (PMNS) matrix~\cite{Olive:2016xmw} and $G^n=\diag(G^n_{e_L},G^n_{\mu_L},G^n_{\tau_L})$. By defining the Wilson coefficients
\begin{equation}
\Delta C_\nu^{(\prime)ij}=\Delta C_\nu^{(\prime)}
(U^\dagger G^n U)^{ij}
\ee
where
\begin{equation}
\Delta C_\nu^{(\prime)}=-\frac{(1-r)\pi g^2 (g^{Z_n}_{b_{L(R)}}-g^{Z_n}_{s_{L(R)}})}{2\sqrt{2}G_F\alpha c_W^2 M_n^2} \,,
\ee
we can write
\be
R_K^{\,\nu}=\frac{\sum_\ell |C_\nu^{\rm SM}+(\Delta C_\nu+\Delta C_\nu^\prime) G_{\ell_L}^n|^2}{3|C_\nu^{\rm SM}|^2}
\ee
where $C_\nu^{\rm SM}=-6.4$. The present experimental bound on the branching ratio is $\mathcal B(\bar B\to \bar K \nu\bar\nu)<3.2\times 10^{-5}$~\cite{Lees:2013kla} at 90\% CL, while the Standard Model prediction is $\mathcal B(\bar B\to \bar K \nu\bar\nu)_{\rm SM}=(4.5\pm 0.7)\times 10^{-6}$. This yields $R_K^{\,\nu}<7.11$ at $90\%$ CL.
\begin{figure}[htb]
\centering
\includegraphics[width=7.7cm]{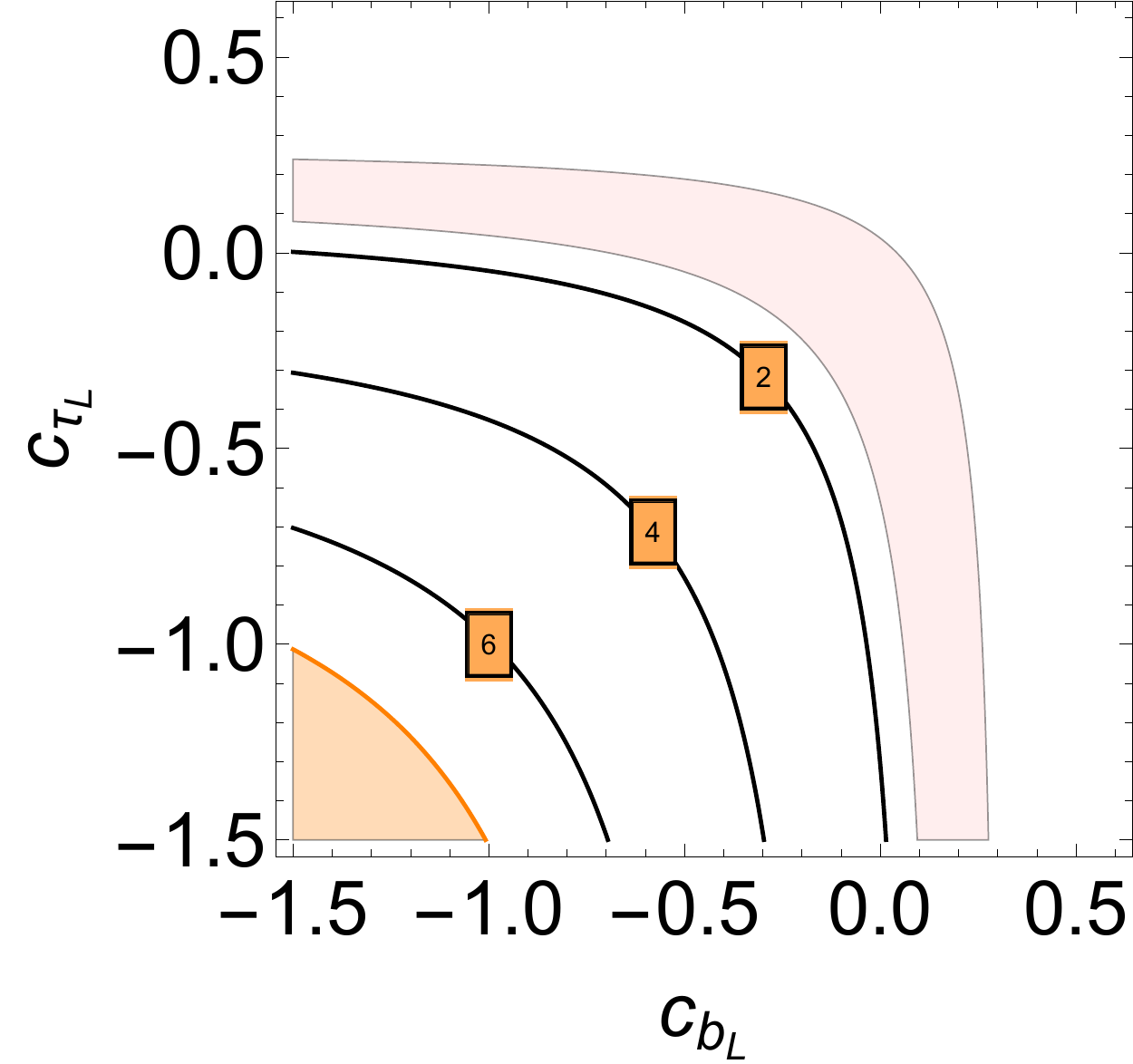} 
\includegraphics[width=7.7cm]{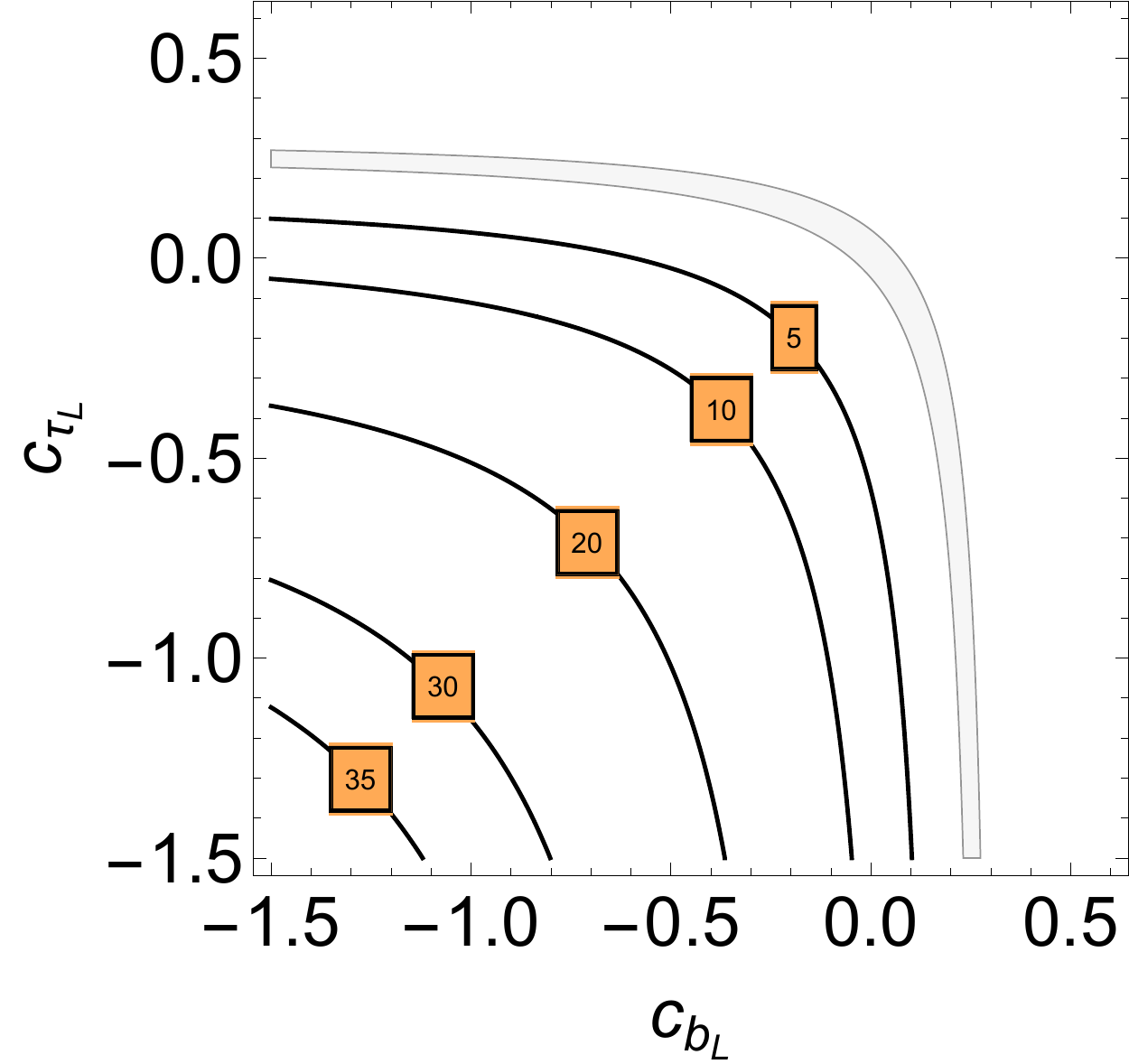} 
\caption{\it Left panel: Region in the plane $(c_{b_L},c_{\tau_L})$ that accommodates $R_K^{\,\nu} < 7.11$ (orange region is the excluded regime). We also display as a red band the interval $0.7 < R_K^{\,\nu} <1.5$, which is a region close to the SM value, and as solid lines the values $R_K^{\,\nu}=2,4,6$. Right panel: Branching ratio $R_K^{\,\tau}$ in the plane $(c_{b_L},c_{\tau_L})$. The gray band corresponds to the interval $0.7<R_K^{\,\tau}<1.5$, which is a region close to the SM value, and the solid lines are the values of $R_K^{\,\tau}$ from $5$ to $35$. We have considered $r=0.75$ and $c_{\mu_L}=0.44$.}
\label{fig:RKtau}
\end{figure} 
In the left panel of Fig.~\ref{fig:RKtau} we show the prediction of $R_K^{\,\nu}$ in the plane $(c_{b_L},c_{\tau_L})$ in our model for $r=0.75$. The orange shadowed region is excluded at $90\%$ CL. The red band is the region for the interval $0.7 < R_K^{\,\nu} <1.5$, corresponding to a possible future measurement of the observable $R_K^{\,\nu}$ close to its Standard Model prediction. Moreover a measurement of $R_K^{\,\nu}$ much larger than the Standard Model prediction would still be possible and be a smoking gun for the model.

Finally, the branching fraction $\mathcal B(\bar B\to\bar K\tau\tau)$, in particular the ratio
\be
R_K^{\,\tau}=\frac{\mathcal B(\bar B\to\bar K \tau\tau)}{\mathcal B(\bar B\to \bar K \tau\tau)_{\rm SM}}=\frac{\left|C_9^{\rm SM}+\Delta C_9^\tau+\Delta C_9^{\prime\,\tau} \right|^2+\left|C_{10}^{\rm SM}+\Delta C_{10}^\tau+\Delta C_{10}^{\prime\,\tau} \right|^2}{2\left|C_9^{\rm SM} \right|^2} 
\ee
 could also be the smoking gun for our model. It has been recently measured by the BaBar Collaboration providing the 90\% CL bound $\mathcal B(\bar B\to\bar K\tau\tau)< 2.25 \times 10^{-3}$~\cite{TheBaBar:2016xwe}, much larger than the Standard Model prediction $\mathcal B(\bar B\to\bar K\tau\tau)_{\rm SM}=(1.44\pm 0.15)\times 10^{-7}$~\cite{Bouchard:2013mia}, and thus leading to the mild bound $R_K^{\,\tau}<1.6\times 10^{4}$. Even the future sensitivity of Belle II $\mathcal B(\bar B\to\bar K\tau\tau)< 2 \times 10^{-4}$~\cite{Bhattacharya:2016mcc} seems to be far away from the Standard Model value. In the right panel of Fig.~\ref{fig:RKtau} we show, in the plane $(c_{b_L},c_{\tau_L})$, contour plots of the ratio $R_K^{\,\tau}$. As we can see the expected Belle II range will not interfere with the allowed region. The gray band corresponds to the interval $0.7<R_k^{\,\tau}<1.5$ corresponding to a possible future measurement of $R_K^{\,\tau}$ close to its Standard Model prediction. Again a hypothetical measurement of $R_K^{\,\tau}$ much larger than the Standard Model prediction would still be possible.

\section{Lepton-flavor universality violation in $R_{D^{(\ast)}}$}
\label{RD}

The charged current decays $\overline{B}\to D^{(\ast)}\ell^- \overline{\nu}_\ell$ have been measured by the BaBar~\cite{Lees:2012xj,Lees:2013uzd}, Belle~\cite{Huschle:2015rga,Sato:2016svk,Abdesselam:2016xqt,Hirose:2016wfn} and LHCb~\cite{Aaij:2015yra} Collaborations. In particular they measure the quantities
\be
R_{D^{(\ast)}}\equiv R_{D^{(\ast)}}^{\tau/\ell}=\frac{\mathcal B(\overline{B}\to D^{(\ast)}\tau^- \overline{\nu}_\tau)}{\mathcal B(\overline{B}\to D^{(\ast)}\ell^- \overline{\nu}_\ell)}\quad (\ell=\mu \textrm{ or }e),
\ee
with the experimental result~\cite{Amhis:2016xyh,Papucci} 
\begin{equation}
R_D^{\rm exp}=0.403\pm 0.047, \quad
R_{D^*}^{\rm exp}=0.310\pm 0.017, \quad \rho= -0.23 
\end{equation} 
as averaged by the heavy flavor averaging group (HFAG), which differs from the current Standard Model calculation~\cite{Amhis:2016xyh}
\begin{equation}
R_D^{\rm SM}=0.300\pm 0.011,\quad
R_{D^*}^{\rm SM}=0.254\pm 0.004 
\end{equation} 
by $2.2\sigma$ and $3.3\sigma$, respectively, although the combined deviation is $\gtrsim 4\sigma$. This is exhibited in the plot of Fig.~\ref{fig:RDexp} where we show, in the plane $(R_D,R_{D^*})$, contour lines of $1\sigma$ (solid), $2\sigma$ (dashed), $3\sigma$ (dot-dashed) and $4\sigma$ (dotted), as well as the spot with the Standard Model prediction.
\begin{figure}[htb]
\centering
\includegraphics[width=8.cm]{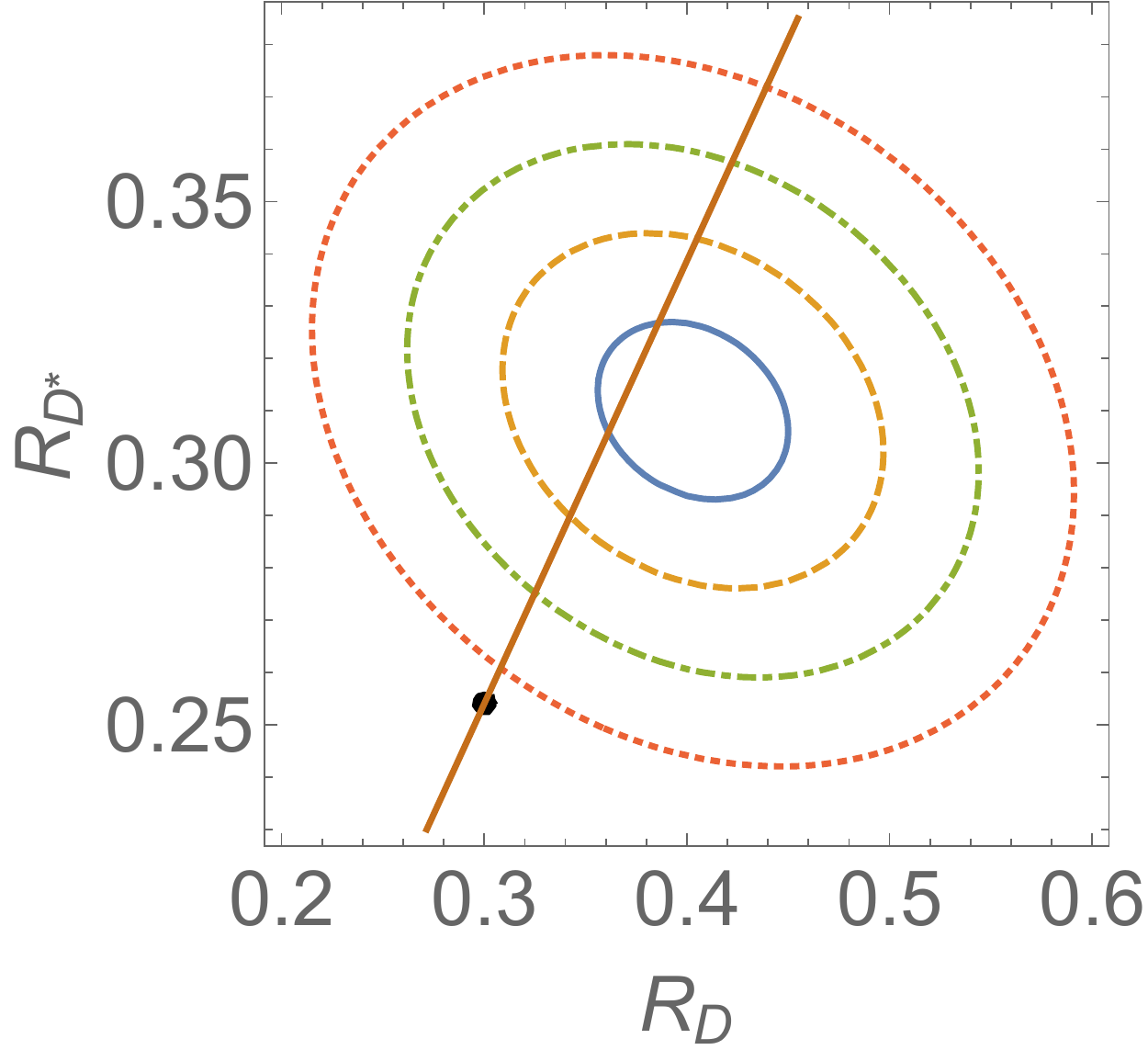}
\caption{\it Experimental region in the $(R_D,R_{D^*})$ plane from 1$\sigma$ (inner ellipse)--4$\sigma$ (outer ellipse). The black spot is the Standard Model prediction. The straight line is where our model prediction lies.}
\label{fig:RDexp}
\end{figure} 

The 4D charged current interaction Lagrangian of the KK modes $W^{(n)}_\mu$ with quarks and leptons can be written, in the mass eigenstate basis, as
\begin{align}
\mathcal L&= \frac{g}{\sqrt{2}}\sum_nW_\mu^{(n)}\bar u_i \left[G_{d_{i_L}}^n V_{ik}+(V_{u_L}^\dagger)_{ij}(G_{d_{j_L}}^n-G_{d_{i_L}}^n)(V_{d_L})_{jk}
\right]\gamma^\mu P_L d_k\nonumber\\
&
+\frac{g}{\sqrt{2}}\sum_nW_\mu^{(n)}\bar\ell_i G_{\ell_{i_L}}^n U_{ij}\gamma^\mu P_L \nu_{j} \,,
\end{align}
where $i,j,\dots$ are flavor indices, and $V_{u_L}$ ($V_{d_L}$) is the unitary matrix diagonalizing the up (down) quark mass matrix~\footnote{To \textit{prevent lepton flavor violation} in our theory, we are assuming that the 5D Yukawa couplings $\widehat Y_\ell$ are such that the charged leptons are diagonal in the interaction basis, so that $V_{\ell_{L,R}}\simeq 1$.}. 

After integrating out the KK modes $W^{(n)}$ one obtains the effective Lagrangian
\be
\mathcal L_{eff}=-\frac{4G_F}{\sqrt{2}} V_{cb}\sum_n\left[ C^{\tau}_n(\bar c\gamma^\nu P_L b)\bar\tau\gamma_\nu (U\nu)_\tau+ C^{\mu}_n(\bar c\gamma^\nu P_L b)\bar\mu\gamma_\nu (U\nu)_\mu\right]
\label{effectiveL}
\ee
where the Wilson coefficients $C^{\tau,\mu}_n$ are given by
\be
C^{\tau,\mu}_n = \frac{m_W^2}{m_{W^{(n)}}^2}\left[ G_{s_L}^n+\frac{(V_{u_L}^\dagger)_{21} (V_{d_L})_{13}}{V_{cb}}(G_{d_L}^n-G_{s_L}^n)+  \frac{(V_{u_L}^\dagger)_{23} (V_{d_L})_{33}}{V_{cb}}(G_{b_L}^n-G_{s_L}^n)
\right]G_{\tau_L,\mu_L}^n  \,.
\ee

In case the first and second generation quarks respect the universality condition, the Wilson coefficients can be written as
\be
C^{\tau,\,\mu}_n=\frac{m_W^2}{m_{W^{(n)}}^2}\left[ G_{q_L}^n+ r(G_{b_L}^n-G_{q_L}^n)
\right]G_{\tau_L,\,\mu_L}^n\ 
\label{Csimp}
\ee
and the coefficient $r$ is given by the ratio
\be
r=\frac{(V_{u_L}^*)_{32} (V_{d_L})_{33}}{V_{cb}}\ .
\label{coefr}
\ee


The corrections to the $R_{D^{(\ast)}}$ observables from the effective operators are given, in terms of the Wilson coefficients, as~\cite{Bhattacharya:2016mcc}~\footnote{We are keeping here the leading contribution from the first KK mode $(n=1)$ with mass $m_{W^{(1)}}\equiv m_{KK}$ and will suppress the KK-index $n$. Also notice that the normalization is such that, for the Standard Model, $C^{\tau,\mu}_{\rm SM}=1$.}
\be
R_{D^{(\ast)}}(C^\tau,C^\mu)=2 R_{D^{(\ast)}}^{\rm SM}\frac{\left|1+
C^\tau  \right|^2
}{1+\left|1+
C^\mu  \right|^2 
}\ .
\label{expresion}
\ee
%
%
\begin{figure}[!htb]
\centering
\includegraphics[width=7.3cm]{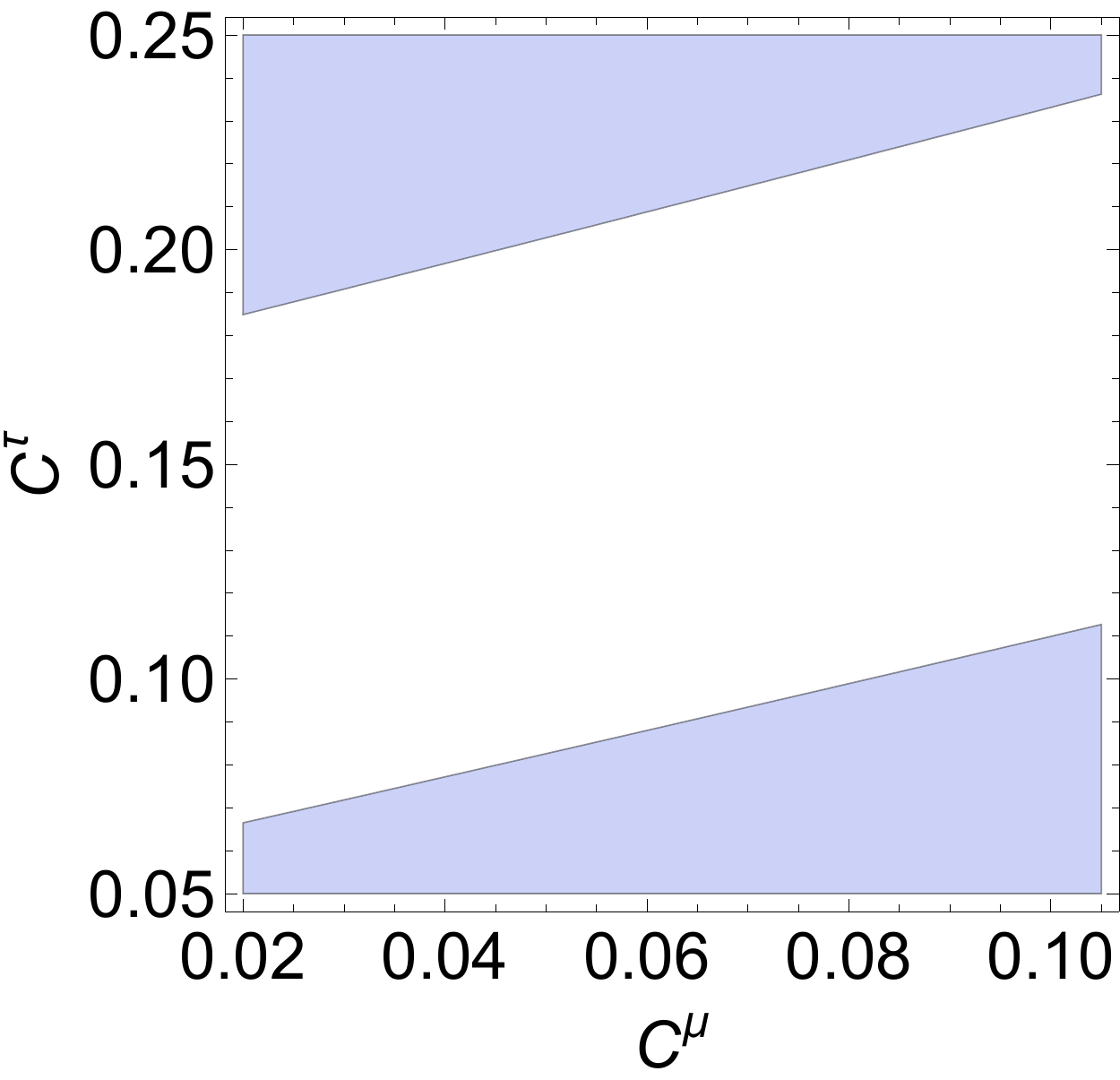} 
\includegraphics[width=7.5cm]{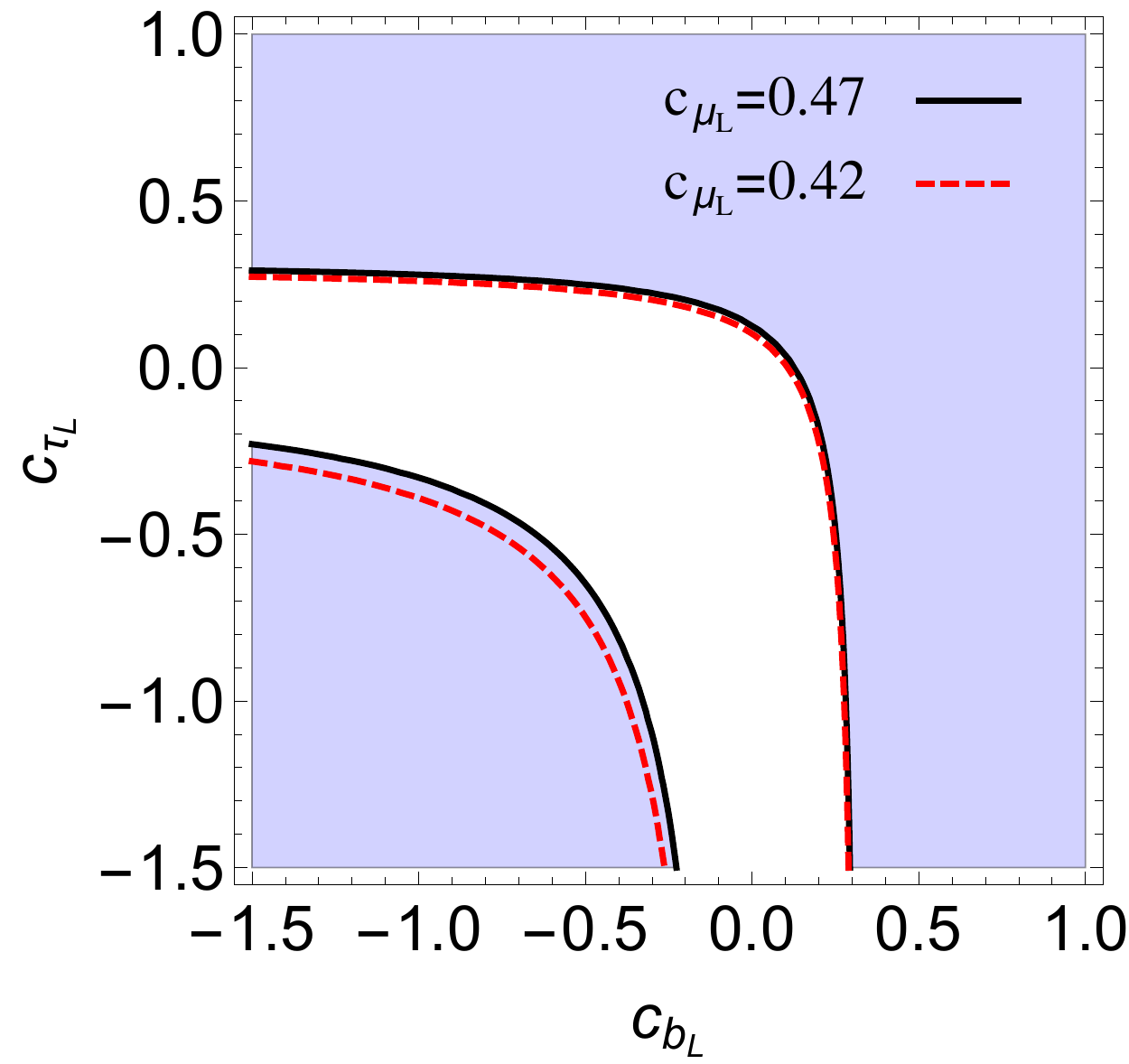} 
\caption{\it 
Left panel: Allowed region in the plane $(C^\mu,C^\tau)$ at the 95\% CL.
Right panel: Corresponding allowed region in the $(c_{b_L},c_{\tau_L})$ plane at the 95\% CL coming from the $(R_D,R_{D^*})$ observables. We have considered $c_{\mu_L}=0.47$ (black lines and shaded area). We also display the limit of the allowed region for the case $c_{\mu_L}=0.42$ (dashed red lines). We have considered $r=0.75$, $m_{KK} = 2$ TeV, and $c_{q_L}=0.8$.
}
\label{fig:RD2}
\end{figure} 
This gives the model prediction along the straight line in Fig.~\ref{fig:RDexp}. Eq.~(\ref{expresion}) translates into the allowed region at the 95\% CL shown in the left panel of Fig.~\ref{fig:RD2}.

The relevant functions in the definition of $C^{\tau,\,\mu}$, $G_{b_L}$, $G_{\tau_L}$ and $G_{\mu_L}$, depend on the three constants $c_{b_L},c_{\tau_L},c_{\mu_L}$, which in turn determine the localization of the third generation left-handed quark doublet and third and second generation of left-handed lepton doublets, respectively. Therefore using Eq.~(\ref{expresion}) we get that the model predictions for $R_{D^{(\ast)}}$ do depend on the constants $c_{b_L},c_{\tau_L},c_{\mu_L}$. The corresponding 95\% CL allowed region in the plane $(c_{b_L},c_{\tau_L})$ is shown in the right panel of Fig.~\ref{fig:RD2} for the two chosen values $c_{\mu_L}=0.47,0.42$. We can see in the plot a mild dependence on the value of the parameter $c_{\mu_L}$. A pretty clear consequence of the plot in the right panel of Fig.~\ref{fig:RD2} is that both $b_L$ and $\tau_L$ fermions are localized towards the IR and thus show an important degree of compositeness in the dual theory. In particular we can see that $c_{b_L}\lesssim 0.29$ and $c_{\tau_L}\lesssim 0.29$. Notice that there is no problem to adjust their masses, for $\mathcal O(1)$ values of the (dimensionless) 5D Yukawa couplings $\sqrt{k}\widehat Y_{b,\tau}$, provided that their right-handed partners $b_R$ and $\tau_R$ are mostly elementary fermions and thus localized towards the UV brane, as we are assuming in this paper.

\subsection{Ditau resonance from bottom-bottom fusion}
\begin{figure}[!htb]
\centering
\includegraphics[width=7.5cm]{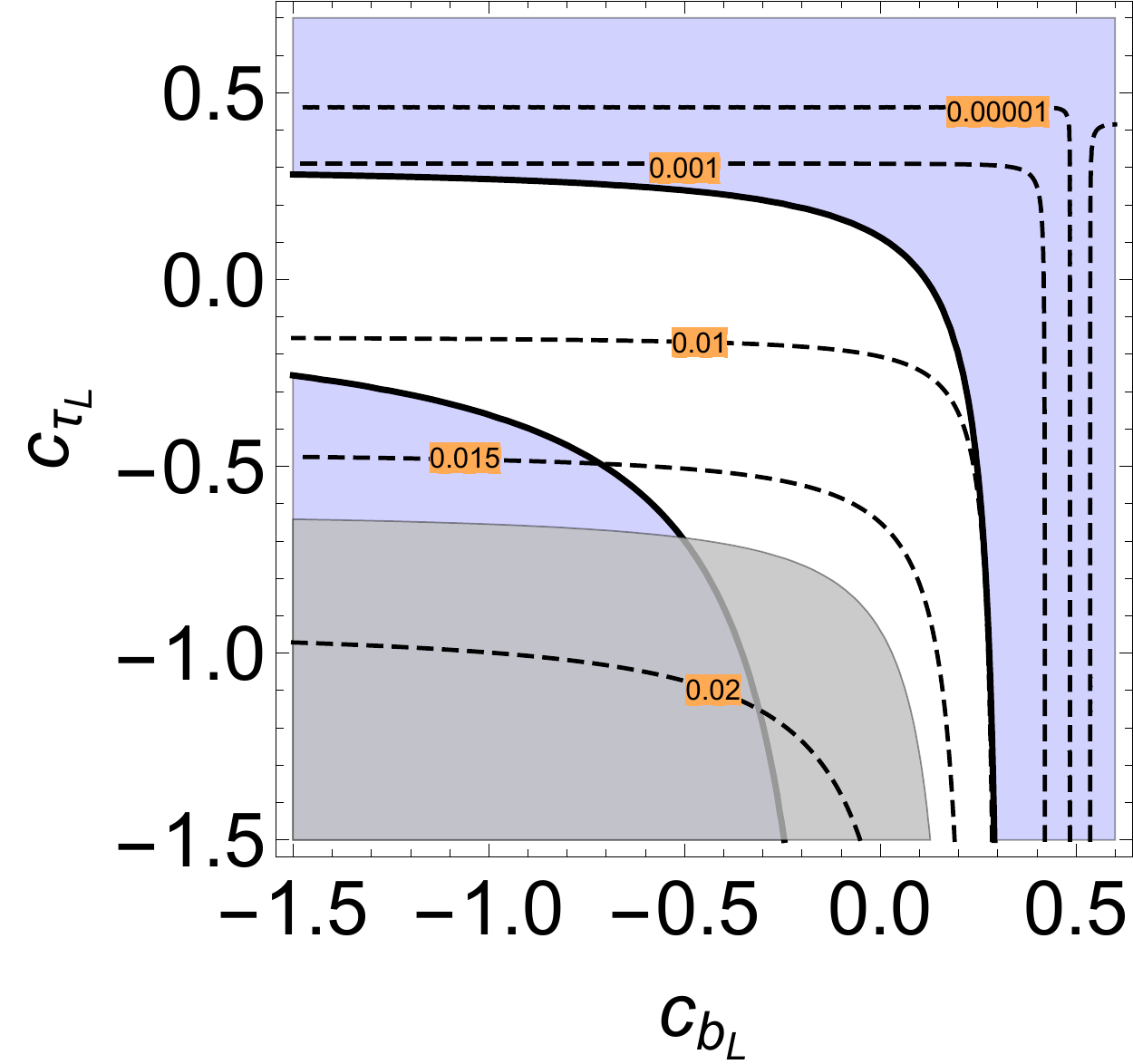} 
\caption{\it Contour plot of $\sigma \cdot \mathcal{B}[Z^n \to \tau^+ \tau^-]$ for $n=1$ (in $pb$) in the plane $(c_{b_L},c_{\tau_L})$.  We show as a gray band in the bottom part of the figure the experimentally excluded region $\sigma \cdot \mathcal{B} > 0.017$. We overlap as well the allowed region coming from the $(R_D,R_{D^*})$ observables. We have considered $c_{\mu_L} = 0.44$.
}
\label{fig:ditaus}
\end{figure}
The strong constraints imposed by the $R_{D^{(\ast)}}$ observables on the parameters $(c_{b_L},c_{\tau_L})$ make the $pp\to\tau\tau$ production from bottom-bottom fusion (especially for IR localized left-handed bottom quarks and UV localized first and second generation quarks) relevant in spite of the suppression of heavy flavors in PDFs.  The analysis has been done in Ref.~\cite{Faroughy:2016osc} and we will follow here the same lines as for the dimuon production from bottom-bottom fusion, as constrained by the $R_K$ anomaly. The cross-section for production of $b\bar b\to Z^1$ is given by the plot in the left panel of Fig.~\ref{fig:dimuons}. Using this information we show in the plot of Fig.~\ref{fig:ditaus} contour lines of $\sigma\cdot \mathcal B(Z^1\to\tau\tau)$. The bounds from the CMS ditau searches at $\sqrt{s}=13$ TeV and 2.2 fb$^{-1}$~\cite{Khachatryan:2016qkc} yield for a 2 TeV vector resonance the 95\% CL bound $\sigma\mathcal B(Z^1\to\tau\tau)\lesssim 0.017$ pb. The corresponding excluded region (the grey area) is shown, in the $(c_{b_L},c_{\tau_L})$ plane,  in the plot of Fig.~\ref{fig:ditaus} which we overlap with the allowed region by the $R_{D^{(\ast)}}$ anomaly. As we can see part of (but not all) the region allowed by $R_{D^{(\ast)}}$ (the part of the parameter region where $b_L$ and/or $\tau_L$ are mostly localized toward the IR) is already excluded by LHC data on ditau production. However the most interesting region, where $c_{\tau_L}>c_{b_L}$, is entirely allowed. 

\subsection{Lepton flavor universality tests}

The anomaly on the experimental values of $R_{D^{(\ast)}}^{\tau/\ell}$ also has to be contrasted with the non-observation of flavor universality violation effects in the $\mu/e$ sector and with lepton flavor universality tests in tau decays~\footnote{We thank Paride Paradisi for pointing out the corresponding observables, which were missing from the first version of the paper, to us.}. In particular in the $\mu/e$ sector, the non-observation of flavor universality violation at the $2\%$ level translates into the condition $R_{D^{(\ast)}}^{\mu/e}\lesssim 1.02$~\cite{Greljo:2015mma,Feruglio:2016gvd}, with
\begin{figure}[htb]
\centering
\vspace{0.1cm}
\includegraphics[width=7.cm]{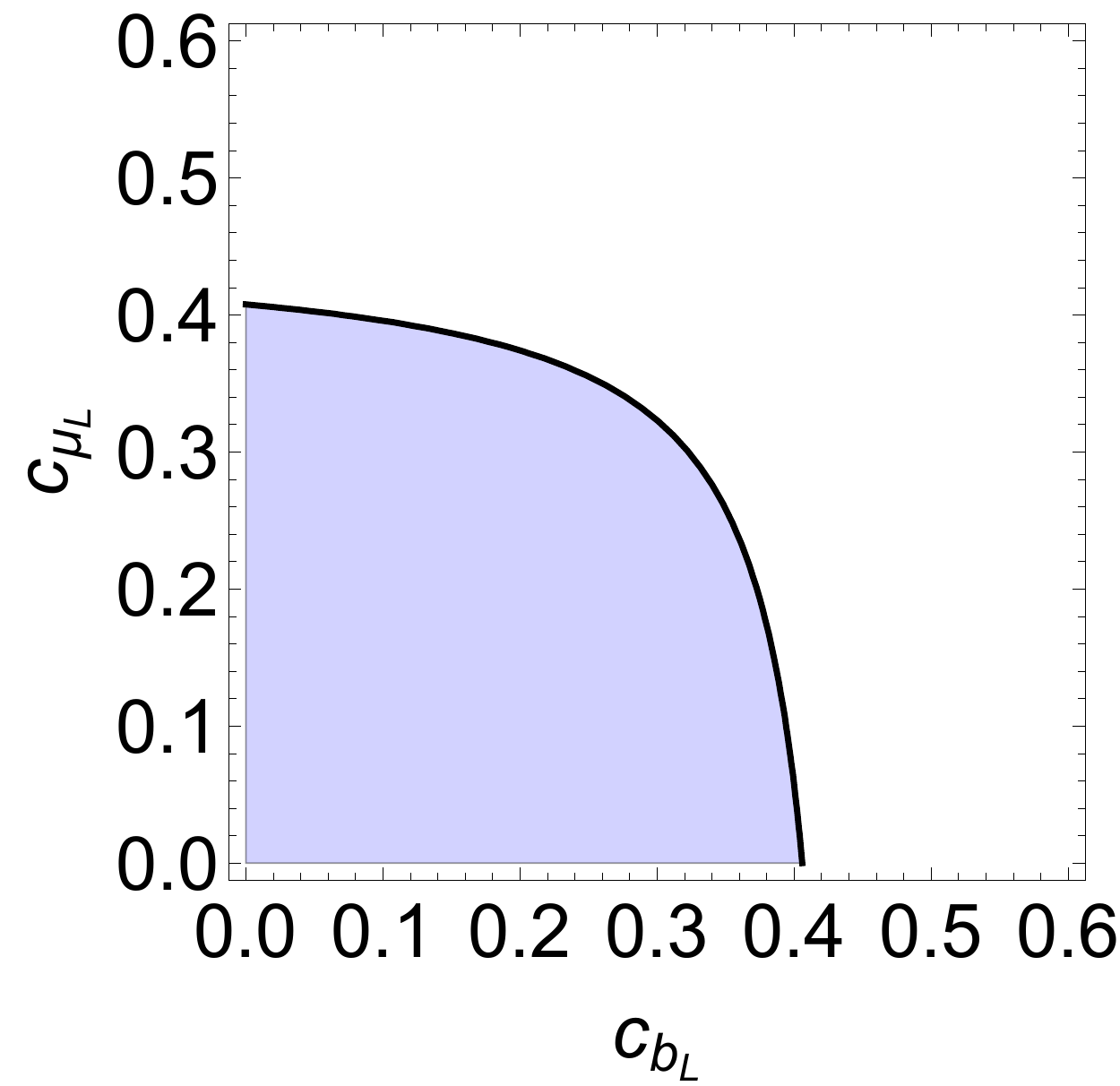}
\includegraphics[width=7cm]{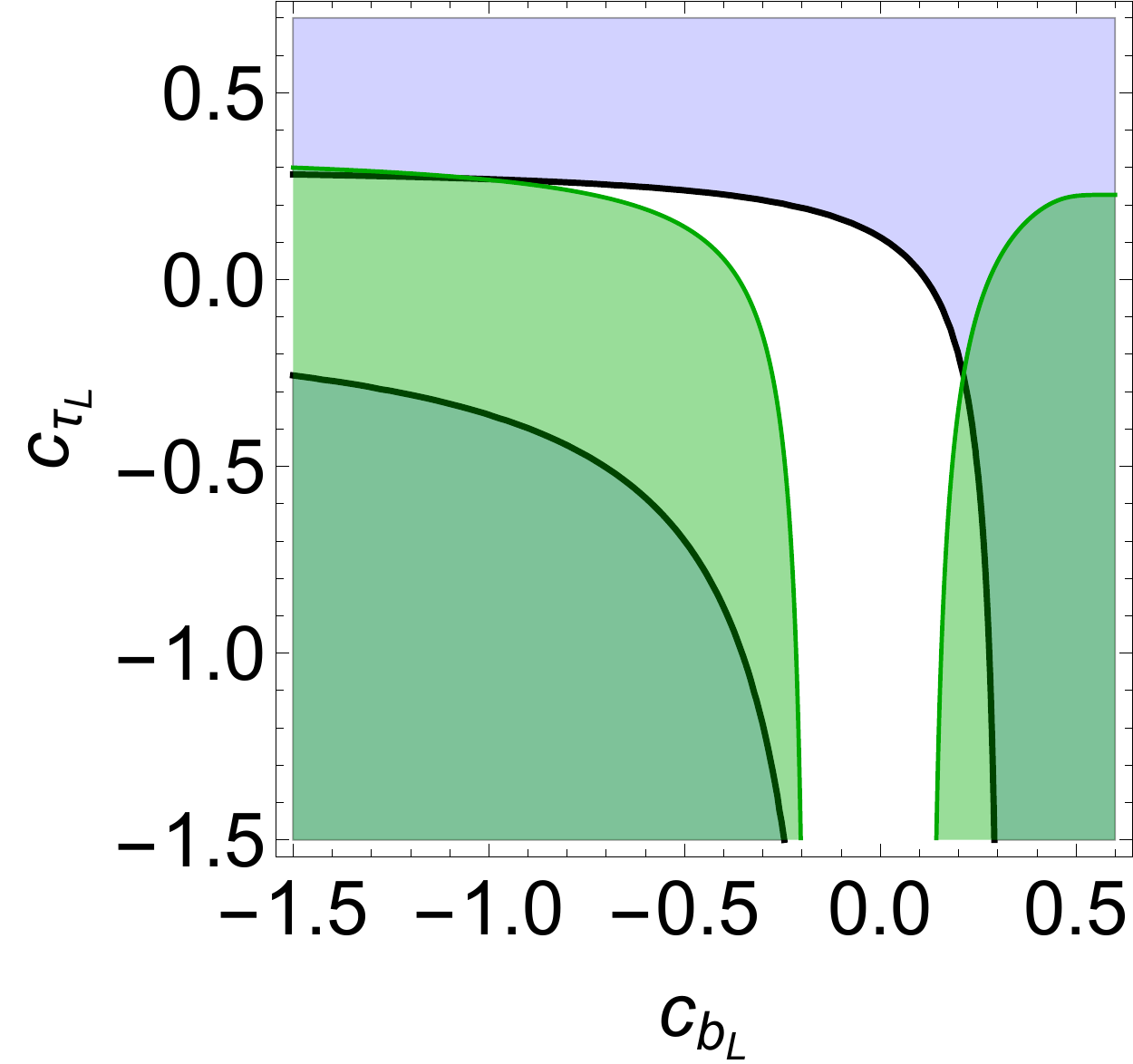}
\caption{\it Left panel: Exclusion (shadowed) region in the plane $(c_{b_L},c_{\mu_L})$ by the condition $R_{D^{(\ast)}}^{\mu/e}\lesssim 1.02$. Right panel: We display in green the excluded region corresponding to $R_\tau^{\tau/\mu} \in [0.996,1.008]$ in the plane $(c_{b_L},c_{\tau_L})$.  We overlap as well the allowed region coming from the $(R_D,R_{D^*})$ observables. We have considered $c_{\mu_L}=0.44$. We have used in both plots $r=0.75$ and $m_{KK}=2$ TeV.}
\label{fig:Cmu}
\end{figure} 
\be
R_{D^{(\ast)}}^{\mu/e}=\frac{\mathcal B(\overline{B}\to D^{(\ast)}\mu^- \overline{\nu}_\mu)}{\mathcal B(\overline{B}\to D^{(\ast)}e^- \overline{\nu}_e)}=\left|1+
C^\mu  \right|^2
\label{expresion2}
\ee
where we have assumed that $c_{e_L}=0.5$ and so $C^e=0$. 
Consequently the condition $R_{D^{(\ast)}}^{\mu/e}\lesssim 1.02$ translates into $C^\mu\lesssim 0.010$. As $C^\mu$ is a function of $(c_{b_L},c_{\mu_L})$ we plot in the left panel of Fig.~\ref{fig:Cmu} the exclusion condition which corresponds to the shadowed area. As we can see from the right plot of Fig.~\ref{fig:RD2}, and the left panel of Fig.~\ref{fig:Cmu}, the bound $c_{b_L}\lesssim 0.29$ would translate into the bound $c_{\mu_L}\gtrsim 0.33$ which is perfectly consistent with the amount of lepton flavor universality breaking obtained in this paper.

Finally the $R_{D^{(\ast)}}$ anomaly, and its corresponding lepton flavor universality violation in the $\tau/\mu$ sector, also has to agree with flavor universality tests performed at the per mille level in tau decays. In particular the observables
\be
R_\tau^{\tau/\ell}=\frac{\mathcal B(\tau\to\ell\nu\bar\nu)/\mathcal B(\tau\to\ell\nu\bar\nu)_{\rm SM}}{\mathcal B(\mu\to e\nu\bar\nu)/\mathcal B(\mu\to e\nu\bar\nu)_{\rm SM}},\quad (\ell=\mu,e)
\ee  
are subject to the experimental bounds~\cite{Pich:2013lsa,Feruglio:2016gvd}, $R_\tau^{\tau/\mu}\in [0.996,1.008]$ and $R_\tau^{\tau/e}\in [1.000,1.012]$ at 95\% CL. In our model, fixing $c_{e_L}=0.5$ implies that $R_\tau^{\tau/e}=1$ while, including the relevant one-loop radiative corrections~\cite{Feruglio:2016gvd}, we can write the $R_\tau^{\tau/\mu}$ observable as
\be
R_\tau^{\tau/\mu} = 1 + 2 \frac{m_W^2}{m^2_{W^{(n)}}} G^n_{\tau_L} (G^n_{\mu_L} - 0.065 G^n_{b_L})  \,.  \label{eq:Rtau1loop}
\ee
One can see that radiative effects proportional to $G_{b_L}^n$ (coming from closing the $b\,\bar c$ quark line which contributes to $R_{D^{(\ast)}}$ by emitting a $W$-gauge boson) with loop suppression factors, compete with tree-level effects proportional to $G_{\mu_L}^n$, 
as accommodation of the $R_{D^{(\ast)}}$ anomaly implies $G_{b_L}^n\gg G_{\mu_L}^n$. This competition produces a partial cancellation and the result leaves more available space than any of the individual effects~\footnote{We thank Paride Paradisi for pointing out this effect to us.}, without introducing any fine-tuning.
The allowed region in the plane $(c_{b_L},c_{\tau_L})$ is shown in the right panel of Fig.~\ref{fig:Cmu}. The green region is excluded from Eq.~(\ref{eq:Rtau1loop}) for $c_{\mu_L}=0.44$, a value consistent with the $R_K$ anomaly from Fig.~\ref{fig:RK2}. The plot from the $R_{D^{(\ast)}}$ anomaly is superimposed and the white region is allowed by both.

The short conclusion in this section is that lepton flavor universality tests can easily agree with the experimental value of the $R_{D^{(\ast)}}$ anomaly.

\subsection{The $Z\overline\tau\tau$ coupling}
Finally the $R_{D^{(\ast)}}$ anomaly has to be contrasted with radiative and KK corrections to the $Z\bar\tau\tau$ coupling. We will do it following the formalism of Sec.~\ref{radiative} and using the experimental value from the fit of Ref.~\cite{ALEPH:2005ab} 
\begin{equation}
g_{\tau_L}^Z=-0.26930\pm 0.00058 \,,
\end{equation}
which leads to the result~\footnote{
The recent fit from Ref.~\cite{Falkowski:2017pss} yields $\Delta g_{\tau_L}^Z+\delta g_{\tau_L}^Z=(0.18\pm 0.59)\times 10^{-3}$ consistent with Eq.~(\ref{eq:vtau}).
}

\be
\Delta g_{\tau_L}^Z+\delta g_{\tau_L}^Z=(0.09\pm 0.58)\times 10^{-3} \,,
\label{eq:vtau}
\ee
where $\Delta g_{\tau_L}^Z$ is given by Eq.~(\ref{eq:Delta_g}) and $\delta g_{\tau_L}^Z$ by Eq.~(\ref{eq:delta_g}).
The allowed region at $2\sigma$ is shown in the plot of Fig.~\ref{fig:Deltagtau}.
\begin{figure}[htb]
\centering
\includegraphics[width=7.5cm]{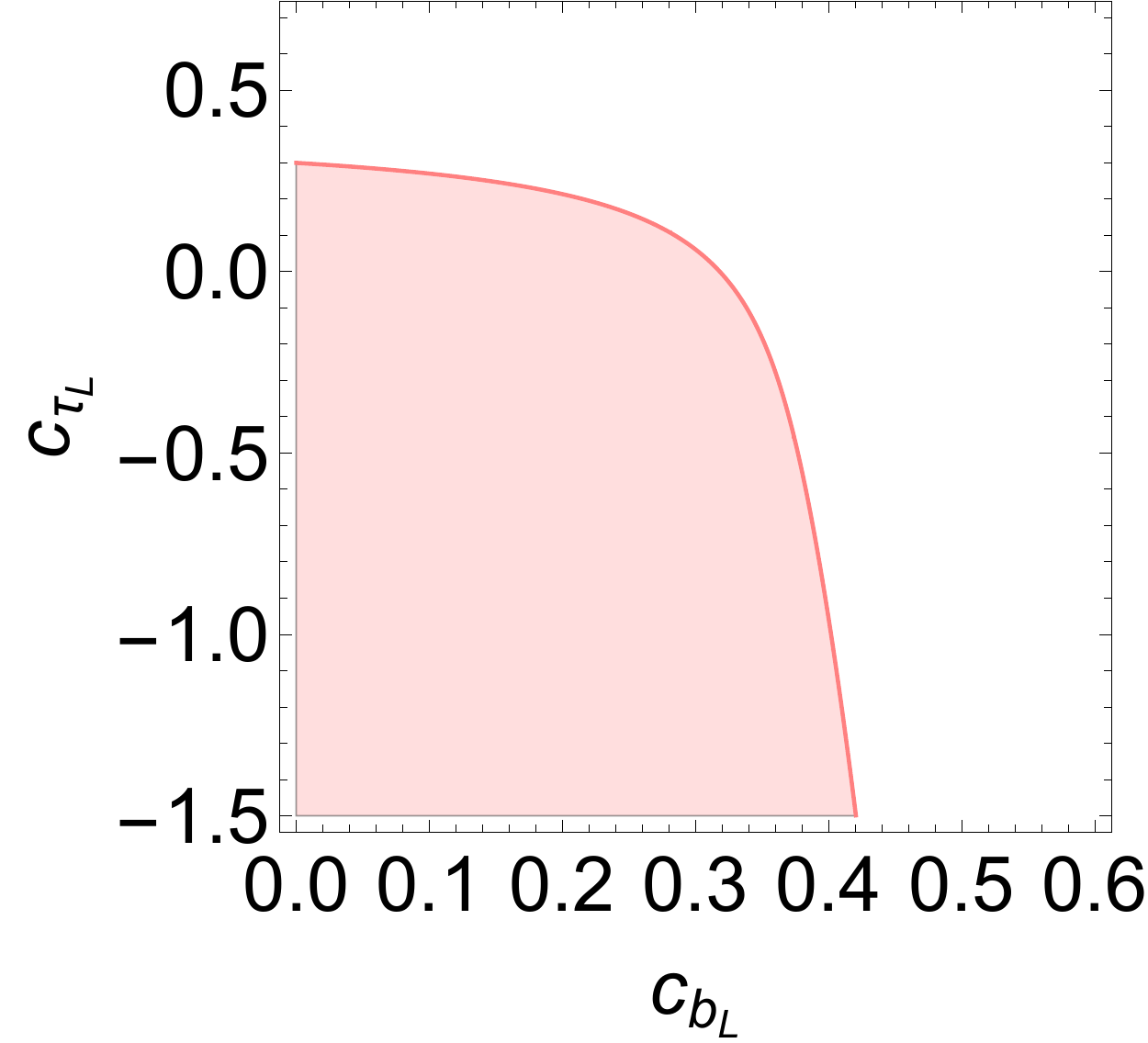}
\caption{\it Region at $2\sigma$ in the plane $(c_{b_L},c_{\tau_L})$ that accommodates the constraint $\Delta g_{\tau_L}^Z+\delta g_{\tau_L}^Z$ (white region) as shown in Eq.~(\ref{eq:vtau}).
}
\label{fig:Deltagtau}
\end{figure} 

To conclude this section and as we can see by comparison of Figs.~\ref{fig:RD2} and \ref{fig:Deltagtau}, there is a tension between data from the $R_{D^{(\ast)}}$ anomaly and electroweak observables, in particular the $Z^\mu\tau_L \gamma_\mu\tau_L$ coupling, $g_{\tau_L}^Z$, because the electroweak corrections to the effective operator $(\bar t_L\gamma^\mu t_L)(\bar\ell_L\gamma_\mu\ell_L)$, give rise to the operator $(H^\dagger D_\mu H)(\bar\ell_L\gamma^\mu\ell_L)$ and thus trigger, after electroweak breaking, a correction to the $Z\bar\tau_L \tau_L$ coupling proportional to $h_t^2$. In short, assuming a Wolfenstein-like structure for the unitary transformations $V_{u_{L(R)}}$, $V_{d_{L(R)}}$ [i.e.~$r\lesssim 1$ in Eq.~(\ref{Vd})] we find that the $R_{D^{(\ast)}}$ anomaly is only satisfied by very composite fermions $(b_L,\tau_L)$ which are in tension with the experimental value of $g_{\tau_L}^Z$.

\section{Conclusions and outlook}
\label{conclusions}
In this paper we have tried to accommodate present data on lepton flavor universality violation in a model with a warped extra dimension, where the Standard Model fields propagate, and which is basically in agreement with electroweak precision observables thanks to a strong deformation of conformality of the metric near the IR brane. Every fermion field $f_{L,R}$ in the model is characterized by a five-dimensional Dirac mass parametrized by a real constant $c_{f_{L,R}}$ which controls its localization or, equivalently in the dual theory, its degree of compositeness. Fermions with $c_f>0.5$ ($c_f<0.5$) are localized toward the UV (IR) brane and correspond in the dual theory to mostly elementary (composite) fields. The coupling of gauge boson KK-modes with fermion $f$ essentially depend on the value of $c_f$: it is very small for elementary fermions and large for composite fermions. In this way the basic elements of lepton flavor universality violation through the exchange of KK gauge bosons is built in \textit{ab initio}, and controlled in the theory by the different values of $c_f$. In particular it is very easy to generate lepton flavor universality violation for electrons, muons and taus by just assuming that electrons are elementary fermions while muons and taus have a certain degree of compositeness. 

The results in this paper depend, to some extent, on the five-dimensional Yukawa matrices $Y_{u,d}^{5D}\equiv\sqrt{k}\widehat Y_{u,d}$ which in turn determine, along with the constants $c_{f_{L,R}}$, the unitary transformations $V_{u_{L,R}}$ and  $V_{d_{L,R}}$. In the absence of a UV theory for the Yukawa couplings $\widehat Y_{u,d}$ we have considered arbitrary matrices $V_{u_{L,R}}$ and $V_{d_{L,R}}$ satisfying the Wolfenstein parametrization, and such that $V_{u_L}^\dagger V_{d_L}=V$, the CKM matrix. As those matrices depend on a number of parameters we have considered generic values for their entries, satisfying the Wolfenstein parametrization and leading to strong bounds in the down-quark sector from $\Delta m_K$ and $\epsilon_K$ and in the up-quark sector from $\Delta m_D$ and $\phi_D$.  An analysis for different values of the parameters, in case they would be provided by particular UV completions of the present model, should be readily done along similar lines as in the present paper. 

Moreover our theory is \textit{lepton flavor conserving}, as we have considered in the charged lepton sector models where the 5D Yukawa matrix $\widehat Y_{\ell}$ is already in diagonal form, i.e.~$V_{\ell_{L,R}}=1_3$, thus avoiding strong constraints from lepton flavor violation. Had we considered models with more generic Wolfenstein-like matrices in the charged lepton sector $V_{\ell_{L,R}}$, bounds on lepton flavor violating processes, as e.g.~$\tau\to 3\mu$ or $\mu\to e\gamma$, would have imposed very strong constraints on the off-diagonal elements of $V_{\ell_{L,R}}$. We postpone the study of this class of models for future investigation.
\begin{figure}[htb]
\centering
\includegraphics[width=7.0cm]{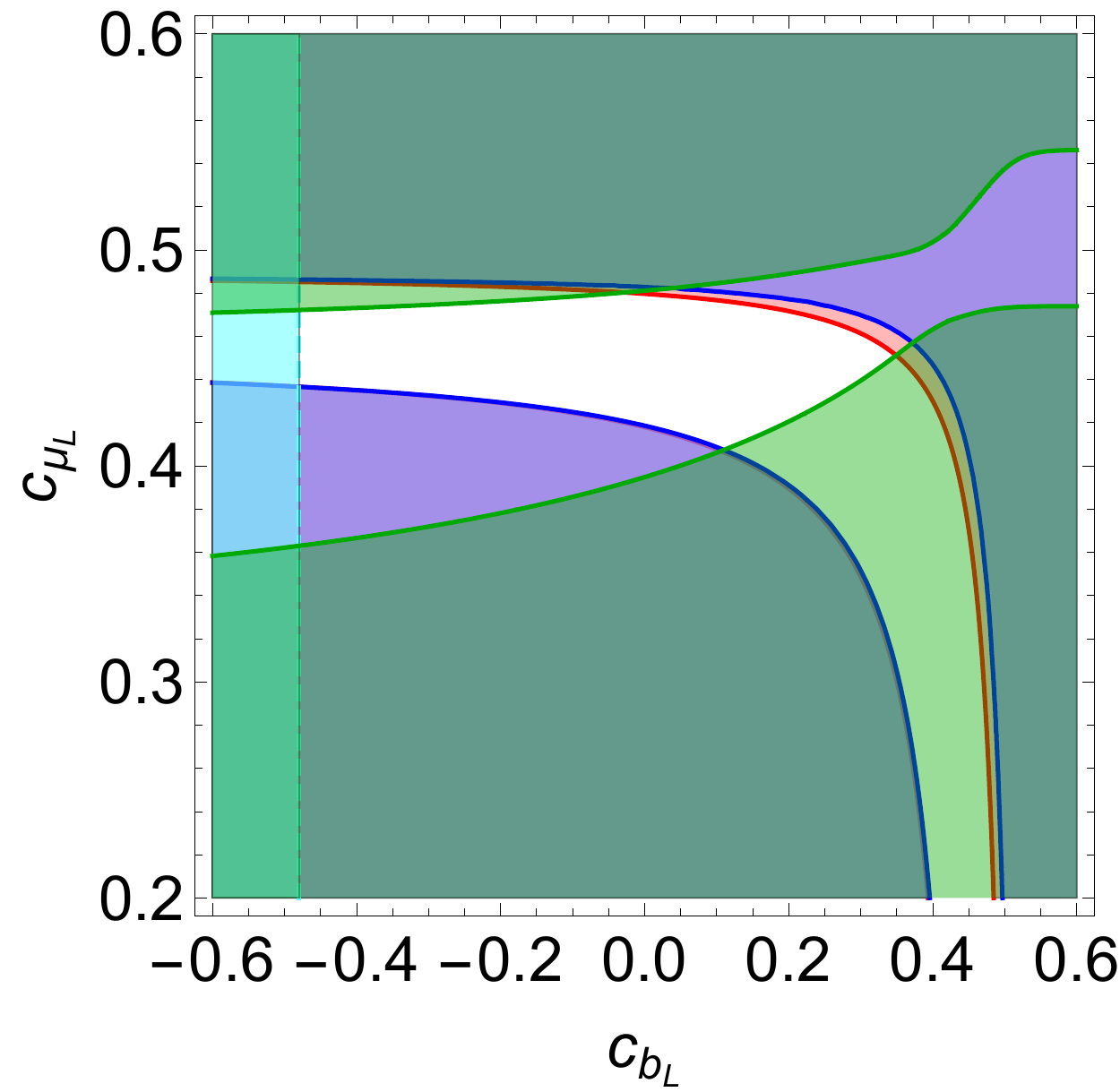} \hspace{1.0cm}
\includegraphics[width=7.0cm]{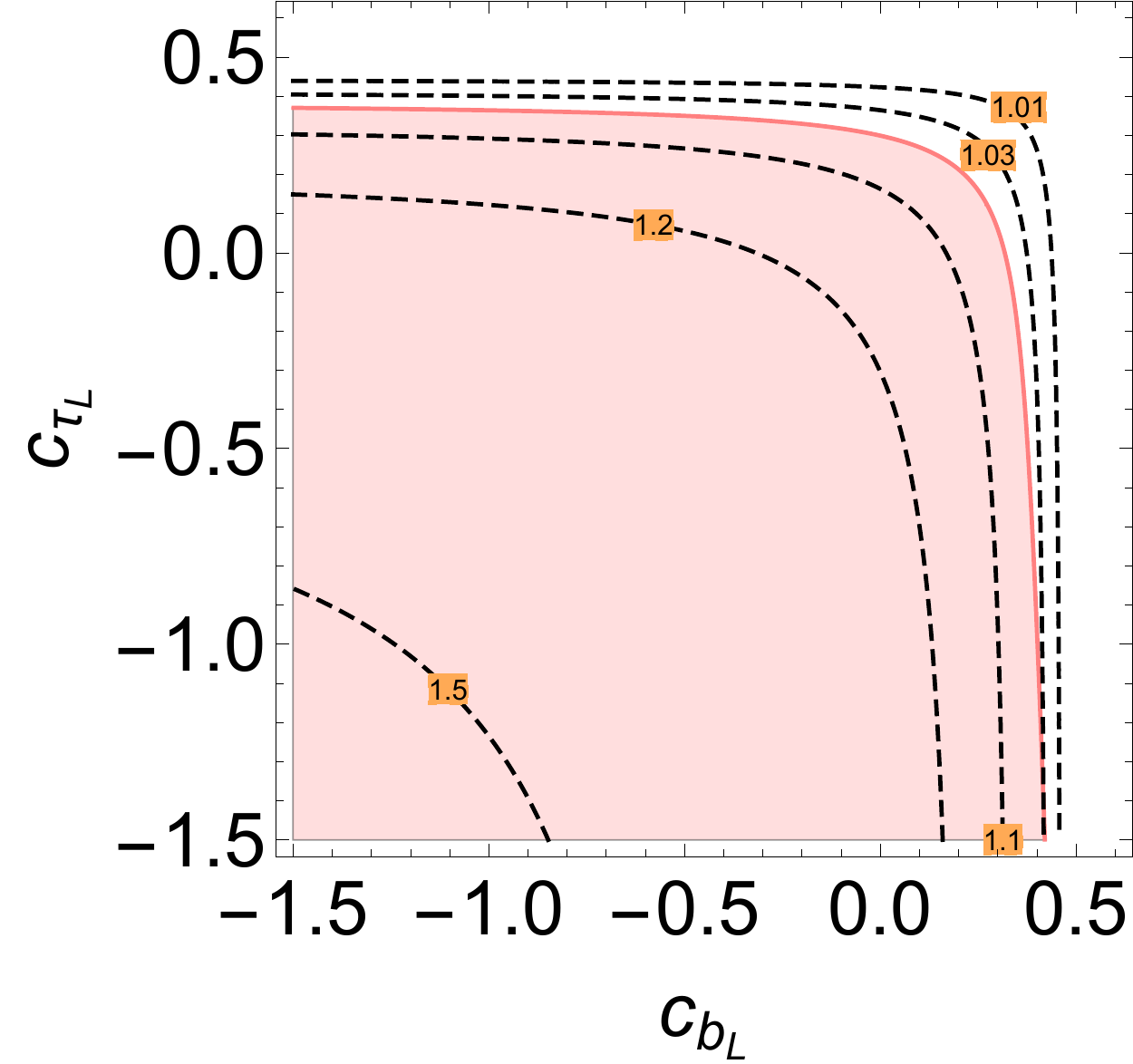} 
\caption{\it 
Left panel: Region in the $(c_{b_L},c_{\mu_L})$ plane that accommodates $R_K$ (solid red line) and $R_0$ (blue line). We also overlap the flavor constraints region $c_{b_L} > -0.48$. We display as green band the excluded region corresponding to $R_\tau^{\tau/\mu}$ [cf. Fig.~\ref{fig:Cmu} (right panel)]. The white region is allowed for any value of $c_{\tau_L}$. 
Right panel: Contour plot in the $(c_{b_L},c_{\tau_L})$ plane of the $R_{D^{(\ast)}}/R_{D^{(\ast)}}^{\rm SM}$. The red shaded area corresponds to the bound from $g_{\tau_L}^Z$ (see Fig.~\ref{fig:Deltagtau}).
}
\label{fig:final}
\end{figure} 

Using the above ideas it is straightforward to accommodate the present flavor universality violations in the observables $R_K$, as well as the rest of observables depending on $b\to s \ell^+\ell^-$ and $b\to \ s \nu\bar\nu$, processes. The summary results from $R_K$ are given in the left panel plot of Fig.~\ref{fig:final} where we show the allowed regions in the plane $(c_{b_L},c_{\mu_L})$, taking into account all different constraints obtained by electroweak observables, direct LHC searches and flavor observables. 
We also have included the green region which is excluded from Eq.~(\ref{eq:Rtau1loop}) for values of $c_{\tau_L}$ below the bound in the plot of Fig.~\ref{fig:Deltagtau}. 
All of them put together leave the approximate allowed region 
\be
0.41\lesssim c_{\mu_L}\lesssim 0.48 \,,\quad -0.48\lesssim c_{b_L}\lesssim 0.35 \,,
\label{secuencia}
\ee
which translate into pretty composite left-handed bottom quarks, and slightly composite left-handed muon leptons. The sequence in Eq.~(\ref{secuencia}) is roughly, within factors of $\mathcal O$(few), in agreement with their relative masses, whose absolute values can be easily fixed with appropriate values of the right-handed component parameters, $c_{b_R}$, and $c_{\mu_R}$, and natural values of the five-dimensional Yukawa couplings.

On the other hand, trying to accommodate the present flavor universality violations in the $R_{D^{(\ast)}}$ observables generates a tension with electroweak observables, in particular with the $Z^\mu \bar\tau_L\gamma_\mu\tau$ coupling as can be seen from the right panel of Fig.~\ref{fig:final} where we gather the allowed region by the $g_{\tau_L}^Z$ coupling, and contour plots of the observables $R_{D^{(\ast)}}/R_{D^{(\ast)}}^{SM}$ with experimental values
\be
\frac{R_{D}^{\rm exp}}{R_{D}^{SM}}=1.37\pm 0.17,\quad \frac{R_{D^{\ast}}^{\rm exp}}{R_{D^{\ast}}^{SM}}=1.28\pm 0.08 \,.
\ee
As we can see from the plot, deviations from one of $R_{D^{(\ast)}}/R_{D^{(\ast)}}^{SM}$ are constrained by $g_{\tau_L}^Z$ to values $\lesssim 5\%$. A possible way out is to allow some (small) departure of the matrices $V_{d_L}$ and $V_{u_L}$ from the Wolfenstein pattern, in particular by allowing that $V_{cb}\ll (V_{u_L})_{32}\ll 1$ which implies in particular that $r\gtrsim 1$. As we can see from Eq.~(\ref{Csimp}) this would strengthen the value of $R_{D^{(\ast)}}$ with less composite $\tau_L$ leptons, which in turn unfasten the tension with the experimental value of $g_{\tau_L}^Z$. Of course the price to pay for this ``solution" is introducing some degree of fine-tuning for the fixing of the small CKM unitary matrix entries from matrices $V_{d_L}$ and $V_{u_L}$ with larger entries. This \textit{little fine-tuned} solution will be worked out elsewhere.

The remaining lepton-flavor universality violation is the anomalous magnetic moment of the muon $a_\mu=(g_\mu-2)/2$, which deviates with respect to the Standard Model prediction $a_\mu^{\rm SM}$ by $\sim 3.6\sigma$, while the corresponding observable for the electron, $a_e$, is in very good agreement with the Standard Model. Our theory has the required ingredients to trigger a sizeable correction to the muon anomalous magnetic moment through the mixing (induced by the muon Yukawa coupling) between left and right-handed muon $n$-KK modes and the corresponding zero modes. However as the mixing is controlled by the experimental bounds $\left|\delta g_{L,R}/g_{L,R}\right|\lesssim 10^{-3}$, it does not have enough power to trigger a large effect, and extra physics should be introduced in the model to encompass explanation of anomalous magnetic moment of the muon. In the context of warped theories, a possibility was already presented in Ref.~\cite{Megias:2017dzd} where heavy vector-like leptons, with the quantum numbers of the Standard Model muons, were introduced and conveniently mix with them through appropriate Yukawa couplings. As it was proven in Ref.~\cite{Megias:2017dzd} the explanation of this effect is consistent with all electroweak and flavor observables, and direct searches of heavy leptons, and implies a high degree of compositeness for vector-like leptons which could be detected at present and future colliders.  

\section*{Acknowledgments}

We thank Paride Paradisi for useful observations on the observables $R_{D^{(\ast)}}^{\mu/e}$ and $R_\tau^{\tau/\ell}$ constraining our model parameters.
L.S. is supported by a \textit{Beca Predoctoral Severo Ochoa del Ministerio de
Econom\'{\i}a y Competitividad} (SVP-2014-068850), and E.M. is
supported by the \textit{Universidad del Pa\'{\i}s Vasco}
UPV/EHU, Bilbao, Spain, as a Visiting Professor. The work of M.Q. and
L.S.~is also partly supported by Spanish MINECO under Grant
CICYT-FEDER-FPA2014-55613-P, by the Severo Ochoa Excellence Program of
MINECO under the grant SO-2012-0234, by \textit{Secretaria d'Universitats i
Recerca del Departament d'Economia i Coneixement de la Generalitat de
Catalunya} under Grant 2014 SGR 1450, and by the CERCA
Program/Generalitat de Catalunya. The research of E.M. is also partly
supported by Spanish MINECO under Grant FPA2015-64041-C2-1-P, by the
Basque Government under Grant IT979-16, and by the Spanish Consolider
Ingenio 2010 Programme CPAN (CSD2007-00042).

\bibliographystyle{JHEP}
\bibliography{refs}

\providecommand{\href}[2]{#2}\begingroup\raggedright\begin{thebibliography}{10}

\bibitem{Randall:1999ee}
L.~Randall and R.~Sundrum, \emph{{A Large mass hierarchy from a small extra
  dimension}}, \href{http://dx.doi.org/10.1103/PhysRevLett.83.3370}{\emph{Phys.
  Rev. Lett.} {\bfseries 83} (1999) 3370--3373},
  [\href{https://arxiv.org/abs/hep-ph/9905221}{{\ttfamily hep-ph/9905221}}].

\bibitem{Cabrer:2009we}
J.~A. Cabrer, G.~von Gersdorff and M.~Quiros, \emph{{Soft-Wall Stabilization}},
  \href{http://dx.doi.org/10.1088/1367-2630/12/7/075012}{\emph{New J. Phys.}
  {\bfseries 12} (2010) 075012},
  [\href{https://arxiv.org/abs/0907.5361}{{\ttfamily 0907.5361}}].

\bibitem{Cabrer:2010si}
J.~A. Cabrer, G.~von Gersdorff and M.~Quiros, \emph{{Warped Electroweak
  Breaking Without Custodial Symmetry}},
  \href{http://dx.doi.org/10.1016/j.physletb.2011.01.058}{\emph{Phys. Lett.}
  {\bfseries B697} (2011) 208--214},
  [\href{https://arxiv.org/abs/1011.2205}{{\ttfamily 1011.2205}}].

\bibitem{Cabrer:2011fb}
J.~A. Cabrer, G.~von Gersdorff and M.~Quiros, \emph{{Suppressing Electroweak
  Precision Observables in 5D Warped Models}},
  \href{http://dx.doi.org/10.1007/JHEP05(2011)083}{\emph{JHEP} {\bfseries 05}
  (2011) 083}, [\href{https://arxiv.org/abs/1103.1388}{{\ttfamily 1103.1388}}].

\bibitem{Cabrer:2011vu}
J.~A. Cabrer, G.~von Gersdorff and M.~Quiros, \emph{{Improving Naturalness in
  Warped Models with a Heavy Bulk Higgs Boson}},
  \href{http://dx.doi.org/10.1103/PhysRevD.84.035024}{\emph{Phys. Rev.}
  {\bfseries D84} (2011) 035024},
  [\href{https://arxiv.org/abs/1104.3149}{{\ttfamily 1104.3149}}].

\bibitem{Cabrer:2011mw}
J.~A. Cabrer, G.~von Gersdorff and M.~Quiros, \emph{{Warped 5D Standard Model
  Consistent with EWPT}},
  \href{http://dx.doi.org/10.1002/prop.201100054}{\emph{Fortsch. Phys.}
  {\bfseries 59} (2011) 1135--1138},
  [\href{https://arxiv.org/abs/1104.5253}{{\ttfamily 1104.5253}}].

\bibitem{Carmona:2011ib}
A.~Carmona, E.~Ponton and J.~Santiago, \emph{{Phenomenology of Non-Custodial
  Warped Models}}, \href{http://dx.doi.org/10.1007/JHEP10(2011)137}{\emph{JHEP}
  {\bfseries 10} (2011) 137},
  [\href{https://arxiv.org/abs/1107.1500}{{\ttfamily 1107.1500}}].

\bibitem{Cabrer:2011qb}
J.~A. Cabrer, G.~von Gersdorff and M.~Quiros, \emph{{Flavor Phenomenology in
  General 5D Warped Spaces}},
  \href{http://dx.doi.org/10.1007/JHEP01(2012)033}{\emph{JHEP} {\bfseries 01}
  (2012) 033}, [\href{https://arxiv.org/abs/1110.3324}{{\ttfamily 1110.3324}}].

\bibitem{Quiros:2013yaa}
M.~Quiros, \emph{{Higgs Bosons in Extra Dimensions}},
  \href{http://dx.doi.org/10.1142/S021773231540012X}{\emph{Mod. Phys. Lett.}
  {\bfseries A30} (2015) 1540012},
  [\href{https://arxiv.org/abs/1311.2824}{{\ttfamily 1311.2824}}].

\bibitem{Megias:2015ory}
E.~Megias, O.~Pujolas and M.~Quiros, \emph{{On dilatons and the LHC diphoton
  excess}}, \href{http://dx.doi.org/10.1007/JHEP05(2016)137}{\emph{JHEP}
  {\bfseries 05} (2016) 137},
  [\href{https://arxiv.org/abs/1512.06106}{{\ttfamily 1512.06106}}].

\bibitem{Megias:2015qqh}
E.~Megias, O.~Pujolas and M.~Quiros, \emph{{On light dilaton extensions of the
  Standard Model}},
  \href{http://dx.doi.org/10.1051/epjconf/201612605010}{\emph{EPJ Web Conf.}
  {\bfseries 126} (2016) 05010},
  [\href{https://arxiv.org/abs/1512.06702}{{\ttfamily 1512.06702}}].

\bibitem{Megias:2016jcw}
E.~Megias, G.~Panico, O.~Pujolas and M.~Quiros, \emph{{Light dilatons in warped
  space: Higgs boson and LHCb anomalies}},
  \href{http://dx.doi.org/10.1016/j.nuclphysbps.2016.12.037}{\emph{Nucl. Part.
  Phys. Proc.} {\bfseries 282-284} (2017) 194--198},
  [\href{https://arxiv.org/abs/1609.01881}{{\ttfamily 1609.01881}}].

\bibitem{Davoudiasl:2009cd}
H.~Davoudiasl, S.~Gopalakrishna, E.~Ponton and J.~Santiago, \emph{{Warped
  5-Dimensional Models: Phenomenological Status and Experimental Prospects}},
  \href{http://dx.doi.org/10.1088/1367-2630/12/7/075011}{\emph{New J. Phys.}
  {\bfseries 12} (2010) 075011},
  [\href{https://arxiv.org/abs/0908.1968}{{\ttfamily 0908.1968}}].

\bibitem{Peskin:1991sw}
M.~E. Peskin and T.~Takeuchi, \emph{{Estimation of oblique electroweak
  corrections}}, \href{http://dx.doi.org/10.1103/PhysRevD.46.381}{\emph{Phys.
  Rev.} {\bfseries D46} (1992) 381--409}.

\bibitem{Olive:2016xmw}
{\scshape Particle Data Group} collaboration, C.~Patrignani et~al.,
  \emph{{Review of Particle Physics}},
  \href{http://dx.doi.org/10.1088/1674-1137/40/10/100001}{\emph{Chin. Phys.}
  {\bfseries C40} (2016) 100001}.

\bibitem{Megias:2016bde}
E.~Megias, G.~Panico, O.~Pujolas and M.~Quiros, \emph{{A Natural origin for the
  LHCb anomalies}},
  \href{http://dx.doi.org/10.1007/JHEP09(2016)118}{\emph{JHEP} {\bfseries 09}
  (2016) 118}, [\href{https://arxiv.org/abs/1608.02362}{{\ttfamily
  1608.02362}}].

\bibitem{Lees:2012xj}
{\scshape BaBar} collaboration, J.~P. Lees et~al., \emph{{Evidence for an
  excess of $\bar{B} \to D^{(*)} \tau^-\bar{\nu}_\tau$ decays}},
  \href{http://dx.doi.org/10.1103/PhysRevLett.109.101802}{\emph{Phys. Rev.
  Lett.} {\bfseries 109} (2012) 101802},
  [\href{https://arxiv.org/abs/1205.5442}{{\ttfamily 1205.5442}}].

\bibitem{Lees:2013uzd}
{\scshape BaBar} collaboration, J.~P. Lees et~al., \emph{{Measurement of an
  Excess of $\bar{B} \to D^{(*)}\tau^- \bar{\nu}_\tau$ Decays and Implications
  for Charged Higgs Bosons}},
  \href{http://dx.doi.org/10.1103/PhysRevD.88.072012}{\emph{Phys. Rev.}
  {\bfseries D88} (2013) 072012},
  [\href{https://arxiv.org/abs/1303.0571}{{\ttfamily 1303.0571}}].

\bibitem{Huschle:2015rga}
{\scshape Belle} collaboration, M.~Huschle et~al., \emph{{Measurement of the
  branching ratio of $\bar{B} \to D^{(\ast)} \tau^- \bar{\nu}_\tau$ relative to
  $\bar{B} \to D^{(\ast)} \ell^- \bar{\nu}_\ell$ decays with hadronic tagging
  at Belle}}, \href{http://dx.doi.org/10.1103/PhysRevD.92.072014}{\emph{Phys.
  Rev.} {\bfseries D92} (2015) 072014},
  [\href{https://arxiv.org/abs/1507.03233}{{\ttfamily 1507.03233}}].

\bibitem{Sato:2016svk}
{\scshape Belle} collaboration, Y.~Sato et~al., \emph{{Measurement of the
  branching ratio of $\bar{B}^0 \rightarrow D^{*+} \tau^- \bar{\nu}_{\tau}$
  relative to $\bar{B}^0 \rightarrow D^{*+} \ell^- \bar{\nu}_{\ell}$ decays
  with a semileptonic tagging method}},
  \href{http://dx.doi.org/10.1103/PhysRevD.94.072007}{\emph{Phys. Rev.}
  {\bfseries D94} (2016) 072007},
  [\href{https://arxiv.org/abs/1607.07923}{{\ttfamily 1607.07923}}].

\bibitem{Abdesselam:2016xqt}
A.~Abdesselam et~al., \emph{{Measurement of the $\tau$ lepton polarization in
  the decay ${\bar B} \rightarrow D^* \tau^- {\bar \nu_{\tau}}$}},
  \href{https://arxiv.org/abs/1608.06391}{{\ttfamily 1608.06391}}.

\bibitem{Hirose:2016wfn}
{\scshape Belle} collaboration, S.~Hirose et~al., \emph{{Measurement of the
  $\tau$ lepton polarization and $R(D^*)$ in the decay $\bar{B} \to D^* \tau^-
  \bar{\nu}_\tau$}},
  \href{http://dx.doi.org/10.1103/PhysRevLett.118.211801}{\emph{Phys. Rev.
  Lett.} {\bfseries 118} (2017) 211801},
  [\href{https://arxiv.org/abs/1612.00529}{{\ttfamily 1612.00529}}].

\bibitem{Aaij:2015yra}
{\scshape LHCb} collaboration, R.~Aaij et~al., \emph{{Measurement of the ratio
  of branching fractions $\mathcal{B}(\bar{B}^0 \to
  D^{*+}\tau^{-}\bar{\nu}_{\tau})/\mathcal{B}(\bar{B}^0 \to
  D^{*+}\mu^{-}\bar{\nu}_{\mu})$}},
  \href{http://dx.doi.org/10.1103/PhysRevLett.115.159901,
  10.1103/PhysRevLett.115.111803}{\emph{Phys. Rev. Lett.} {\bfseries 115}
  (2015) 111803}, [\href{https://arxiv.org/abs/1506.08614}{{\ttfamily
  1506.08614}}].

\bibitem{Aaij:2014ora}
{\scshape LHCb} collaboration, R.~Aaij et~al., \emph{{Test of lepton
  universality using $B^{+}\rightarrow K^{+}\ell^{+}\ell^{-}$ decays}},
  \href{http://dx.doi.org/10.1103/PhysRevLett.113.151601}{\emph{Phys. Rev.
  Lett.} {\bfseries 113} (2014) 151601},
  [\href{https://arxiv.org/abs/1406.6482}{{\ttfamily 1406.6482}}].

\bibitem{Altmannshofer:2013foa}
W.~Altmannshofer and D.~M. Straub, \emph{{New Physics in $B \to K^*\mu\mu$?}},
  \href{http://dx.doi.org/10.1140/epjc/s10052-013-2646-9}{\emph{Eur. Phys. J.}
  {\bfseries C73} (2013) 2646},
  [\href{https://arxiv.org/abs/1308.1501}{{\ttfamily 1308.1501}}].

\bibitem{Gauld:2013qba}
R.~Gauld, F.~Goertz and U.~Haisch, \emph{{On minimal $Z'$ explanations of the
  $B\to K^*\mu^+\mu^-$ anomaly}},
  \href{http://dx.doi.org/10.1103/PhysRevD.89.015005}{\emph{Phys. Rev.}
  {\bfseries D89} (2014) 015005},
  [\href{https://arxiv.org/abs/1308.1959}{{\ttfamily 1308.1959}}].

\bibitem{Sierra:2015fma}
D.~Aristizabal~Sierra, F.~Staub and A.~Vicente, \emph{{Shedding light on the
  $b\to s$ anomalies with a dark sector}},
  \href{http://dx.doi.org/10.1103/PhysRevD.92.015001}{\emph{Phys. Rev.}
  {\bfseries D92} (2015) 015001},
  [\href{https://arxiv.org/abs/1503.06077}{{\ttfamily 1503.06077}}].

\bibitem{Crivellin:2015era}
A.~Crivellin, L.~Hofer, J.~Matias, U.~Nierste, S.~Pokorski and J.~Rosiek,
  \emph{{Lepton-flavour violating $B$ decays in generic $Z'$ models}},
  \href{http://dx.doi.org/10.1103/PhysRevD.92.054013}{\emph{Phys. Rev.}
  {\bfseries D92} (2015) 054013},
  [\href{https://arxiv.org/abs/1504.07928}{{\ttfamily 1504.07928}}].

\bibitem{Celis:2015ara}
A.~Celis, J.~Fuentes-Martin, M.~Jung and H.~Serodio, \emph{{Family nonuniversal
  Z′ models with protected flavor-changing interactions}},
  \href{http://dx.doi.org/10.1103/PhysRevD.92.015007}{\emph{Phys. Rev.}
  {\bfseries D92} (2015) 015007},
  [\href{https://arxiv.org/abs/1505.03079}{{\ttfamily 1505.03079}}].

\bibitem{Falkowski:2015zwa}
A.~Falkowski, M.~Nardecchia and R.~Ziegler, \emph{{Lepton Flavor
  Non-Universality in B-meson Decays from a U(2) Flavor Model}},
  \href{http://dx.doi.org/10.1007/JHEP11(2015)173}{\emph{JHEP} {\bfseries 11}
  (2015) 173}, [\href{https://arxiv.org/abs/1509.01249}{{\ttfamily
  1509.01249}}].

\bibitem{Descotes-Genon:2015uva}
S.~Descotes-Genon, L.~Hofer, J.~Matias and J.~Virto, \emph{{Global analysis of
  $b\to s\ell\ell$ anomalies}},
  \href{http://dx.doi.org/10.1007/JHEP06(2016)092}{\emph{JHEP} {\bfseries 06}
  (2016) 092}, [\href{https://arxiv.org/abs/1510.04239}{{\ttfamily
  1510.04239}}].

\bibitem{Allanach:2015gkd}
B.~Allanach, F.~S. Queiroz, A.~Strumia and S.~Sun, \emph{{$Z′$ models for the
  LHCb and $g-2$ muon anomalies}},
  \href{http://dx.doi.org/10.1103/PhysRevD.93.055045}{\emph{Phys. Rev.}
  {\bfseries D93} (2016) 055045},
  [\href{https://arxiv.org/abs/1511.07447}{{\ttfamily 1511.07447}}].

\bibitem{Buttazzo:2016kid}
D.~Buttazzo, A.~Greljo, G.~Isidori and D.~Marzocca, \emph{{Toward a coherent
  solution of diphoton and flavor anomalies}},
  \href{http://dx.doi.org/10.1007/JHEP08(2016)035}{\emph{JHEP} {\bfseries 08}
  (2016) 035}, [\href{https://arxiv.org/abs/1604.03940}{{\ttfamily
  1604.03940}}].

\bibitem{Biancofiore:2013ki}
P.~Biancofiore, P.~Colangelo and F.~De~Fazio, \emph{{On the anomalous
  enhancement observed in $B \to D^{(*)}\tau{\bar \nu}_\tau$ decays}},
  \href{http://dx.doi.org/10.1103/PhysRevD.87.074010}{\emph{Phys. Rev.}
  {\bfseries D87} (2013) 074010},
  [\href{https://arxiv.org/abs/1302.1042}{{\ttfamily 1302.1042}}].

\bibitem{Descotes-Genon:2016hem}
S.~Descotes-Genon, L.~Hofer, J.~Matias and J.~Virto, \emph{{The $b \to s l^+
  l^-$ anomalies and their implications for new physics}},  in \emph{{51st
  Rencontres de Moriond on EW Interactions and Unified Theories La Thuile,
  Italy, March 12-19, 2016}}, 2016.
\newblock \href{https://arxiv.org/abs/1605.06059}{{\ttfamily 1605.06059}}.

\bibitem{Kosnik:2012dj}
N.~Kosnik, \emph{{Model independent constraints on leptoquarks from $b \to s
  \ell^+ \ell^-$ processes}},
  \href{http://dx.doi.org/10.1103/PhysRevD.86.055004}{\emph{Phys. Rev.}
  {\bfseries D86} (2012) 055004},
  [\href{https://arxiv.org/abs/1206.2970}{{\ttfamily 1206.2970}}].

\bibitem{Sakaki:2013bfa}
Y.~Sakaki, M.~Tanaka, A.~Tayduganov and R.~Watanabe, \emph{{Testing leptoquark
  models in $\bar B \to D^{(*)} \tau \bar\nu$}},
  \href{http://dx.doi.org/10.1103/PhysRevD.88.094012}{\emph{Phys. Rev.}
  {\bfseries D88} (2013) 094012},
  [\href{https://arxiv.org/abs/1309.0301}{{\ttfamily 1309.0301}}].

\bibitem{Hiller:2014yaa}
G.~Hiller and M.~Schmaltz, \emph{{$R_K$ and future $b \to s \ell \ell$ physics
  beyond the standard model opportunities}},
  \href{http://dx.doi.org/10.1103/PhysRevD.90.054014}{\emph{Phys. Rev.}
  {\bfseries D90} (2014) 054014},
  [\href{https://arxiv.org/abs/1408.1627}{{\ttfamily 1408.1627}}].

\bibitem{Gripaios:2014tna}
B.~Gripaios, M.~Nardecchia and S.~A. Renner, \emph{{Composite leptoquarks and
  anomalies in $B$-meson decays}},
  \href{http://dx.doi.org/10.1007/JHEP05(2015)006}{\emph{JHEP} {\bfseries 05}
  (2015) 006}, [\href{https://arxiv.org/abs/1412.1791}{{\ttfamily 1412.1791}}].

\bibitem{Sahoo:2015wya}
S.~Sahoo and R.~Mohanta, \emph{{Scalar leptoquarks and the rare $B$ meson
  decays}}, \href{http://dx.doi.org/10.1103/PhysRevD.91.094019}{\emph{Phys.
  Rev.} {\bfseries D91} (2015) 094019},
  [\href{https://arxiv.org/abs/1501.05193}{{\ttfamily 1501.05193}}].

\bibitem{Becirevic:2015asa}
D.~Bečirević, S.~Fajfer and N.~Košnik, \emph{{Lepton flavor nonuniversality
  in b→sℓ$^+$ℓ$^-$ processes}},
  \href{http://dx.doi.org/10.1103/PhysRevD.92.014016}{\emph{Phys. Rev.}
  {\bfseries D92} (2015) 014016},
  [\href{https://arxiv.org/abs/1503.09024}{{\ttfamily 1503.09024}}].

\bibitem{Alonso:2015sja}
R.~Alonso, B.~Grinstein and J.~Martin~Camalich, \emph{{Lepton universality
  violation and lepton flavor conservation in $B$-meson decays}},
  \href{http://dx.doi.org/10.1007/JHEP10(2015)184}{\emph{JHEP} {\bfseries 10}
  (2015) 184}, [\href{https://arxiv.org/abs/1505.05164}{{\ttfamily
  1505.05164}}].

\bibitem{Dumont:2016xpj}
B.~Dumont, K.~Nishiwaki and R.~Watanabe, \emph{{LHC constraints and prospects
  for $S_1$ scalar leptoquark explaining the $\bar B \to D^{(*)} \tau \bar\nu$
  anomaly}}, \href{http://dx.doi.org/10.1103/PhysRevD.94.034001}{\emph{Phys.
  Rev.} {\bfseries D94} (2016) 034001},
  [\href{https://arxiv.org/abs/1603.05248}{{\ttfamily 1603.05248}}].

\bibitem{Das:2016vkr}
D.~Das, C.~Hati, G.~Kumar and N.~Mahajan, \emph{{Towards a unified explanation
  of $R_{D^{(\ast)}}$, $R_{K}$ and $(g-2)_{\mu}$ anomalies in a left-right
  model with leptoquarks}},
  \href{http://dx.doi.org/10.1103/PhysRevD.94.055034}{\emph{Phys. Rev.}
  {\bfseries D94} (2016) 055034},
  [\href{https://arxiv.org/abs/1605.06313}{{\ttfamily 1605.06313}}].

\bibitem{Sahoo:2016pet}
S.~Sahoo, R.~Mohanta and A.~K. Giri, \emph{{Explaining the $R_{K}$ and
  $R_{D^{(*)}}$ anomalies with vector leptoquarks}},
  \href{http://dx.doi.org/10.1103/PhysRevD.95.035027}{\emph{Phys. Rev.}
  {\bfseries D95} (2017) 035027},
  [\href{https://arxiv.org/abs/1609.04367}{{\ttfamily 1609.04367}}].

\bibitem{Bhattacharya:2016mcc}
B.~Bhattacharya, A.~Datta, J.-P. Guévin, D.~London and R.~Watanabe,
  \emph{{Simultaneous Explanation of the $R_K$ and $R_{D^{(*)}}$ Puzzles: a
  Model Analysis}},
  \href{http://dx.doi.org/10.1007/JHEP01(2017)015}{\emph{JHEP} {\bfseries 01}
  (2017) 015}, [\href{https://arxiv.org/abs/1609.09078}{{\ttfamily
  1609.09078}}].

\bibitem{Alonso:2016oyd}
R.~Alonso, B.~Grinstein and J.~Martin~Camalich, \emph{{The lifetime of the
  $B_c^-$ meson and the anomalies in $B\to D^{(*)}\tau\nu$}},
  \href{http://dx.doi.org/10.1103/PhysRevLett.118.081802}{\emph{Phys. Rev.
  Lett.} {\bfseries 118} (2017) 081802},
  [\href{https://arxiv.org/abs/1611.06676}{{\ttfamily 1611.06676}}].

\bibitem{Celis:2016azn}
A.~Celis, M.~Jung, X.-Q. Li and A.~Pich, \emph{{Scalar contributions to $b\to c
  (u) \tau \nu$ transitions}},
  \href{http://dx.doi.org/10.1016/j.physletb.2017.05.037}{\emph{Phys. Lett.}
  {\bfseries B771} (2017) 168--179},
  [\href{https://arxiv.org/abs/1612.07757}{{\ttfamily 1612.07757}}].

\bibitem{Altmannshofer:2017wqy}
W.~Altmannshofer, C.~Niehoff and D.~M. Straub, \emph{{$B_s\to\mu^+\mu^-$ as
  current and future probe of new physics}},
  \href{http://dx.doi.org/10.1007/JHEP05(2017)076}{\emph{JHEP} {\bfseries 05}
  (2017) 076}, [\href{https://arxiv.org/abs/1702.05498}{{\ttfamily
  1702.05498}}].

\bibitem{Chen:2017hir}
C.-H. Chen, T.~Nomura and H.~Okada, \emph{{Excesses of muon $g-2$,
  $R_{D^{(\ast)}}$, and $R_K$ in a leptoquark model}},
  \href{https://arxiv.org/abs/1703.03251}{{\ttfamily 1703.03251}}.

\bibitem{Greljo:2015mma}
A.~Greljo, G.~Isidori and D.~Marzocca, \emph{{On the breaking of Lepton Flavor
  Universality in B decays}},
  \href{http://dx.doi.org/10.1007/JHEP07(2015)142}{\emph{JHEP} {\bfseries 07}
  (2015) 142}, [\href{https://arxiv.org/abs/1506.01705}{{\ttfamily
  1506.01705}}].

\bibitem{Alok:2016qyh}
A.~K. Alok, D.~Kumar, S.~Kumbhakar and S.~U. Sankar, \emph{{D* polarization as
  a probe to discriminate new physics in B --> D* tau nubar}},
  \href{https://arxiv.org/abs/1606.03164}{{\ttfamily 1606.03164}}.

\bibitem{Ivanov:2016qtw}
M.~A. Ivanov, J.~G. Körner and C.-T. Tran, \emph{{Analyzing new physics in the
  decays $\bar{B}^0 \to D^{(\ast)}\tau^-\bar\nu_{\tau}$ with form factors
  obtained from the covariant quark model}},
  \href{http://dx.doi.org/10.1103/PhysRevD.94.094028}{\emph{Phys. Rev.}
  {\bfseries D94} (2016) 094028},
  [\href{https://arxiv.org/abs/1607.02932}{{\ttfamily 1607.02932}}].

\bibitem{Faroughy:2016osc}
D.~A. Faroughy, A.~Greljo and J.~F. Kamenik, \emph{{Confronting lepton flavor
  universality violation in B decays with high-$p_T$ tau lepton searches at
  LHC}}, \href{http://dx.doi.org/10.1016/j.physletb.2016.11.011}{\emph{Phys.
  Lett.} {\bfseries B764} (2017) 126--134},
  [\href{https://arxiv.org/abs/1609.07138}{{\ttfamily 1609.07138}}].

\bibitem{Feruglio:2016gvd}
F.~Feruglio, P.~Paradisi and A.~Pattori, \emph{{Revisiting Lepton Flavor
  Universality in B Decays}},
  \href{http://dx.doi.org/10.1103/PhysRevLett.118.011801}{\emph{Phys. Rev.
  Lett.} {\bfseries 118} (2017) 011801},
  [\href{https://arxiv.org/abs/1606.00524}{{\ttfamily 1606.00524}}].

\bibitem{Biancofiore:2014wpa}
P.~Biancofiore, P.~Colangelo and F.~De~Fazio, \emph{{Rare semileptonic $B\to
  K^* \ell^+ \ell^- $ decays in RS$_c$ model}},
  \href{http://dx.doi.org/10.1103/PhysRevD.89.095018}{\emph{Phys. Rev.}
  {\bfseries D89} (2014) 095018},
  [\href{https://arxiv.org/abs/1403.2944}{{\ttfamily 1403.2944}}].

\bibitem{Biancofiore:2014uba}
P.~Biancofiore, P.~Colangelo, F.~De~Fazio and E.~Scrimieri, \emph{{Exclusive $b
  \to s \nu \bar \nu$ induced transitions in RS$_c$ model}},
  \href{http://dx.doi.org/10.1140/epjc/s10052-015-3353-5}{\emph{Eur. Phys. J.}
  {\bfseries C75} (2015) 134},
  [\href{https://arxiv.org/abs/1408.5614}{{\ttfamily 1408.5614}}].

\bibitem{Carmona:2015ena}
A.~Carmona and F.~Goertz, \emph{{Lepton Flavor and Nonuniversality from Minimal
  Composite Higgs Setups}},
  \href{http://dx.doi.org/10.1103/PhysRevLett.116.251801}{\emph{Phys. Rev.
  Lett.} {\bfseries 116} (2016) 251801},
  [\href{https://arxiv.org/abs/1510.07658}{{\ttfamily 1510.07658}}].

\bibitem{GarciaGarcia:2016nvr}
I.~Garc\'{\i}a~Garc\'{\i}a, \emph{{LHCb anomalies from a natural perspective}},
  \href{http://dx.doi.org/10.1007/JHEP03(2017)040}{\emph{JHEP} {\bfseries 03}
  (2017) 040}, [\href{https://arxiv.org/abs/1611.03507}{{\ttfamily
  1611.03507}}].

\bibitem{Bobeth:2007dw}
C.~Bobeth, G.~Hiller and G.~Piranishvili, \emph{{Angular distributions of
  $\bar{B} \to \bar{K} \ell^+\ell^-$ decays}},
  \href{http://dx.doi.org/10.1088/1126-6708/2007/12/040}{\emph{JHEP} {\bfseries
  12} (2007) 040}, [\href{https://arxiv.org/abs/0709.4174}{{\ttfamily
  0709.4174}}].

\bibitem{Bordone:2016gaq}
M.~Bordone, G.~Isidori and A.~Pattori, \emph{{On the Standard Model predictions
  for $R_K$ and $R_{K^*}$}},
  \href{http://dx.doi.org/10.1140/epjc/s10052-016-4274-7}{\emph{Eur. Phys. J.}
  {\bfseries C76} (2016) 440},
  [\href{https://arxiv.org/abs/1605.07633}{{\ttfamily 1605.07633}}].

\bibitem{Das:2014sra}
D.~Das, G.~Hiller, M.~Jung and A.~Shires, \emph{{The $ \overline{B}\to
  \overline{K}\pi \ell \ell $ and $ {\overline{B}}_s\ \to \overline{K}K\ell
  \ell $ distributions at low hadronic recoil}},
  \href{http://dx.doi.org/10.1007/JHEP09(2014)109}{\emph{JHEP} {\bfseries 09}
  (2014) 109}, [\href{https://arxiv.org/abs/1406.6681}{{\ttfamily 1406.6681}}].

\bibitem{Aaij:2015oid}
{\scshape LHCb} collaboration, R.~Aaij et~al., \emph{{Angular analysis of the
  $B^{0} \to K^{*0} \mu^{+} \mu^{-}$ decay using 3 fb$^{-1}$ of integrated
  luminosity}}, \href{http://dx.doi.org/10.1007/JHEP02(2016)104}{\emph{JHEP}
  {\bfseries 02} (2016) 104},
  [\href{https://arxiv.org/abs/1512.04442}{{\ttfamily 1512.04442}}].

\bibitem{CMS:2014xfa}
{\scshape LHCb, CMS} collaboration, V.~Khachatryan et~al., \emph{{Observation
  of the rare $B^0_s\to\mu^+\mu^-$ decay from the combined analysis of CMS and
  LHCb data}}, \href{http://dx.doi.org/10.1038/nature14474}{\emph{Nature}
  {\bfseries 522} (2015) 68--72},
  [\href{https://arxiv.org/abs/1411.4413}{{\ttfamily 1411.4413}}].

\bibitem{Bobeth:2013uxa}
C.~Bobeth, M.~Gorbahn, T.~Hermann, M.~Misiak, E.~Stamou and M.~Steinhauser,
  \emph{{$B_{s,d} \to l^+ l^-$ in the Standard Model with Reduced Theoretical
  Uncertainty}},
  \href{http://dx.doi.org/10.1103/PhysRevLett.112.101801}{\emph{Phys. Rev.
  Lett.} {\bfseries 112} (2014) 101801},
  [\href{https://arxiv.org/abs/1311.0903}{{\ttfamily 1311.0903}}].

\bibitem{Hurth:2016fbr}
T.~Hurth, F.~Mahmoudi and S.~Neshatpour, \emph{{On the anomalies in the latest
  LHCb data}},
  \href{http://dx.doi.org/10.1016/j.nuclphysb.2016.05.022}{\emph{Nucl. Phys.}
  {\bfseries B909} (2016) 737--777},
  [\href{https://arxiv.org/abs/1603.00865}{{\ttfamily 1603.00865}}].

\bibitem{Mahmoudi:2016mgr}
F.~Mahmoudi, T.~Hurth and S.~Neshatpour, \emph{{Present Status of $b \to s
  \ell^+ \ell^−$ Anomalies}},
  \href{http://dx.doi.org/10.1016/j.nuclphysbps.2017.03.008}{\emph{Nucl. Part.
  Phys. Proc.} {\bfseries 285-286} (2017) 39--44},
  [\href{https://arxiv.org/abs/1611.05060}{{\ttfamily 1611.05060}}].

\bibitem{Capdevila:2016ivx}
B.~Capdevila, S.~Descotes-Genon, J.~Matias and J.~Virto, \emph{{Assessing
  lepton-flavour non-universality from $B\to K^*\ell\ell$ angular analyses}},
  \href{http://dx.doi.org/10.1007/JHEP10(2016)075}{\emph{JHEP} {\bfseries 10}
  (2016) 075}, [\href{https://arxiv.org/abs/1605.03156}{{\ttfamily
  1605.03156}}].

\bibitem{Capdevila:2016fhq}
B.~Capdevila, S.~Descotes-Genon, L.~Hofer, J.~Matias and J.~Virto, \emph{{$B
  \to K^*(\to K\pi) \ell^+\ell^-$ theory and the global picture: What's
  next?}}, {\emph{PoS} {\bfseries LHCP2016} (2016) 073},
  [\href{https://arxiv.org/abs/1609.01355}{{\ttfamily 1609.01355}}].

\bibitem{Altmannshofer:2017fio}
W.~Altmannshofer, C.~Niehoff, P.~Stangl and D.~M. Straub, \emph{{Status of the
  $B\rightarrow K^*\mu ^+\mu ^-$ anomaly after Moriond 2017}},
  \href{http://dx.doi.org/10.1140/epjc/s10052-017-4952-0}{\emph{Eur. Phys. J.}
  {\bfseries C77} (2017) 377},
  [\href{https://arxiv.org/abs/1703.09189}{{\ttfamily 1703.09189}}].

\bibitem{ALEPH:2005ab}
{\scshape SLD Electroweak Group, DELPHI, ALEPH, SLD, SLD Heavy Flavour Group,
  OPAL, LEP Electroweak Working Group, L3} collaboration, S.~Schael et~al.,
  \emph{{Precision electroweak measurements on the $Z$ resonance}},
  \href{http://dx.doi.org/10.1016/j.physrep.2005.12.006}{\emph{Phys. Rept.}
  {\bfseries 427} (2006) 257--454},
  [\href{https://arxiv.org/abs/hep-ex/0509008}{{\ttfamily hep-ex/0509008}}].

\bibitem{Falkowski:2017pss}
A.~Falkowski, M.~Gonz‡lez-Alonso and K.~Mimouni, \emph{{Compilation of
  low-energy constraints on 4-fermion operators in the SMEFT}},
  \href{https://arxiv.org/abs/1706.03783}{{\ttfamily 1706.03783}}.

\bibitem{Aaboud:2016cth}
{\scshape ATLAS} collaboration, M.~Aaboud et~al., \emph{{Search for high-mass
  new phenomena in the dilepton final state using proton-proton collisions at
  $\sqrt{s}=13$ TeV with the ATLAS detector}},
  \href{http://dx.doi.org/10.1016/j.physletb.2016.08.055}{\emph{Phys. Lett.}
  {\bfseries B761} (2016) 372--392},
  [\href{https://arxiv.org/abs/1607.03669}{{\ttfamily 1607.03669}}].

\bibitem{Khachatryan:2016qkc}
{\scshape CMS} collaboration, V.~Khachatryan et~al., \emph{{Search for heavy
  resonances decaying to tau lepton pairs in proton-proton collisions at $
  \sqrt{s}=13 $ TeV}},
  \href{http://dx.doi.org/10.1007/JHEP02(2017)048}{\emph{JHEP} {\bfseries 02}
  (2017) 048}, [\href{https://arxiv.org/abs/1611.06594}{{\ttfamily
  1611.06594}}].

\bibitem{Aad:2015fna}
{\scshape ATLAS} collaboration, G.~Aad et~al., \emph{{A search for $
  t\overline{t} $ resonances using lepton-plus-jets events in proton-proton
  collisions at $ \sqrt{s}=8 $ TeV with the ATLAS detector}},
  \href{http://dx.doi.org/10.1007/JHEP08(2015)148}{\emph{JHEP} {\bfseries 08}
  (2015) 148}, [\href{https://arxiv.org/abs/1505.07018}{{\ttfamily
  1505.07018}}].

\bibitem{Lillie:2007yh}
B.~Lillie, L.~Randall and L.-T. Wang, \emph{{The Bulk RS KK-gluon at the LHC}},
  \href{http://dx.doi.org/10.1088/1126-6708/2007/09/074}{\emph{JHEP} {\bfseries
  09} (2007) 074}, [\href{https://arxiv.org/abs/hep-ph/0701166}{{\ttfamily
  hep-ph/0701166}}].

\bibitem{Chatrchyan:2013lca}
{\scshape CMS} collaboration, S.~Chatrchyan et~al., \emph{{Searches for new
  physics using the $t\bar{t}$ invariant mass distribution in pp collisions at
  $\sqrt{s}$=8  TeV}},
  \href{http://dx.doi.org/10.1103/PhysRevLett.111.211804,
  10.1103/PhysRevLett.112.119903}{\emph{Phys. Rev. Lett.} {\bfseries 111}
  (2013) 211804}, [\href{https://arxiv.org/abs/1309.2030}{{\ttfamily
  1309.2030}}].

\bibitem{Agashe:2006hk}
K.~Agashe, A.~Belyaev, T.~Krupovnickas, G.~Perez and J.~Virzi, \emph{{LHC
  Signals from Warped Extra Dimensions}},
  \href{http://dx.doi.org/10.1103/PhysRevD.77.015003}{\emph{Phys. Rev.}
  {\bfseries D77} (2008) 015003},
  [\href{https://arxiv.org/abs/hep-ph/0612015}{{\ttfamily hep-ph/0612015}}].

\bibitem{Greljo:2017vvb}
A.~Greljo and D.~Marzocca, \emph{{High-$p_T$ dilepton tails and flavour
  physics}},  \href{https://arxiv.org/abs/1704.09015}{{\ttfamily 1704.09015}}.

\bibitem{Isidori:2015oea}
G.~Isidori, \emph{{Flavour Physics and Implication for New Phenomena}},
  \href{http://dx.doi.org/10.1142/9789814733519_0017}{\emph{Adv. Ser. Direct.
  High Energy Phys.} {\bfseries 26} (2016) 339--355},
  [\href{https://arxiv.org/abs/1507.00867}{{\ttfamily 1507.00867}}].

\bibitem{Lees:2013kla}
{\scshape BaBar} collaboration, J.~P. Lees et~al., \emph{{Search for $B \to
  K^{(*)} \nu \overline \nu$ and invisible quarkonium decays}},
  \href{http://dx.doi.org/10.1103/PhysRevD.87.112005}{\emph{Phys. Rev.}
  {\bfseries D87} (2013) 112005},
  [\href{https://arxiv.org/abs/1303.7465}{{\ttfamily 1303.7465}}].

\bibitem{TheBaBar:2016xwe}
{\scshape BaBar} collaboration, J.~P. Lees et~al., \emph{{Search for
  $B^{+}\rightarrow K^{+} \tau^{+}\tau^{-}$ at the BaBar experiment}},
  \href{http://dx.doi.org/10.1103/PhysRevLett.118.031802}{\emph{Phys. Rev.
  Lett.} {\bfseries 118} (2017) 031802},
  [\href{https://arxiv.org/abs/1605.09637}{{\ttfamily 1605.09637}}].

\bibitem{Bouchard:2013mia}
{\scshape HPQCD} collaboration, C.~Bouchard, G.~P. Lepage, C.~Monahan, H.~Na
  and J.~Shigemitsu, \emph{{Standard Model Predictions for $B \to K \ell^+
  \ell^-$ with Form Factors from Lattice QCD}},
  \href{http://dx.doi.org/10.1103/PhysRevLett.112.149902,
  10.1103/PhysRevLett.111.162002}{\emph{Phys. Rev. Lett.} {\bfseries 111}
  (2013) 162002}, [\href{https://arxiv.org/abs/1306.0434}{{\ttfamily
  1306.0434}}].

\bibitem{Amhis:2016xyh}
Y.~Amhis et~al., \emph{{Averages of $b$-hadron, $c$-hadron, and $\tau$-lepton
  properties as of summer 2016}},
  \href{https://arxiv.org/abs/1612.07233}{{\ttfamily 1612.07233}}.

\bibitem{Papucci}
F.~U. Bernlochner, Z.~Ligeti, M.~Papucci and D.~J. Robinson, \emph{{Combined
  analysis of semileptonic $B$ decays to $D$ and $D^*$: $R(D^{(*)})$,
  $|V_{cb}|$, and new physics}},
  \href{http://dx.doi.org/10.1103/PhysRevD.95.115008}{\emph{Phys. Rev.}
  {\bfseries D95} (2017) 115008},
  [\href{https://arxiv.org/abs/1703.05330}{{\ttfamily 1703.05330}}].

\bibitem{Pich:2013lsa}
A.~Pich, \emph{{Precision Tau Physics}},
  \href{http://dx.doi.org/10.1016/j.ppnp.2013.11.002}{\emph{Prog. Part. Nucl.
  Phys.} {\bfseries 75} (2014) 41--85},
  [\href{https://arxiv.org/abs/1310.7922}{{\ttfamily 1310.7922}}].

\bibitem{Megias:2017dzd}
E.~Megias, M.~Quiros and L.~Salas, \emph{{$g_\mu-2$ from Vector-Like Leptons in
  Warped Space}}, \href{http://dx.doi.org/10.1007/JHEP05(2017)016}{\emph{JHEP}
  {\bfseries 05} (2017) 016},
  [\href{https://arxiv.org/abs/1701.05072}{{\ttfamily 1701.05072}}].

\end{thebibliography}\endgroup


\end{document}